\documentclass[10pt]{article}

\usepackage{vmargin}   
\setmarginsrb{3.2cm}{2.5cm}{3.2cm}{2.5cm}{1cm}{1cm}{1cm}{1.6cm}   
\usepackage[fleqn]{amsmath}   
\usepackage{amssymb,amsthm}  
\usepackage{wasysym}

\newcommand{\ev}{\mathrm{ev}}
\newcommand{\Ad}{\mathrm{Ad}}
\newcommand{\ad}{\mathrm{ad}}
\newcommand{\dd}{\mathrm{d}}
\newcommand{\hyprs}{{}_r\phi_s \Big( 
          \begin{matrix}
            \, a_1, \ldots ,  a_r \\
            \,  b_1,\ldots,b_s
          \end{matrix} \; \Big|\; 
	  q , z \Big) 
	  }
\newcommand{\hyp}[5]{{}_2\phi_1 \Big( 
          \begin{matrix}
            \, #1 \, , \,  #2 \\
             #3
          \end{matrix} \; \Big|\; 
	  #4 , #5 \Big) 
	  }
\newcommand{\hyper}[3]{{}_0\phi_1 \Big( 
          \begin{matrix}
            \, - \\
            \, #1
          \end{matrix} \; \Big|\; 
	  #2 , #3 \Big) 
	  }
\newcommand{\W}[5]{W \Big[ 
          \,
          \begin{matrix}
             #1 & #2 \\
             #3 & #4
          \end{matrix}  \,\Big| 
	   \,  #5 \, \Big] 
	  }
\def\eps{\epsilon}
\newcommand{\IRF}{\mathrm{IRF}}
\newcommand{\Ver}{\mathrm{8V}}
\newcommand{\re}{\mathrm{re}}
\newcommand{\im}{\mathrm{im}}

\newcommand{\Vect}{\mathrm{Vect}}
\newcommand{\Aut}{\mathrm{Aut}}

\newcommand{\utheta}{\underline\theta}
\newcommand{\uphi}{\underline\varphi}
\newcommand{\upsi}{\underline\psi}
\newcommand{\uS}{\underline{S}}
\newcommand{\uW}{\underline{W}}
\newcommand{\uP}{\underline{P}}
\newcommand{\uQ}{\underline{Q}}

\newcommand{\uR}{\underline{R}}
\newcommand{\mC}{\mathfrak{C}}

\def\onehalf{\frac{1}{2}}
\def\g{\mathfrak g}
\def\n{\mathfrak n}
\def\h{\mathfrak h}
\def\b{\mathfrak b}
\def\l{\mathfrak l}
\def\C{\mathbb C}
\def\N{\mathbb N}
\def\Z{\mathbb Z}
\def\A{\mathbb A}

\newtheorem{definition}{\sf Definition}
\newtheorem{proposition}{\sf Proposition}
\newtheorem{theorem}{\sf Theorem}
\newtheorem{lemma}{\sf Lemma}
\newtheorem{conjecture}{\sf Conjecture}

\newcounter{Prob}
\newcounter{Prop}
\newcounter{Theo}
\newcounter{Def}
\newcounter{Lemm} 
   
{\theoremstyle{remark}
\newtheorem{rem}{Remark}[section]}

\def\proof{\medskip\noindent {\sf{\underline{Proof}:\ }}}
\def\cqfd{\hfill\nobreak\hbox{$\Box$}\par\bigbreak}

\def\id{{\rm id}}
\def\End{{\rm End}}

\def\deg{{\rm deg}}

\title{\bf Universal Vertex-IRF  Transformation for Quantum Affine Algebras}
\author{
E. Buffenoir\thanks{Universit\'e Montpellier 2, CNRS,  LPTA, UMR 5207, France, Eric.Buffenoir@lpta.univ-montp2.fr}, 
Ph. Roche\thanks{Universit\'e Montpellier 2, CNRS,  LPTA, UMR 5207, France, Philippe.Roche@lpta.univ-montp2.fr},
V. Terras\thanks{Universit\'e de Lyon, ENS Lyon, CNRS, Laboratoire de Physique, UMR 5672, France, on leave of absence from  Universit\'e Montpellier 2, CNRS,  LPTA, UMR 5207, France, Veronique.Terras@lpta.univ-montp2.fr}
}  
\date{\today}

\begin{document}

\maketitle  

\begin{abstract} 
We construct a universal Vertex-IRF transformation between   Vertex
type universal solution and   Face type universal solution of the quantum
dynamical Yang-Baxter equation. This universal Vertex-IRF
transformation satisfies the generalized coBoundary equation and is an
extension  of our previous work to the quantum affine
$U_q(A^{(1)}_r)$ case. This solution  has a simple Gauss decomposition
which is constructed using Sevostyanov's characters  of twisted quantum Borel algebras.
We show that the evaluation of this universal solution in the evaluation representation of 
$U_q(A_1^{(1)})$ gives the standard Baxter's transformation between the 8-Vertex model and the IRF height model.
\end{abstract}

\section{Introduction}
A fundamental result in the study of  solutions of the Quantum Yang-Baxter Equation (QYBE)  is  the work of Belavin-Drinfeld \cite{BD}. This fundamental theorem classifies the solutions of the Classical Yang-Baxter Equation (CYBE) with and without spectral parameter. 
Among these solutions one can single out the elliptic solution associated to 
$A_r$, which admits an elliptic parametrization in term of the spectral
 parameter. 

An important problem is to construct {\it universal} solutions of the QYBE  associated 
to these classical solutions, such that the matricial solutions are just obtained 
by evaluation of the universal solution under a representation. 
The $(r+1)\times(r+1)$ elliptic matricial solutions is known as the Baxter-Belavin 
solution \cite{Bax0, Bel}. It took 15 years of work from the Belavin-Drinfeld 
result  to obtain an explicit construction of universal solutions. In order to do 
so, one has to enlarge the picture to also include  solutions of the Quantum 
Dynamical Yang-Baxter Equation (QDYBE). 

The basis of this previous  subject was settled in the work of G. Felder \cite{Fe}. There, it was recognized that the known  matricial solutions of Interaction-Round-Faces 
(IRF) statistical models were matricial solutions of the QDYBE. 
Among these models, the archetypal ones were defined by Baxter \cite{Bax} and 
Jimbo, Miwa, Okado \cite{JMO}. These models also have an elliptic parametrization.

The analog of the work of Belavin-Drinfeld in the dynamical case was
accomplished by Etingof-Varchenko \cite{EV} and O. Schiffmann
\cite{Sch}, who classified the solutions of the Classical Dynamical Yang-Baxter Equation (CDYBE). 
The construction of the universal solutions associated to some of these solutions 
were done in  \cite{Bab}, \cite{JKOS}, \cite{ABRR}, \cite{ESS}:
\cite{Bab} contains the first construction of the universal solution of the QDYBE in the $sl(2)$ case; \cite{JKOS} contains the construction, in the affine case, of universal solutions of the QDYBE associated to the elliptic IRF models  and the Belavin-Baxter solutions;  \cite{ABRR} introduces a linear equation as a tool for solving the QDYBE; \cite{ESS}  obtains a quantization of all solutions of the CDYBE of \cite{Sch} using a modified linear equation. This last work solves in particular the problem of finding explicit universal solutions of the QYBE associated to the classical solutions of the Belavin-Drinfeld theorem without spectral parameter. 

A question which is finally little understood is the study of the QDYBE up to {\it all} allowed  dynamical gauge transformations. For example, if $R(x)$ is a solution of the QDYBE with dynamics $\h$ and is $\h$-invariant,     
$\underline{R}(x)=M_1(x)^{-1} M_2(xq^{h_1})^{-1} {R}(x)\,M_1(xq^{h_2})\,M_2(x)$  also satisfies the QDYBE if $M(x)$ commutes
with $\h,$ but what happens if we do not impose  this last assumption?  

{}From \cite{BDF}, one knowns which standard solution  of the CDYBE, in
the finite dimensional case, can be   dynamically  gauge transformed to   a non dynamical
solution of the CYBE: this is possible only in the case of $A_r$, and the non
dynamical solution has to be of Cremmer-Gervais's type.
In a previous work \cite{BRT}, we studied this particular problem in the quantum
case and gave a construction of the universal solution $M(x),$ that we
called Quantum Dynamical coBoundary Element.

Such type of result exists also in the affine case for matrix solutions. 
It is known \cite{Bax,JMO2} that, in the $A_r^{(1)}$ case, a  dynamical gauge 
matrix relates the $A_r^{(1)}$ height model of \cite{JMO}  (the dynamics of which is the whole Cartan algebra) to the Belavin-Baxter solution \cite{Bel} (the 
dynamics of which is only along the line generated by the central element). 
This Vertex-IRF transformation is important because it maps the Belavin-Baxter model which has no $\h$-invariance to a face model which has an $\h$-invariance and for which Bethe Ansatz techniques can then be applied.

For example in the case where $r=1,$ the Belavin-Baxter solution reduces to the $8$-Vertex $R$-matrix ${\bf R}^{\Ver}(z_1,z_2;p)$, where $z_1,z_2$ are the spectral parameters and $p$ is the elliptic nome. The IRF (Interaction-Round-Faces) matrix solution ${\bf R}^{\IRF}(z;p,w)$, in which $w$ is the extra dynamical parameter, corresponds to the Andrews-Baxter-Forrester height model \cite{ABF}, or solid-on-solid (SOS) model: this model is defined on a square lattice to each site of which one associates a height $s$ such that, if $s_1$ and $s_2$ correspond to adjacent sites, one has $s_1-s_2=\pm 1$. Interactions are described by weights $\W{s_1}{s_2}{s_3}{s_4}{z}$ (with $|s_1-s_2|=|s_2-s_3|=|s_3-s_4|=|s_4-s_1|=1$) associated to faces and given by the matrix elements of ${\bf R}^{\IRF}$:
\begin{equation*} 
  {\bf R}^{\IRF}(z;p,wq^s)
   =\sum_{i_1,j_1,i_2,j_2=1}^2 \W{s}{s+\eps'_1}{s+\eps_2}{s+\eps_1+\eps_2}{z}\;
   E_{i_1, j_1}\otimes E_{i_2, j_2},
\end{equation*} 
with $\eps_\alpha=3-2 i_\alpha$, $\eps'_\alpha=3-2 j_\alpha$, for $\alpha=1,2$ ($\eps_\alpha,\eps'_\alpha\in\{1,-1\}$).
In this framework, the
Quantum Dynamical Yang-Baxter Equation for ${\bf R}^{\IRF}(z;p,w)$ translates into Baxter's star-triangle relation for the Boltzmann weights $W$ as shown in \cite{Fe}.

In this context, the Vertex-IRF transformation is a $2\times 2$ matrix 
$\mathbf{S}(z;p,w)$ which satisfies 
\begin{equation*}
     \mathbf{S}_1 (z_1;p,w)\,\mathbf{S}_2 (z_2;p,wq^{h_1})\,
        \mathbf{R}^{\IRF}(z_1/z_2;p,w) 
     ={\bf R}^{\Ver}(z_1,z_2;p)\,
        \mathbf{S}_2 (z_2;p,w)\,\mathbf{S}_1 (z_1;p,wq^{h_2}).
\end{equation*} 
 This transformation is usually written in the following form, involving the column vectors $\Phi^{(+)}(z;p,w)$ and $\Phi^{(-)}(z;p,w)$ of the matrix $\mathbf{S}(z;p,w)$:
\begin{multline}\label{vertex-IRF}
   {\bf R}^{\Ver}(z_1,z_2;p)\, \big[ \Phi^{(l'-l)}\big(z_1;p,w q^{m'-l'}\big)
              \otimes\Phi^{(m'-l')} (z_2;p,w )\big]\\
   =\sum_m  \W{l}{l'}{m}{m'}{\frac{z_1}{z_2}}\,
     \big[\Phi^{(m'-m)} (z_1;p,w )\otimes
                   \Phi^{(m-l)}\big(z_2;p,w q^{m'-m}\big)\big],
\end{multline}
for all values of the heights $l$, $l'$, $m'$ such that $|l-l'|=|m'-l'|=1$, where the summation is over values of the height $m$ such that $|m-l|=|m-m'|=1$. 
This transformation can similarly be expressed in terms of the inverse $\mathbf{M} (z;p,w)$ of the matrix $\mathbf{S}(z;p,w)$,
\begin{equation*}
  \mathbf{R}^{\IRF}(z_1/z_2;p,w) \,\mathbf{M}_1 (z_1;p,wq^{h_2})\,\mathbf{M}_2 (z_2;p,w)
     =\mathbf{M}_2 (z_2;p,wq^{h_1})\,\mathbf{M}_1 (z_1;p,w)\,{\bf R}^{\Ver}(z_1,z_2;p),
\end{equation*} 
which is  usually written in terms of the line vectors $\widetilde\Phi^{(+)}(z;p,w)$ and $\widetilde\Phi^{(-)}(z;p,w)$ of the matrix $\mathbf{M}(z;p,w)$,
\begin{multline} \label{vertex-IRF-dual}
  \big[ \widetilde\Phi^{(m'-l')}\big(z_1;p,w \big)
              \otimes\widetilde\Phi^{(l'-l)} (z_2;p,w q^{m'-l'} )\big]\,{\bf R}^{\Ver}(z_1,z_2;p)\\
  =\sum_{m}  \W{l}{m}{l'}{m'}{\frac{z_1}{z_2}}\,
     \big[\widetilde\Phi^{(m-l)} (z_1;p,w q^{m'-m})\otimes
          \widetilde\Phi^{(m'-m)}\big(z_2;p,w \big)\big],
\end{multline}
where the summation is taken as in \eqref{vertex-IRF}.
This vertex-IRF transformation, first obtained in \cite{Bax}, is the core of Baxter's solution of the 8-Vertex model.

In the present work, we construct the universal dynamical gauge
transformation which relates the universal solutions associated to
these two models. The result that we obtain is an extension of our
previous work to the affine case.

\bigskip

Our article is organized as follows.
 
In Section 2 we recall some results on quantum affine algebras.

Section 3 contains results on Dynamical Quantum Groups. We
introduce the notion of generalized translation datum, and associate
to it a linear equation, the triangular solutions of which satisfy the Quantum 
Dynamical coCycle equation. We formulate the problem of Vertex-IRF
transformation and the related Generalized Dynamical coBoundary
notion.

In Section 4 we study the universal Vertex-IRF transformation $M(x)$.
It admits a Gauss decomposition which can be expressed in terms of infinite
products constructed from  basic elements $\mathfrak{C}^{[+]}(x)$ and $\mathfrak{C}^{[-]}(x).$ 
We give sufficient conditions on  $\mathfrak{C}^{[\pm]}(x)$ for  $M(x)$
to be  a Vertex-IRF transformation, and even stronger to be a
generalized Quantum Dynamical coBoundary.

In the case of $A_r^{(1)}$ and for the Belavin-Drinfeld solution,
the element $\mathfrak{C}^{[\pm]}(x)$ are constructed in Section 5 using  particular
types of Sevostyanov's characters \cite{Sev} of twisted version of quantum Borel algebras.
Among the sufficient conditions, the ``hexagonal relation'' requires
special attention. Its proof needs the explicit computation of 
 $\mathfrak{C}^{\pm}(x).$
We provide this explicit computation and shows that the hexagonal
relation is satisfied.

The last part of this section is devoted to the computation in the
$A_1^{(1)}$ case of the universal Vertex-IRF transformation
represented in the evaluation representation. As expected, one recovers
the known Vertex-IRF transformation between the 8-Vertex model and the IRF height model.

\section{Results on  Quantum Affine Universal Envelopping Algebras}
\label{sectionQAUEA}

\subsection{Definitions and generalities}

Let  $\g$ be  a Kac-Moody Lie algebra of simply laced finite type or non twisted simply laced affine type. To  $\g$ we  associate, as usual, a finite dimensional simple simply laced Lie algebra  denoted $\overset{\circ}{\g}:$ 
in the finite dimensional case $\overset{\circ}{\g}=\g$,  and in the affine case we have the usual realisation
${\g}= \overset{\circ}{\g}[t,t^{-1}]\oplus \C d \oplus \C c.$ 

 We denote $\overset{\circ}{\g}=\overset{\circ}{\n}_-
\oplus \overset{\circ}{\h} \oplus \overset{\circ}{\n}_+$ a Cartan decomposition.
Let $\overset{\circ}{\Gamma}=\{\alpha_1,\ldots,\alpha_r\}$ be the Dynkin diagram, 
$\overset{\circ}{\Delta}\subset\overset{\circ}{\h}{}^*$ the root system and  
$\overset{\circ}{\Delta}{}_+$ the set of positive roots. 
We denote $\theta \in \overset{\circ}{\Delta}{}_+$ the maximal root, it defines positive integers $a_i$ by $\theta=\sum_{i=1}^r a_i\alpha_i \in \overset{\circ}{\Delta}{}_+.$  
As usual, we define
$\overset{\circ}{\rho}=\frac{1}{2}\sum_{\alpha\in 
\overset{\circ}{\Delta}{}^+} \alpha$. 

Let $(\cdot,\cdot)$ be the  non degenerate invariant bilinear form on $\overset{\circ}{\g}$,
normalized in such a way that 
the induced form on $\overset{\circ}{\h}{}^*$, also denoted  $(\cdot,\cdot)$,
satisfies the equation $(\alpha,\alpha)=2$ for all roots $\alpha$.
This bilinear form induces an isomorphism $\overset{\circ}{\nu}:\overset{\circ}{\h}\rightarrow
\overset{\circ}{\h}{}^*.$
To a root $\alpha\in\overset{\circ}{\h}{}^*$ one associates the coroot 
$\alpha^{\vee}=\frac{2}{(\alpha,\alpha)}\overset{\circ}{\nu}{}^{-1}(\alpha)=\overset{\circ}{\nu}{}^{-1}(\alpha)$, and denote $h_{\alpha_i}=\alpha_i^{\vee}.$
Let $\lambda_1,\ldots,\lambda_r\in\overset{\circ}{\h}{}^*$ be the fundamentals weights, {\it i.e.} $(\lambda_i)_i$ is the dual basis to $(h_{\alpha_i})_i.$ We denote $\overset{\circ}{\zeta}{}^{\alpha_i}=\overset{\circ}{\nu}{}^{-1}(\lambda_i).$
Finally, let  $\Omega_{\overset{\circ}{\h}}\in\overset{\circ}{\h}\otimes\overset{\circ}{\h}$ 
be the inverse element to the form 
$(\cdot,\cdot)$ on $\overset{\circ}{\h}$; we have $\Omega_{\overset{\circ}{\h}}=\sum_{i=1}^r \overset{\circ}{\zeta}{}^{\alpha_i}\otimes h_{\alpha_i}.$

\bigskip

If $\g$ is a non twisted affine Lie algebra, we have  ${\g}=\overset{\circ}{\g}[t,t^{-1}]\oplus \C d
\oplus \C c$ as a vector space,  where $c$ is the central element and $d$ is the grading
element. The Lie algebra structure is defined such that for Laurent homogeneous polynomials $a(t)=a\otimes t^m,b(t)=b\otimes t^n$, one has
\begin{equation*}
  [a(t),b(t)]=[a,b]\otimes t^{m+n}+\text{Res}_0(a'(t),b(t))c,
\end{equation*}
and $[d,a(t)]=ta'(t)$, $[d,c]=0$.
The derived subalgebra ${\g}'$ is equal to
$\g[t,t^{-1}]\oplus \C c.$

The Cartan subalgebra of ${\g}$ is ${\h}=\overset{\circ}{\h} \oplus 
\C c \oplus\C d$.
The dual of the Cartan subalgebra is given by 
${\h}^*=\overset{\circ}{\h}{}^* \oplus \C \gamma \oplus \C \delta$, where $ \gamma,\delta$ are defined by the relations $\langle \gamma,
\overset{\circ}{\h}\rangle=\langle \gamma,d\rangle=\langle\delta,\overset{\circ}{\h}\rangle=\langle
\delta,c\rangle=0$ and $\langle \gamma,c \rangle=\langle \delta,d
\rangle=1$.
We denote $ \alpha_0=\delta-\theta$, and $h_{\alpha_0}=c-\theta^{\vee}$.
We also define the Cartan matrix of 
${\g}$, with elements $a_{ij}=\langle h_{\alpha_i}, \alpha_{j}\rangle$, $i,j=0,\ldots,r $, 
and denote  $g=1+\sum_{i=1}^r a_i$ the dual  Coxeter number of $\g.$
We denote $Q=\oplus_{i=0}^r
{\mathbb Z}\alpha_i$ the root lattice and $Q^+=\oplus_{i=0}^r
{\mathbb Z}^+\alpha_i.$ 

$(\cdot,\cdot)$ can be extended to a non degenerate symmetric bilinear form on
${\h}$ by
$(h_{\alpha_i},h_{\alpha_j})=\langle\alpha_i,h_{\alpha_j}\rangle$, $(d,h_{\alpha_i})=\delta_{i,0}$, and $(d,d)=0.$ 
It defines an isomorphism  ${\nu}:{\h}\rightarrow
{\h}^{*}.$ 

Let $\Lambda_0=\gamma,\,\Lambda_1,\ldots, \Lambda_r\in \h^*$ be the fundamental weights,  {\it i.e.}
$\langle \Lambda_i,h_{\alpha_j}\rangle=\delta_{ij}$, and
$\langle \Lambda_i,d\rangle=0.$ 
We define ${\rho}=\sum_{i=0}^r\Lambda_i$;
we have ${\rho}=\overset{\circ}{\rho}+g \delta.$
We denote ${\zeta}^{\alpha_i}={\nu}^{-1}(\Lambda_i)$, 
${\zeta}^d={\nu}^{-1}(\delta)$ and  $\varpi={\nu}^{-1}
({\rho}).$
By construction, $({\zeta}^{\alpha_0},\ldots, {\zeta}^{\alpha_r}, {\zeta}^d)$ is the dual basis of $(h_{\alpha_0},\ldots, h_{\alpha_r},d)$ with respect to 
$(.,. )$, and
we have ${\zeta}^{\alpha_0}=d$, ${\zeta}^{\alpha_i}=\overset{\circ}{\zeta}{}^{\alpha_i}+a_i d$,
${\zeta}^d=c.$
We denote $\Omega_{{\h}}$ the   
 inverse element to the form 
$(\cdot,\cdot)$ on ${\h},$ which is given as $\Omega_{{\h}}=
\sum_{i=0}^r {\zeta}^{\alpha_i}\otimes h_{\alpha_i}+{\zeta}^d\otimes d=
c\otimes d+d\otimes c+\Omega_{\overset{\circ}{\h}}.$

Let ${\Gamma}$ and ${\Delta}$ be the
Dynkin diagram and root system of $\g$. 
Thus ${\Gamma}=
\overset{\circ}{\Gamma} \cup
\{\alpha_0\},$ ${\Delta}=
(\overset{\circ}{\Delta}+\Z \delta) \cup (\Z\setminus\{0\}) \delta.$ 
  
We denote ${\Delta}_{\re}$ the set of real roots 
and ${\Delta}_{\im}$  the set of imaginary roots. 
The set of positive real roots is
 ${\Delta}_{\re}^+=\overset{\circ}{\Delta}{}^+\cup (\overset{\circ}{\Delta}+\Z^{+*} \delta)$
and the set of positive imaginary roots is ${\Delta}_{\im}^+=\Z^{+*} \delta$.
The multiplicity of a real root is $1$  whereas the multiplicity of an imaginary root is $\dim \h$.
We therefore define   $\overline{\Delta}_{\im}^+={\Delta}_{\im}^+\times \{1,\ldots,\dim \h \},$ and  $\overline{\Delta}^+={\Delta}_{\re}^+\cup \overline{\Delta}_{\im}^+.$

Note that in the case where $\g$ is of finite type all the roots are real and the set of imaginary roots 
is empty.  
\bigskip

Let $q$ be a nonzero complex number such that $|q|<1$.
Let $U_q({\g})$ be the quantized affine algebra corresponding to $\g$: 
it is the  algebra over $\C$ with generators $e_{i}$, $f_{i}$, $i=0,1,\ldots,r$ 
and $q^{hh'}=q^{h'h}$, $h,h' \in {\h}\oplus\C$, and with relations
\begin{alignat}{2}
  &q^{(x+y)z}=q^{xz}q^{yz}, \quad x,y,z\in {\h}, & &q^0=1,
                        \\
  &q^{hh'} e_{i} =e_i \, q^{(h+\alpha_i(h))(h'+\alpha_i(h'))}, & 
             &q^{hh'}  f_{i} =f_{i}\, q^{(h-\alpha_i(h))(h'-\alpha_i(h'))},
               \displaybreak[0]\\
  &e_{i}\,f_{j}-f_{j}\, e_{i}=\delta_{ij}
            \frac{q^{h_{\alpha_i}}-q^{-h_{\alpha_i}}}{q-q^{-1}}, & &
                \displaybreak[0]\\
  &\sum_{k=0}^{1-a_{ij}} (-1)^k \bmatrix 1-a_{ij} \\ k \endbmatrix_{q} 
                       e_{i}^{1-a_{ij}-k}e_{j}e_{i}^k=0, & &\quad i\ne j,\\
  &\sum_{k=0}^{1-a_{ij}} (-1)^k \bmatrix 1-a_{ij} \\ k \endbmatrix_{q} 
                       f_{i}^{1-a_{ij}-k}f_{j}f_{i}^k=0, & &\quad i\ne j,
\end{alignat}
where
\begin{equation*} 
   \bmatrix n \\ k \endbmatrix_q=\frac{[n]_q!}{[k]_q! [n-k]_q!}, \qquad 
   [n]_q! = [1]_q[2]_q\cdots [n]_q, \qquad 
   [n]_q= \frac {q^n - q^{-n}}{q-q^{-1}}.
\end{equation*}
Here we have enlarged the usual version of $U_q(\g)$ by adjoining the  elements 
$q^{hh'}$, $h,h'\in \h,$  and have extended the action of the roots on $\h\oplus \C$ by $\alpha(h+\lambda 1)=\alpha(h).$

The subalgebra of $U_q(\g)$ generated by $q^h$, $h\in\h$ and $e_i,\, f_i$, $i=0,1,\ldots,r$, is a Hopf algebra with comultiplication $\Delta$  given by 
\begin{equation}
  \Delta(e_{i}) = e_{i}\otimes q^{h_{\alpha_i}} + 1\otimes e_{i}, \quad
  \Delta(f_{i}) = f_{i}\otimes 1 + q^{-h_{\alpha_i}}\otimes f_{i},\quad 
  \Delta(q^h) = q^h \otimes q^h.
\end{equation}
Note that $U_q(\g)$ itself is not a  Hopf algebra  because formally
$\Delta(q^{h^2})=(q^{h^2}\otimes q^{h^2})\, q^{2h\otimes h}.$
Therefore one extends the algebra $U_q(\g)\otimes U_q(\g)$ by adjoining the elements $q^{h\otimes h'}, h,h'\in \h$ (whose commutations relations with the generators of $U_q(\g)\otimes U_q(\g)$ are straightforward to define) and we denote $U_q(\g)\widehat{\otimes }U_q(\g)$ this algebra.  We hence obtain a well defined map, morphism of algebra, $\Delta: U_q(\g)\rightarrow   U_q(\g)\widehat{\otimes } U_q(\g).$ Following the usage in the litterature we will still name  $U_q(\g)$ a Hopf algebra.

$U_q(\g)$ is a $Q$-graded algebra, we will associate to an homogeneous element $x\in U_q(\g)$ its weight $\underline{x}\in Q.$
We define the principal gradation of $U_q(\g)$ by setting $\deg(e_i)=-\deg(f_i)=1$, $\deg(q^{hh'})=0.$
$U_q(\g)$ is therefore a $\Z$-graded algebra. An homogeneous element $x$ is of degree $\deg(x)=(\underline{x},\rho).$

The quantum group $U_q\big(\overset{\circ}{\g}\big)$ is the Hopf subalgebra of
$U_q({\g})$ generated by 
$e_i$, $f_i$, $i\ge 1$, and $q^{hh'}$, $h,h'\in \overset{\circ}{\h}\oplus\C$.  

The quantum group $U_q({\g}')$ is the Hopf subalgebra of $U_q({\g})$ generated by 
$e_i$, $f_i$, $i\ge 0$, and $q^{hh'}$, 
$h,h'\in \overset{\circ}{\h}\oplus {\mathbb C}c\oplus\C$.  

\bigskip

It is sometimes convenient to view $q$ as a formal parameter and to define an 
antimorphism of algebra of $U_q(\g)$ as 
\begin{equation}
   q^*=q^{-1},\quad (e_i)^*=f_i,\quad (f_i)^*=e_i,\quad 
          (h_{\alpha_i})^*=h_{\alpha_i},\quad d^*=d.\label{defstar}
\end{equation}
(This can be made precise by working with $U_q(\g)$ as being a Hopf algebra over $\C(q,q^{-1}).$)

Let $\g$ be of finite or affine type, we now define different notions associated to the  polarisation of $U_q({\mathfrak g}).$ 
Let $U_q(\h)$ be the subalgebra generated by $q^{hh'}$, $h,h'\in \h\oplus\C,$
let $U_q(\b_+)$ (resp. $U_q(\b_-)$) be the sub Hopf algebra generated by $U_q(\h)$, $e_{\alpha}$, (resp.  $U_q(\h)$, $f_{\alpha}$),  $\alpha\in \Gamma$,
and let $U_q(\n_+)$ (resp. $U_q(\n_-)$) be the subalgebra of 
$U_q(\mathfrak{g})$ generated
by $e_{\alpha}$, $\alpha\in \Gamma$ (resp. $f_{\alpha}$, $\alpha\in \Gamma$).
We have  $U_q(\b_+)=U_q(\n_+)\otimes U_q({\mathfrak h})$ as a vector space,
as well as  $U_q(\b_-)= U_q({\mathfrak h})\otimes U_q(\n_-)$.
We denote by $\iota_{\pm}:U_q(\b_{\pm})\rightarrow U_q({\mathfrak h})$
the associated projections on the zero-weight subspaces. $\iota_{\pm}$ are morphisms of algebra, and we define  the ideals   
$U^{\pm}_q({\mathfrak g})=\ker ( \iota_{\pm} )$. 

We will need special types of completions of $U_q(\g)$ and $U_q(\b_\pm)$ which will be defined  in the next subsection. In our previous work \cite{BRT} we defined these completion using the category of finite dimensional $U_q(\g)$-modules when $\g$ is finite dimensional. In the case where $\g$ is an affine Lie algebra we cannot proceed this way because this would amount to divide out by the relation $c=0.$

\bigskip

Let us finally end this subsection by recalling  the definition of some functions that we will use throughout the article:
the infinite product
\begin{equation}
   (z_1,\ldots,z_p;q_1,\ldots,q_n)_\infty 
   =\prod_{k=1}^p\prod_{l_1,\ldots,l_n=0}^{\infty} (1-z_k q_1^{l_1}\ldots q_n^{l_n}), \qquad (\vert q_1\vert<1,\ldots,\vert q_n\vert<1) 
\end{equation}
the $q$-theta function,
\begin{equation}
   \Theta_q(z)=(z,q z^{-1},q;q)_\infty,
\end{equation}
and the $q$-exponential function (which is meromorphic in $z$)
\begin{equation}
\exp_q(z)=e_{q}^z=\sum_{n\geq 0}^{+\infty}\frac{z^n}{(n)_q !}=\frac{1}{((1-q^2)z;q^2)_{\infty}}\qquad 
         \text{with}\quad (n)_q=q^{n-1}[n]_q,
\end{equation}

and which inverse is given by the entire function
\begin{equation}
(\exp_q(z))^{-1}=e_{q^{-1}}^{-z}=((1-q^2)z;q^2)_{\infty}. 
\end{equation}

We will also sometimes  write  $\log_q(A)=B$ instead of $A=q^B$.

\subsection{PBW basis and $R$-matrix}

$U_q(\g)$ admits a Poincar\'e-Birkhoff-Witt (PBW) basis constructed through the use of a normal order of the positive roots.  This procedure is recalled in Appendix~\ref{append-PBW}, we use  the convention and the method  of \cite{KS}. 

Let $<$ be a normal (also called convex)  order   on the set ${\Delta}^{+}$ in the sense of \cite{KS},  this means that:

\begin{definition}{\em Normal order}

1). each non simple root $\alpha+\beta\in \Delta^{+}$, with  $\alpha,\beta\in{\Delta}^+$ not colinear, satisfies $\alpha <\alpha+\beta< \beta$,

2). for any simple roots $\alpha_i,\alpha_j\in \overset{\circ}{\Delta}{}^+$
and $l,n\geq 0$, $k>0$, $\ \alpha_i+n\delta< k\delta<\delta-\alpha_j+l\delta$.

(in the finite type case one has only to consider axiom  1.)
\end{definition}

We can extend this strict  order to a total  order on the set ${\Delta}^{+}$ by
\begin{equation}
  \alpha\leq \beta \Leftrightarrow  \alpha=\beta \;\text{or}\; \alpha<\beta,
\end{equation}
and extend it on  $\overline{\Delta}^+$ as follows:
\begin{align}
  &\forall \alpha \in \Delta^{+}_{\re},\ \forall (k\delta,i) \in \overline{\Delta}^{+}_{\im},
   \quad
   \alpha\leq (k\delta, i) \Leftrightarrow \alpha < k\delta,\\
  &\forall \, (k\delta,i),\, (k'\delta,j) \in \overline{\Delta}^{+}_{\im},\quad
   (k\delta, i) \leq (k'\delta,j)\Leftrightarrow k\delta \leq k'\delta \;\text{or}\; (k=k'\;\text{and}\; i<j).
\end{align}

To each $\alpha\in \overline{\Delta}^+$ one associates an element 
$e_{\alpha}\in U_q(\n_+)$ as explained in Appendix~\ref{append-PBW}.
Let $P$ be the set of finite increasing sequences of elements of $\overline{\Delta}^+$ ;  if $\gamma\in  P$, $\gamma=(\gamma_1,\ldots,\gamma_n)$ with $\gamma_1\leq\cdots\leq \gamma_n,$ we denote $e_{\gamma}=\prod_{k=1}^n e_{\gamma_k}, (e_{\emptyset}=1).$
We have $U_q(\n_+)=\oplus_{\gamma\in P}\C e_{\gamma},$ {\it i.e.}
 $e_{\alpha}, \alpha\in {\overline{\Delta }}{}^{+}$ is a PBW basis of   $U_q(\n_+).$
 We will denote $\underline {\gamma}=\underline{e_{\gamma}}=\sum_{k=1}^n\gamma_k.$

The structure coefficients of   $U_q(\n_+)$ are defined by   $e_{\gamma}e_{\gamma'}=\sum_{\gamma''}C_{\gamma,\gamma'}^{\gamma''}e_{\gamma''}$, where
 $C_{\gamma,\gamma'}^{\gamma''}\in \C(q)$, and satisfy the property that for fixed $\gamma''$ there is only a finite number of couples $(\gamma,\gamma')$ such that $C_{\gamma,\gamma'}^{\gamma''}\not=0$ (this is implied by the property that the number of $\beta,\beta'\in Q^+$ satisfying 
$\beta+\beta'=\underline{\gamma}''$ is finite).

We recall that $U_q(\b_+)=U_q(\n_+)\otimes U_q(\h)$ as a vector space.
$U_q(\h)$ is defined with a structure of $U_q(\n_+)$ right-module, by 
\begin{equation}
  q^{hh'}\lhd y= q^{(h+\underline{y}(h))(h'+\underline{y}(h'))}
\end{equation}
if $y$ is an homogeneous element of $U_q(\n_+)$  and $.\lhd y$ is a morphism of the algebra $U_q(\h).$
The algebra structure of $U_q(\b_+)$ can be defined as:
\begin{equation}
(x\otimes a)(y\otimes b)=
xy\otimes (a\lhd y)b.
\end{equation} 
 We will denote $U_q(\n_+)\widetilde{\otimes}U_q(\h)$ this algebra. 
This motivates the following definition of completion of $U_q(\b_+)$:
we define a completion $(U_q(\n_+)\widetilde{\otimes} U_q(\h))^c$ of $U_q(\n_+)\widetilde{\otimes} U_q(\h)$ as the vector space of maps from $P$ to $U_q(\h);$
an element $x\in (U_q(\n_+)\widetilde{\otimes} U_q(\h))^c$ is written as 
$x=\sum_{\gamma} e_{\gamma}\otimes x_{\gamma}$, $x_\gamma\in U_q(\h).$

$(U_q(\n_+)\widetilde{\otimes} U_q(\h))^c$ can be endowed with a structure of associative algebra by
\begin{equation}
\sum_{\gamma\in P} e_{\gamma}\otimes  x_\gamma\cdot
\sum_{\gamma'\in P} e_{\gamma'}\otimes y_{\gamma'}=
\sum_{\gamma''\in P}\Big(\sum_{\gamma,\gamma'\in P} C_{\gamma,\gamma'}^{\gamma''}e_{\gamma''}\otimes  (x_{\gamma}\lhd e_{\gamma'})y_{\gamma'} \Big) . 
\end{equation}
We define $(U_q(\n_+))^c$ as the subalgebra of elements $x\in (U_q(\b_+))^c$ such that 
$x_\gamma\in \C.$

Similarly we  define $f_{\gamma}=e_{\gamma}^*,\ \gamma\in P$ where $*$ is defined by  \eqref{defstar}. Then 
$U_q(\n_-)=\oplus_{\gamma\in P} \C f_\gamma,$ and $f_{\gamma}f_{\gamma'}=\sum_{\gamma''}
(C_{\gamma' \gamma}^{\gamma''})^* f_{\gamma''}.$
We define a left action of $U_q(\n_-)$ on $U_q(\h)$ as
\begin{equation}
 y\rhd q^{hh'}= q^{(h+\underline{y}(h))(h'+\underline{y}(h'))},
\end{equation}
with $y$ homogeneous element and $y\rhd.$ morphism of the algebra $U_q(\h).$

We  define analogously the  completion $(U_q(\h)\widetilde{\otimes} U_q(\n_-))^c$ as the vector space of maps from $P$ to $U_q(\h)$, an element  $x\in (U_q(\h)\widetilde{\otimes} U_q(\n_-))^c$ being  written as 
$x=\sum_{\gamma} x_{\gamma}\otimes f_{\gamma}$, $x_\gamma\in U_q(\h),$
with 
\begin{equation*}
\sum_{\gamma\in P}   x_\gamma \otimes  f_{\gamma}\cdot
\sum_{\gamma'\in P}  y_{\gamma'}\otimes f_{\gamma'}=
\sum_{\gamma''\in P}\Big(\sum_{\gamma,\gamma'\in P} (C_{\gamma',\gamma}^{\gamma''
})^*
x_{\gamma} (f_{\gamma}\rhd y_{\gamma'})\otimes f_{\gamma''}\Big).
\end{equation*}
Similarly we define $(U_q(\n_-))^c$ as the subalgebra of elements $x\in (U_q(\h)\otimes U_q(\n_-))^c$ such that $x_\gamma\in \C.$

Let $\gamma\in P,$ we define
$\iota^{+}_{\gamma}:(U_q(\n_+)\widetilde{\otimes} U_q(\h))^c\rightarrow U_q(\h),\ 
x\mapsto x_\gamma,$ and 
$\iota^{-}_{\gamma}:( U_q(\h)\widetilde{\otimes} U_q(\n_-))^c\rightarrow U_q(\h),\ 
x\mapsto x_\gamma.$ The previously defined maps  $\iota^{\pm}$ are 
$\iota^{\pm}=\iota^{\pm}_{\emptyset}\vert_{U_q(\b_\pm)}.$

The map $U_q(\n_+)\otimes U_q(\h)\otimes U_q(\n_-)\rightarrow U_q(\g),\ x\otimes y\otimes z\mapsto xyz$ being  an isomorphism of vector space,  an element $x$ of $U_q(\g)$ is uniquely written as $x=\sum_{\gamma,\gamma'}e_\gamma x_{\gamma,\gamma'}f_{\gamma'}$ where $x_{\gamma,\gamma'}$ is not zero  just for a finite number of $(\gamma,\gamma').$
This suggests the following definition:
we define $(U_q(\g))^c$ as the vector space of maps 
$P^{\times 2}\rightarrow U_q(\h),$ an element 
$x\in (U_q(\g))^c$ will be expressed  as the series $x=\sum_{\gamma,\gamma'}e_\gamma x_{\gamma,\gamma'}f_{\gamma'}.$ $(U_q(\g))^c$ is naturally endowed with a structure of left-right $U_q(\b_+)^c$-$U_q(\b_-)^c$ bimodule by multiplication.
Note that it is also endowed with a natural structure of left- right $U_q(\b_-)$-$U_q(\b_+)$ module by multiplication.

Note also that  $U_q(\n_-)\otimes U_q(\h)\otimes U_q(\n_+)\rightarrow U_q(\g),\ x\otimes y\otimes z\mapsto xyz$ is an isomorphism, therefore we can define 
 $U_q(\g)^{c(op)}$ as the vector space of maps $P^{\times 2}\rightarrow U_q(\h).$  
 An element 
$x\in (U_q(\g))^{c(op)}$ will be written as the series 
$x=\sum_{\gamma,\gamma'}f_\gamma x_{\gamma,\gamma'}e_{\gamma'}.$ $(U_q(\g))^{c(op)}$ is naturally endowed with a structure of left-right $U_q(\b_-)^c$-$U_q(\b_+)^c$ bimodule by multiplication.
It is also endowed with a structure of  left-right  $U_q(\b_+)$-$U_q(\b_-)$ module by multiplication.

We extend the definition of $\iota^{\pm}_{\gamma}$ to $(U_q(\g))^{c(op)}$ as 
\begin{align}
 &\iota^{\pm}_{\gamma}:(U_q(\g))^{c(op)}\rightarrow (U_q(\b_\mp))^c \nonumber\\
 &\iota^{+}_{\gamma}(x)=\sum_{\gamma'}f_{\gamma'} x_{\gamma',\gamma},\quad
 \iota^{-}_{\gamma}(x)=\sum_{\gamma'} x_{\gamma,\gamma'}e_{\gamma'}.
\end{align}


Note however that there is no canonical structure of algebra on $(U_q(\g))^c$ which  would contains $U_q(\g),(U_q(\b_+))^c, (U_q(\b_-))^c$ as  subalgebras.

\bigskip

In order to construct the space where the $R$ matrix lies, we need a completion of 
 $U_q(\b_+)\otimes U_q(\b_-)=U_q(\n_+)\widetilde{\otimes} (U_q(\h)\otimes  U_q(\h))
\widetilde{\otimes} U_q(\n_-).$

The algebra $U_q(\n_+)\widetilde{\otimes} U_q(\h)^{\otimes 2}\widetilde{\otimes}  U_q(\n_-)$ admits the following completion: we define the algebra  $(U_q(\n_+)\widetilde{\otimes} U_q(\h)^{\otimes 2} \widetilde{\otimes}  U_q(\n_-))^c$ as being the set of maps
 $P^{\times 2}\rightarrow U_q(\h)^{\widehat{\otimes} 2}.$
An element $x\in (U_q(\n_+)\widetilde{\otimes} U_q(\h)^{\otimes 2} \widetilde{\otimes}  U_q(\n_-))^c$ is written as $x=\sum_{\gamma,\gamma'}
(e_{\gamma}\otimes 1)(x_{\gamma,\gamma'})(1\otimes f_{\gamma'}),$ and the algebra  law is defined as:
\begin{multline}
 \sum_{\gamma,\gamma'}
(e_{\gamma}\otimes 1)(x_{\gamma,\gamma'})(1\otimes f_{\gamma'}).\sum_{\beta,\beta'}
(e_{\beta}\otimes 1)(y_{\beta,\beta'})(1\otimes f_{\beta'})\\
 =\sum_{\gamma,\gamma',\beta,\beta'}(e_{\gamma}e_{\beta}\otimes 1)
(x_{\gamma,\gamma'}\lhd (e_{\beta}\otimes 1))
((1\otimes f_{\gamma'})\rhd y_{\beta,\beta'})(1\otimes f_{\gamma'} f_{\beta'}).\label{bplusbmoinsc}
\end{multline}
\bigskip

$U_q(\g)$ is a quasitriangular Hopf algebra: there exists 
$R\in (U_q(\n_+)\widetilde{\otimes} U_q({\h})^{\widehat{\otimes} 2}\widetilde{\otimes}  U_q(\n_-))^c$ satisfying to the axioms of quasitriangularity.
The explicit expression of the 
$R$-matrix in the finite or affine case is given in term of a PBW basis of $U_q(\g)$ constructed through the use  of a normal order of the roots. In the finite case the expression is given in \cite{KT0}. 
In the affine case the expression of $R$  is given in \cite{KT,D}.

If $\g$ is of affine or finite type, we define  $K=q^{\Omega_{{\h}}}\in U_q({\h})^{\widehat{\otimes} 2}$ and  denote $k=q^{\frac{1}{2}
m(\Omega_{{\h}})}\in U_q(\h)$ where $m$ is the multiplication.
We have $\Delta(k)=K k_1k_2.$ 
When $\g$ is of finite type,  the expression of the $R-$matrix of $U_q(\g)$ is given by
\begin{equation}
   R=K\,\widehat{R},\;\;
   \widehat{R}=\stackrel{<}{\prod_{\alpha\in {\Delta}^+}}\widehat{R}_{\alpha}   \qquad\text{where}\quad
  \widehat{R}_{\alpha}=
        \exp_{q^{-1}}\big[(q-q^{-1})\, e_{\alpha}\otimes f_{\alpha}\big].
\end{equation}

When $\g$ is of  affine type, the  expression of the $R$-matrix of $U_q(\g)$ is given by:
\begin{equation}
   R=K\,\widehat{R}
    =K\, \widehat{R}_{\re}^+ \,\widehat{R}_{\im}\, \widehat{R}_{\re}^-
\end{equation}
with 
\begin{equation} 
  \widehat{R}_{\re}^+=\stackrel{<}{\prod_{\alpha\in \Delta_{\re}^+\atop\alpha<\delta}}\widehat{R}_{\alpha},
          \qquad
  \widehat{R}_{\re}^-= \stackrel{<}{\prod_{\alpha\in \Delta_{\re}^+\atop\alpha>\delta}}\widehat{R}_{\alpha}.
          \end{equation}
and  
\begin{equation}
   \widehat{R}_{\im}=\exp\bigg[ (q-q^{-1})
       \sum^{r}_{i,j=1} \sum_{n>0}\, c_{ij}^{(n)} \, 
                e_{n\delta}^{(i)}\otimes f_{n\delta}^{(j)}\bigg],
\end{equation}
$\big(c_{ij}^{(n)}\big)$ being the inverse matrix of the matrix  $\big(\frac{[n(\alpha_i,\alpha_j)]_{q}}{n}\big)_{i,j=1,\ldots,r}.$
It is far from trivial to show that a normal order exists in the affine case.
A  construction  of  normal order using the affine Weyl group has been done in \cite{Bec}, and Appendix~\ref{append-PBW} contains  elements of this construction in the $A_r^{(1)}$ case.

\subsection{The $U_q(A_1^{(1)})$ case}

We use here the following normal order on the positive roots of $U_q(A_1^{(1)})$:
\begin{equation}
  \alpha_1<\alpha_1+\delta<\alpha_1+2\delta<\cdots<k\delta<\cdots<
  \alpha_0+2\delta<\alpha_0+\delta<\alpha_0
  \quad(k\in \Z^{+ *}).\label{normalorderaffinesl2}
\end{equation}
Then the expression of the $R$ matrix reads:
\begin{multline}
   R=K\stackrel{\rightarrow}{\prod_{n\geq 0}}
       e_{q^{-1}}^{[(q-q^{-1})\, 
                     e_{\alpha_1+n\delta}\otimes f_{\alpha_1+n\delta}]}\\
         \times
     \exp\bigg[\sum_{n>0} (q-q^{-1})\frac{n}{[2n]_q}\,
                e_{n\delta}\otimes f_{n\delta}\bigg]\,
     \stackrel{\leftarrow}{\prod_{n\geq 0}}
       e_{q^{-1}}^{[(q-q^{-1})\,e_{\alpha_0+n\delta}\otimes f_{\alpha_0+n\delta}]}.
   \label{UniversalRsl2}
\end{multline}

Let $\pi$ be the fundamental two dimensional representation of $U_q(A_1)$ acting on 
$V={\mathbb C}^{ 2}$. 
Let $z\in {\mathbb C}^{\times}$, we define the evaluation representation $\text{ev}_z$ of  $U_q(A_1^{(1)}{}')$ acting on $V$ as:
\begin{alignat*}{3}
  &\text{ev}_z(e_1)=\pi(e_1)=E_{1,2},\quad  &
  &\text{ev}_z(f_1)=\pi(e_1)=E_{2,1},\quad  &
  &\text{ev}_z(h_{\alpha_1})=\pi(h_{\alpha_1})=E_{1,1}-E_{2,2},\\
  &\text{ev}_z(e_0)=z\pi(f_1),\quad & 
  &\text{ev}_z(f_0)=z^{-1}\pi(e_1),\quad & 
  &\text{ev}_z(h_{\alpha_0})=-\pi(h_{\alpha_1})\ \; 
      (\text{{\it i.e.}}\ \text{ev}_z(c)=0).
\end{alignat*}

Using results of Appendix~\ref{append-PBW}, we obtain the following action of the  PBW basis in the evaluation representation:
\begin{alignat*}{2}
  &\text{ev}_z(e_{\alpha_1+n\delta})=(-q^{-1}z)^n E_{1,2},\quad &
  &\text{ev}_z(f_{\alpha_1+n\delta})=(-qz^{-1})^n E_{2,1},\\
  &\text{ev}_z(e_{\alpha_0+n\delta})=z (-q^{-1}z)^n E_{2,1},\quad &
  &\text{ev}_z(f_{\alpha_0+n\delta})=z^{-1} (-qz^{-1})^n E_{1,2},\\
  &\text{ev}_z(e'_{n\delta})=z^n(-q)^{1-n}(E_{1,1}-q^{-2}E_{2,2}),\quad &
  &\text{ev}_z(f'_{n\delta})=z^{-n}(-q)^{n-1}(E_{1,1}-q^{2}E_{2,2}),\\
  &\text{ev}_z(e_{n\delta})=-(-z)^n\frac{[n]_q}{n}(E_{1,1}-q^{-2n}E_{2,2}),
         \quad  &
  &\text{ev}_z(f_{n\delta})=-(-z)^{-n}\frac{[n]_q}{n}(E_{1,1}-q^{2n}E_{2,2}).
\end{alignat*} 

We define $\mathbf{R}(z,z')=(\text{ev}_z\otimes \text{ev}_{z'})(R),$ trigonometric solution of the QYBE, also called $R$-matrix of the $6$-vertex model with spectral parameters  $z,z'.$ Using the universal expression of the $R$-matrix (\ref{UniversalRsl2}) and the expression of the evaluation representation on the PBW basis we obtain:

\begin{equation}
\mathbf{R}(z,z')=q^{\onehalf}f(z/z')\begin{pmatrix}
1& 0 & 0& 0\\
0& \frac{z'-z}{qz'-q^{-1}z}& \frac{(q-q^{-1})z'}{qz'-q^{-1}z}&0\\
0& \frac{(q-q^{-1})z}{qz'-q^{-1}z}&\frac{z'-z}{qz'-q^{-1}z}&0\\
 0 & 0& 0&1
\end{pmatrix}
\end{equation}
with 
\begin{equation}
f(u)=\exp\bigg[\sum_{n>0}\frac{q^n-q^{-n}}{q^n+q^{-n}}\frac{u^n}{n}\bigg]
    =\frac{(u,q^4u;q^4)_{\infty}}{(q^2u;q^4)_{\infty}^2}.
\end{equation}

\subsection{The $U_q(A_r^{(1)})$ case}

We assume now that $\g=A_r^{(1)},$ $r\geq 2$. 

The maximal root of $A_r$ is 
$\theta=\sum_{i=1}^r \alpha_i.$ 
The fundamental representation of $U_q(A_r)$ acting on $V={\mathbb C}^{r+1}$ is defined as:
\begin{equation}
\pi(e_i)=E_{i,i+1},\quad\pi(f_i)=E_{i+1,i},\quad
\pi(h_{\alpha_i})=E_{i,i}-E_{i+1,i+1},\quad i=1,\ldots,r.
\end{equation} 

This representation extends to a representation of $U_q({\g}')$ as follows:
let $z\in{\mathbb C}^{\times},$ we define $\text{ev}_z$ the evaluation representation of $U_q({\g}')$ acting on $V$ by
\begin {alignat}{3}
  &\text{ev}_z(e_i)=\pi(e_i),\quad &
  &\text{ev}_z(f_i)=\pi(f_i),\quad &
  &\text{ev}_z(h_{\alpha_i})=\pi(h_{\alpha_i}),\quad
  i=1,\ldots, r,\\
  &\text{ev}_z(e_0)=z E_{r+1,1},\quad &
  &\text{ev}_z(f_0)=z^{-1} E_{1,r+1},\quad &
  &\text{ev}_z(c)=0.
\end{alignat}
In Appendix~\ref{append-PBW}, we give the image of some elements of the PBW basis of $U_q(A_r^{(1)})$ under  the evaluation.

We will also need the expression of the matrix elements $c_{ij}^{(n)}$, which can be computed exactly.
For this we  use the fact that the inverse of the $r\times r$ matrix
\begin{equation}
 \widetilde{A}=\begin{pmatrix}
   q+q^{-1}& -1 & 0& \hdotsfor{2}&0\\
   -1 &q+q^{-1}&-1&0&\hdotsfor{1}&0\\
   0  &-1 &q+q^{-1}&-1&0& 0\\
   \hdotsfor{6}\\
  0& \hdotsfor{1}&0&-1&q+q^{-1}&-1\\
  0& \hdotsfor{2}&0&-1&q+q^{-1}
\end{pmatrix}
\end{equation} 
is given by
\begin{equation}
  \big(\widetilde{A}^{-1}\big)_{ij}=\frac{[\min(i,j)]_q[r+1-\max(i,j)]_q}{[r+1]_q}.
\end{equation}
Therefore we obtain that
\begin{equation} 
 c_{ij}^{(n)}=\frac{n[\min(i,j)]_{q^n}[r+1-\max(i,j)]_{q^n}}{[n]_q[r+1]_q}=
\frac{n}{[n]_q^2}\frac{[n\min(i,j)]_q[n(r+1-\max(i,j))]_q}{[n(r+1)]_q}.\label{cijk}
\end{equation}

\section{Results on  Dynamical  Quantum Groups}
\label{sectionDAQG}

For $x\in (\C^{\times})^{\dim \h}$, $xq^h\in  (\C^{\times})^{\dim \h}\otimes U_q(\h)$ is defined as:
\begin{equation}
  xq^{h}=\begin{cases}
   (x_1q^{h_{\alpha_1}},\ldots,x_rq^{h_{\alpha_r}})
         &\text{if $\g$ is of finite type,}\\
   (x_0q^{h_{\alpha_0}},x_1q^{h_{\alpha_1}},\ldots,x_rq^{h_{\alpha_r}},x_dq^{d}) 
         &\text{if $\g$ is of affine type}.
        \end{cases}
\end{equation}

\bigskip

Let us first formulate the dynamical Yang-Baxter equation:

\begin{definition}{\em Quantum Dynamical Yang-Baxter Equation (QDYBE)}
\par\noindent Let $V$ be a $\h$-simple finite dimensional $U_q(\g)$-module. 
A meromorphic function
 $R:(\C^{\times})^{\dim \h} \rightarrow \End(V\otimes V)$ 
is said to satisfy the Quantum Dynamical Yang-Baxter Equation  if
 \begin{equation}
  R_{12}(x)\, R_{13}(xq^{h_2})\, R_{23}(x)=
  R_{23}(xq^{h_1})\, R_{13}(x)\, R_{12}(xq^{h_3}). \label{DYBE}
\end{equation}
Let $\l$ be a subspace of ${\h},$
$R$ is said to be ``of effective dynamics $\l$ '' if 
\begin{align}
 & R(x) \text{ is of zero $\l$-weight, {\it i.e.} commutes with the action of $\l$}, \\ 
 &(\id\otimes\id\otimes \nu(t))( R_{12}(xq^{h_3}))=R_{12}(x)\;\; \forall t\in \l^{\perp}.
\end{align}
\end{definition}

We would like to extend this definition to the notion of universal solutions.
This can easily be done when $\g$ is of  finite type using our previous formalism \cite{BRT}. It is more delicate to formulate it rigorously in the affine case. We will only provide a universal formulation of the notion of Quantum Dynamical coCycle which is a closely related concept.

There is however no problem of definition for the classical limit of this equation,  the Classical Dynamical Yang-Baxter Equation, the solutions of which have been completely classified under very mild assumptions.

\subsection{Classical Dynamical r-matrices }
\label{subsec-classdrybe}

We begin with the formulation of the Classical Dynamical Yang-Baxter Equation.
If  $r:({\C^{\times}})^{\dim \h} \rightarrow 
{\g}^{\otimes 2}$ and $g:({\C^{\times}})^{\dim \h} \rightarrow G$, where $Lie(G)=\g$, we denote $d_3 r_{12}=\frac{d}{dt}(r_{12}(xe^{th_3}))_{t=0}$, and
$d_2 g_1=\frac{d}{dt}(g_{1}(xe^{th_2}))_{t=0}.$

\begin{definition}{\em Classical Dynamical Yang-Baxter Equation (CDYBE)}\par\noindent
A  map  $r:({\C^{\times}})^{\dim \h} \rightarrow {\g}^{\otimes 2}$ satisfies the CDYBE  if
\begin{equation}
[[r(x),r(x)]]-d_3r_{12}(x)-d_1 r_{23}(x)+d_2 r_{13}(x)=0,
\end{equation}
where $[[A,A]]=[A_{12},A_{13}+A_{23}]+[A_{13},A_{23}].$\\
$r$ is said to be of effective dynamics $l\subset\h$ if  
 $r$ is $\mathfrak{l}$-invariant and  if $d_3r_{12}\in \g^{\otimes 2}\otimes \l. $\\
$r$ is said to satisfy the unitarity condition if $r_{12}(\lambda)+r_{21}(\lambda)=
\Omega_{{\h}}.$ \\
Let $r,r'$ be solutions of the CDYBE with effective dynamics $\l$, they  are said to be dynamically gauge equivalent if there exists a map $g:({\C^{\times}})^{\dim \h} \rightarrow L$ with $L\subset G$ and $Lie(L)=\l$ such that 
\begin{equation}
r'=r^{g}=(\Ad_{g}\otimes \Ad_{g})(r+g_1^{-1}d_2 g_1-g_{2}^{-1}d_1 g_2).
\end{equation}
\end{definition}

P. Etingof and A. Varchenko have classified in \cite{EV} the unitary solutions of the CDYBE of effective dynamics $\h$ up to automorphism of $\g$ and up
to dynamical  gauge equivalence under  the assumption that $\g$ is a finite dimensional simple Lie algebra. 
The solutions are in bijection with the  subsets $X\subset \Gamma.$ 
The {\it standard} dynamical solution is obtained by taking $X=\Gamma.$

O. Schiffmann \cite{Sch} has classified the unitary solutions 
($\g$ simple finite dimensional) of the
CDYBE of effective dynamics $\l\subset \h$ such that $\l$ contains a regular semi-simple element (this last assumption  forbids to take $\l=\{0\}$ for example). He has shown that the solutions, up to automorphism and dynamical gauge equivalence, are 
classified by generalized Belavin-Drinfeld triples, {\it i.e.} triples of the form $(\Gamma_1, \Gamma_2, T)$, where $\Gamma_1, \Gamma_2\subset\Gamma$ and $T:\Gamma_1 \rightarrow \Gamma_2$, preserving the scalar product with 
$\l=\{\Vect \{\alpha-T(\alpha), \alpha\in\Gamma_1\}\}^{\perp}.$
The classification of \cite{EV} is recovered by taking $X=\Gamma_1=\Gamma_2$ whereas the classification of Belavin-Drinfeld of unitary non dynamical solutions of classical Yang-Baxter equation is recovered when the triple is nilpotent.

We recall here various definitions and results in the finite type case.

\begin{definition}{} Let $\g$ be of finite type,
a Belavin-Drinfeld triple is given by  ${\cal T}=(\Gamma_1,\Gamma_2,T)$, where  $\Gamma_1,\Gamma_2$ are  subsets of $\Gamma,$  and  $T:\Gamma_1\rightarrow \Gamma_2$ is a bijection satisfying:
\begin{align}
 &1.\ (T\alpha,T\alpha')=(\alpha,\alpha'),\qquad
    \forall \alpha,\alpha'\in \Gamma_1.\label{Tisometry}\\
 &2.\ \text{$T$ is ``nilpotent'' {\it i.e.}} \qquad
    \forall \alpha\in \Gamma_1,\; \exists k\in \mathbb{N},\;
    T^{k}(\alpha)\in \Gamma_1,\;  T^{k+1}(\alpha)\notin \Gamma_1.\label{nilpotency}
\end{align}
\par We define $\sigma^\pm:\Gamma\rightarrow \Gamma$  by $\sigma^+(\alpha)=T(\alpha)$, if $\alpha \in \Gamma_1,$  $\sigma^+(\alpha)=0$, if $\alpha \in \Gamma\setminus\Gamma_1$ (resp. $\sigma^-(\alpha)=T^{-1}(\alpha)$, if $\alpha \in \Gamma_2,$  $\sigma^-(\alpha)=0$, if $\alpha \in \Gamma\setminus\Gamma_2$).\\
$\sigma^\pm$  extends to $\sigma^\pm:\mathfrak{b}_\pm \rightarrow \mathfrak{b}_\pm$  morphism of Lie algebra defined by:
$\sigma^+(e_{\alpha})=e_{T(\alpha)}$, $\sigma^+(\overset{\circ}{\zeta}{}^{\alpha})=
\overset{\circ}{\zeta}{}^{T(\alpha)}$ for  $\alpha \in \Gamma_1$, and  $0$ otherwise  (resp. $\sigma^-(f_{\alpha})=f_{T^{-1}(\alpha)}$, $\sigma^-(\overset{\circ}{\zeta}{}^{\alpha})=\overset{\circ}{\zeta}{}^{T^{-1}(\alpha)}$ for  $\alpha \in \Gamma_2$ and  $0$ otherwise).\\
\indent If  ${\cal T}$ is a  Belavin-Drinfeld triple, one defines $A_{{\cal T}}$ to be the subset of   $\bigwedge^2 \mathfrak{h}$ of elements  $s$ satisfying
\begin{equation}
\forall \alpha\in \Gamma_1,\quad
2((T\alpha-\alpha)\otimes \id)(s)=((\alpha+T\alpha)\otimes \id)(\Omega_{\mathfrak{h}})\label{Sbelavindrinfeld}
\end{equation}
$A_{{\cal T}}$ is an affine space of dimension $n_{{\cal T}}=\frac{k(k-1)}{2}$, $k=\vert \Gamma\setminus \Gamma_1\vert.$\\
$({\cal T},s)$ with  $s\in A_{{\cal T}}$ is called  Belavin-Drinfeld quadruple.\\
\indent To such a  quadruple is  associated  an element $r_{T,s}\in \mathfrak{g}^{\otimes 2}$ defined by:
\begin{equation*}
r_{T,s}=r+s+\sum_{\alpha\in \Delta^+}\sum_{l=1}^{+\infty}\frac{(\alpha,\alpha)}{2}
(\sigma^+)^l(e_{\alpha})\wedge f_{\alpha},
\end{equation*}
where $r=\frac{1}{2}\Omega_{{\mathfrak h}}+\sum_{\alpha\in \Delta^+}\frac{(\alpha,\alpha)}{2}
e_{\alpha}\wedge f_{\alpha}$ is the standard classical $r-$matrix.
\end{definition}

We  denote $r^{0}=\frac{1}{2}\Omega_{\mathfrak h}+s$ and $r^0_{\alpha \beta}=(\alpha \otimes \beta)(r^0).$ The constraints imposing $s$ to be in  $A_{{\cal T}}$ translates into 
\begin{alignat}{2}
 &r^0_{\alpha \beta}+r^0_{\beta \alpha}=(\alpha,\beta),\quad & 
 &\forall \alpha,\beta \in \Gamma, \label{r0constraint1}\\
 &r^0_{T(\alpha) \beta}+r^0_{\beta \alpha}=0,\quad &
 &\forall \alpha \in \Gamma_1,\forall \beta \in \Gamma.\label{r0constraint2}
\end{alignat}
Due to (\ref{nilpotency}), every  $\gamma\in \Gamma$ can be expressed uniquely in the form $\gamma=T^{-m}(\alpha)$ for a certain nonnegative  integer $ m$ and  $\alpha \in \Gamma\setminus\Gamma_1.$ 
As a result, if $\delta=T^{-l}(\beta)$ for $0\leq l$, $\beta \in \Gamma\setminus\Gamma_1,$ (\ref{r0constraint1})(\ref{r0constraint2}) implies 
\begin{equation}
 r^{0}_{\gamma \delta}-r^{0}_{\alpha \beta}=
 \begin{cases}
 (\;T^{-1}(\alpha)+\cdots+T^{l-m}(\alpha)\;,\;\beta\;)\qquad 
         &\mbox{if} \quad l<m, \\
 -(\;\beta+T^{-1}(\beta)+\cdots+T^{m-l+1}(\beta)\;,\;\alpha\;)\qquad 
         &\mbox{if}\quad  m<l,\\ 
  0 \qquad &\mbox{if}\quad m=l.
\end{cases}\label{solr0}
\end{equation}
As soon as we have chosen the $n_{{\cal T}}$ numbers $r^0_{\alpha \beta}$ for all $\alpha,\beta \in \Gamma\setminus \Gamma_1$ obeying only to (\ref{r0constraint1}), we can then determine completely the remaining coefficients of $r^{0}$ by using (\ref{solr0}) without any new constraint. The element $r^0$ determined in such a way fullfills necessarily (\ref{r0constraint1}) and (\ref{r0constraint2}). 

\begin{theorem}{\em Belavin-Drinfeld}\par\noindent
1). If $({\cal T}, s)$ is a Belavin-Drinfeld quadruple, $r_{T,s}$ satisfies  the CYBE and  the unitarity relation.\par\noindent
2). $\{ r\in \mathfrak{g}^{\otimes 2}, [[r,r]]=0, r_{12}+r_{21}=\Omega_{\mathfrak g}\}/\Aut(\mathfrak g)$ is  in bijection with  $(\bigcup_{\cal {T}}\{r_{T,s}, s\in A_{\cal{T}}\})/\Aut(\Gamma). $ 
\end{theorem}

The examples of  Belavin-Drinfeld quadruple corresponding to  $k=r$ and  $k=1$ are easily described:

$\bullet$ k=r.  $\Gamma_1=\Gamma_2=\emptyset, A_{\emptyset}=\bigwedge^2 \mathfrak{h}.$

$\bullet$ k=1. In this case  $\mathfrak{g}=A_r$. $\Gamma_{1}=\{\alpha_2,\ldots,\alpha_r\}$, 
$\Gamma_{2}=\{\alpha_1,\ldots,\alpha_{r-1}\}$ and  $T(\alpha_i)=\alpha_{i-1}.$ This  triple is called the shift  and  
$\dim A_{\emptyset}=0.$ The admissibility conditions have a unique solution :
\begin{equation*}
 s=\frac{1}{2}\sum_{j=1}^{r-1}\overset{\circ}{\zeta}{}^{j}\wedge
 \overset{\circ}{\zeta}{}^{j+1}.
\end{equation*}

\begin{definition} 
A generalized Belavin-Drinfeld is  ${\cal T}=(\Gamma_1,\Gamma_2,T)$, where  $\Gamma_1,\Gamma_2$ are subsets 
of  $\Gamma$ and  $T:\Gamma_1\rightarrow \Gamma_2$ is a  bijection satisfying:
$(T\alpha,T\alpha')=(\alpha,\alpha'),\ \forall \alpha,\alpha'\in \Gamma_1.$ The   nilpotence condition is not imposed.
One defines  the space of dynamics  by  
$\l({\cal T})=(\Vect\{\alpha-T(\alpha),\alpha\in \Gamma_1\})^{\bot}.$
\end{definition}

\begin{theorem}{\em (O. Schiffmann)}\par\noindent
To each  generalized Belavin-Drinfeld triple  ${\cal T}$ is  associated a solution  $r_{\cal T}$ of  the CDYBE satisfying the unitarity condition.\par\noindent
One still defines $T:\mathfrak{n}^+\rightarrow \mathfrak{n}^+$ Lie algebra morphism by the same definition as previously but  $T$ is now no more nilpotent.
One defines the Cayley transform of  $T,$ 
$C_T:l^{\bot}\rightarrow l^{\bot},$ by:
\begin{equation*}
 (\alpha-T(\alpha),C_T(y))=(\alpha+T(\alpha),y),\forall \alpha\in \Gamma_1.
\end{equation*}
$C_T$ is  skewsymmetric  and   one can define
\begin{equation*}
 r_{\cal T}(x)=r-\frac{1}{2}(C_T\otimes \id)\Omega_{l^{\bot}}+a(x)-a(x)_{21},
\forall x\in (\C^{\times})^{ \dim \h}
\end{equation*}
with 
\begin{equation*}
 a(x)=\sum_{\alpha\in \Delta^+}\sum_{l=1}^{+\infty}\frac{(\alpha,\alpha)}{2}
(\prod_{j=1}^r x_j^{-2 l\langle\overset{\circ}{\zeta}{}_j,\alpha\rangle})
T^l(e_{\alpha})\wedge f_{\alpha}. 
\end{equation*}
Moreover each unitary solution of the CDYBE of effective dynamics $\l$ with $\l$ containing a semi simple regular element is, up to automorphism of $\g$ and up to dynamical gauge transformation lying in $L\subset G$ with $Lie(L)=\l$,  in the previous list.
\end{theorem}

The solution associated to   ${\cal T}=(\Gamma,\Gamma,\id)$ is called  the  {\it standard} solution of  the CDYBE.

\begin{rem} If  ${\cal T}$ is  nilpotent, then  the sum defining $a(x)$ is finite. In this case one can associate to each element $s\in A_{\cal T}$ a dynamical gauge transformation   $g_s(x)= \prod_{j=1}^r x_j^{2(\overset{\circ}{\zeta}_j\otimes\id)(s)}.$
One has  $r_{\cal T}^{g_s}=r_{{\cal T},s}.$
\end{rem}

\begin{rem}A construction of a solution of the CDYBE associated to any generalized Belavin-Drinfeld triple of a symmetrizable Kac Moody algebra (in particular of affine type) has been done in \cite{ES} (but no complete classification theorem is known in this case).
\end{rem}


\subsection{Quantum Dynamical coCycles  Equation and Quantum Dynamical Yang-Baxter Equation}
\label{subsec-QDCE}

As first understood by O. Babelon \cite{Bab}, a  universal solution of the QDYBE equation can be obtained from a 
solution of the {\em Quantum Dynamical coCycle Equation}.
In the finite type case, we refer to \cite{BRT} for a precise statement.

In the present work we will only define precisely the notion of Universal Quantum Dynamical coCycle Equation.

Let $\g$ be of finite or affine type, we denote ${\cal D}^{\pm}(\h)$ the following commutative algebra:
\begin{equation}
  {\cal D}^{\pm}(\h)=\begin{cases}
   \C[x_1^{\mp 2},\ldots,x_r^{\mp 2} ]
         &\text{if $\g$ is of finite type,}\\
    \C[x_0^{\mp 2},x_1^{\mp 2},\ldots,x_r^{\mp 2}, x_d^{\mp 2}] 
         &\text{if $\g$ is of affine type}.
        \end{cases}
\end{equation}
Let  ${\cal F}^{(n)}(\h)$ be the field of fractions of the commutative algebra 
$U_q(\h)^{\widehat{\otimes} n}\otimes  {\cal D}^{-}(\h).$ 
${\cal F}^{(1)}(\h)$ is endowed with a structure of right $U_q(\n_+)$-module (resp. left  $U_q(\n_-)$-module) algebra by extending the action $\lhd$ of $U_q(\n_+)$  (resp. $\rhd$ of  $U_q(\n_-)$) on $U_q(\h)$ and by acting trivially on ${\cal D}^{-}(\h).$
As a result we can define the extension of algebras
 $(U_q(\n_+)\widetilde{\otimes} {\cal F}^{(1)}(\h))^c=
(U_q(\b_+)\otimes {\cal F}^{(0)}(\h))^c,$ and 
$( {\cal F}^{(1)}(\h)\widetilde{\otimes} U_q(\n_-) )^c=
({\cal F}^{(0)}(\h) \otimes U_q(\b_-))^c.$ 
By extension we denote $(U_q^{\pm}(\g)\otimes  {\cal F}^{(0)}(\h))^c$ the subalgebra  of $(U_q(\b_\pm)\otimes {\cal F}^{(0)}(\h))^c$ which is the kernel of $\iota^{\pm}_{\emptyset}.$
Analogously we can define $(U_q(\g)\otimes {\cal F}^{(0)}(\h))^c$ and 
 $(U_q(\g)\otimes {\cal F}^{(0)}(\h))^{c(op)}.$

 We also define $(U_q(\n_+)\widetilde{\otimes}  {\cal F}^{(2)}(\h)\widetilde{\otimes}  U_q(\n_-))^c$ as the set of maps from $P^{\times 2}\rightarrow  {\cal F}^{(2)}(\h)$ with the same algebra law as (\ref{bplusbmoinsc}).
We will extend the principal gradation on 
$U_q(\b_\pm)\otimes {\cal F}^{(0)}(\h)$ by 
$\deg(x\otimes y)=\deg(x)$, $x\in U_q(\b_\pm)$, $y\in  {\cal F}^{(0)}(\h)\setminus
 \{0\}.$

\begin{theorem}{\em Quantum Dynamical Cocycle Equation (QDCE)\label{TheoremQDCE}}
\par\noindent
Let $F(x)\in ( U_q(\n_+)\widetilde{\otimes} {\cal F}^{(2)}(\h )\widetilde{\otimes}   U_q(\n_-))^c$,
the elements 
$(\iota_{\gamma}^+\otimes\id\otimes\iota_{\gamma'}^-)\big((\Delta\otimes \id)(F(x))\, F_{12}(xq^{h_3})\big)$ and $(\iota_{\gamma}^+\otimes\id\otimes\iota_{\gamma'}^-)\big( (\id\otimes \Delta)(F(x))\, F_{23}(x)\big)$ are well defined for all $\gamma,\gamma'\in P$ and are lying in $(1\otimes U_q(\g)\otimes 1){\cal F}^{(3)}(\h).$\par\noindent
$F(x)$  is a 
{\em Quantum Dynamical coCycle}  if 
$F(x)$ is an invertible element satisfying the  {\em Quantum Dynamical coCycle Equation},
\begin{multline}
(\iota_{\gamma}^+\otimes\id\otimes\iota_{\gamma'}^-)\big((\Delta\otimes \id)(F(x))\, F_{12}(xq^{h_3})\big)\\
 =
(\iota_{\gamma}^+\otimes\id\otimes\iota_{\gamma'}^-)\big( (\id\otimes \Delta)(F(x))\, F_{23}(x)\big), \qquad\forall \gamma,\gamma'\in P.
  \label{eq:s-cocycle}
\end{multline}
Let $\l$ be a subset of $\h$, $F(x)$ is said to be of effective dynamics $\l$ 
if $F(x)$  is $\mathfrak{l}$-invariant and  if 
$(\id\otimes\id\otimes \nu(t))
( F_{12}(xq^{h_3}))=F_{12}(x)\;\;, \forall t\in \l^{\perp}.$
\end{theorem}

If $F(x)$ is a Quantum Dynamical coCycle and $R$ is the standard universal $R$-matrix of  $U_q({\mathfrak{g}})$, then one formally defines 
\begin{equation}
   R(x)=F_{21}(x)^{-1}R_{12}\,F_{12}(x).
   \label{R=FRF}
\end{equation} 
In the case where $\g$ is of finite type, $R(x)$ can be rigourously defined in the sense of \cite{BRT}  and satisfies the universal QDYBE. Indeed, in this case, for every finite dimensional representations $\pi,\pi'$ of $U_q(\g)$ acting on $V,V'$, $(\pi\otimes \pi')(R(x))$ is an element of
 $\End(V\otimes V')\otimes \C(x_1^2,\ldots,x_r^2),$ and the Universal Dynamical Yang-Baxter equation on $R(x)$ makes sense.

In the case where $\g$ is of  affine type, and if $\pi$ is a finite dimensional $\h$-simple representation of $U_q(\g)$ such that $(\pi\otimes \pi)(R(x))$ is a meromorphic function of $x$, then $(\pi\otimes \pi)(R(x))$ satisfies the QDYBE. If $(\pi\otimes \pi)(F(x))$ is of effective dynamics $\l$ then $(\pi\otimes \pi)(R(x))$ is also of effective dynamics $\l.$

The explicit construction, in the finite dimensional case, of   the universal Dynamical coCycle corresponding to the standard solution of the QDYBE has been done  in \cite{ABRR} by means of an auxiliary linear equation, the ABRR equation.
In the affine case, for the example of the standard IRF solution and for the vertex solution of Belavin-Baxter  type (case of $A_r^{(1)}$), the construction of the Dynamical coCycle  has been achieved by M. Jimbo, H. Konno, S. Odake, J. Shiraishi in \cite{JKOS}.
The former construction   has been extended to any generalized Belavin-Drinfeld triple in \cite{ESS},
whereas the latter has been extended to the affine case and to 
any generalized Belavin-Drinfeld triple being an  automorphism of ${\Gamma}$ by
P. Etingof, O. Schiffmann and A. Varchenko in \cite{ESV}. 
All these methods are using  a linear equation of modified 
ABRR type for the construction of  $F(x).$

We provide here a general construction of Quantum Dynamical Cocycles, using a method which generalizes known methods such as \cite{ABRR,JKOS,ESS}.

In the course of the proof we will need additional  spaces that we now 
introduce.
We define $(U_q(\b_+)\otimes U_q(\g)\otimes {\cal F}^{(0)}(\h))^{c(op)}$ as being the vector space of series
 $x=\sum_{\gamma,\gamma',\gamma''\in P^{\times 3}}(e_{\gamma}\otimes 1)x_{\gamma,\gamma',\gamma''}(1\otimes  f_{\gamma'}e_{\gamma''})$ with
 $x_{\gamma,\gamma',\gamma''}\in {\cal F}^{(2)}(\h).$
We define $(U^+_q(\g)\otimes U_q(\g)\otimes {\cal F}^{(0)}(\h))^{c(op)}$ as being the subspace  of elements  $x$ such that $x_{\emptyset,\gamma',\gamma''}=0.$
We can analogously define the vectorspace $(U_q(\b_+)\otimes U_q(\g)^{\otimes p}\otimes {\cal F}^{(0)}(\h))^{c(op)}$ and $(U^+_q(\g)\otimes U_q(\g)^{\otimes p}\otimes {\cal F}^{(0)}(\h))^{c(op)}$

We also analogously define   $(U_q(\g)\otimes U_q(\b_-)\otimes {\cal F}^{(0)}(\h))^{c(op)}$ as being the vector space of series  $\sum_{\gamma,\gamma',\gamma''\in P^{\times 3}}(f_{\gamma}e_{\gamma'}\otimes 1)a_{\gamma,\gamma',\gamma''}(1\otimes f_{\gamma''})$ with
 $a_{\gamma,\gamma',\gamma''}\in {\cal F}^{(2)}(\h).$
We define   $(U_q(\g)\otimes U^-_q(\g)\otimes {\cal F}^{(0)}(\h))^{c(op)}$ as being the subspace of elements $x$  such that $x_{\gamma,\gamma',\emptyset}=0.$

\begin{definition}{\label{def-GTQ} \em Generalized Translation Quadruple}\par\noindent
A generalized translation quadruple is a collection  $(\theta^{+},\theta^{-},\varphi^{0}, S^{(1)})$ such that
\begin{enumerate}
\item 
  $\theta^{\pm}:U_q({\b}_\pm)\rightarrow U_q({\b}_\pm)\otimes U_q(\h)\otimes {\cal D}^{\pm}(\h )$ 
are morphisms of algebra preserving the degree. We can therefore extend $\theta^{\pm}$ to  $(U_q({\b}_\pm)\otimes {\cal F}(\h)^{(0)})^c$ by  ${\cal F}(\h)^{(0)}$-linearity (and continuity).
It will be convenient to denote, for $v \in U_q({\b}_\pm)\otimes {\cal F}(\h)^{(0)}$,   
$\theta^{\pm}_{[x]}(v)=\theta^{\pm}(v)$ this fraction.
 Note that because of the degree preserving property we have  $\theta^{\pm}(U^{\pm}_q(\g))\subset U^{\pm}_q(\g)\otimes 
{\cal F}(\h)^{(0)}$, $\theta^{\pm}(U_q(\h))\subset U_q(\h).$

\item $\forall u \in U_q({\mathfrak h}), \forall v \in U_q({\b}_\pm)$ we have 
\begin{equation}
 [\;(\theta^{\pm}\theta^{\mp}-\id)(u)\;, \;\theta^{\pm}(v)\;]=0. \label{quasiInverse}
\end{equation}
\item 
  $\varphi^{0}$ and $S^{(1)}$ are invertible elements of 
  $U_q({\h})^{\widehat{\otimes} 2}$ such that
 $\log_q(\varphi^{0})$ and $\log_q(S^{(1)})$ belong to $\h^{\otimes 2}.$ 
\item $\theta^\pm$, $\varphi^0$ and $S^{(1)}$ satisfy to the following properties:
\begin{align}
  &\theta^{\pm}_{[xq^{h_2}]1}=(\Ad_{\varphi^0})^{\mp 1}\circ \theta^{\pm}_{[x]1},
              \label{axiomGTDb}\\
  &\theta^{+}_{[x]1}(\widehat{R})
   =\big(\Ad_{\varphi^{0}}\circ \theta^{-}_{[x]2}\big)(\widehat{R}),  
              \label{axiomGTDd}\\
  &\theta^{+}_{[x] 1}(\varphi^{0})=\theta^{-}_{[x] 2}(\varphi^{0})=\varphi^0,  
              \label{axiomGTDe}
\end{align}
as well as 
\begin{alignat}{2}
  &\theta^{+}_{[x] 1}( S^{(1)}_{12})=\theta^{-}_{[x] 2}( S^{(1)}_{12}),  & &
          \label{axiomGTDg}\\
  &\big[\;K_{12}S^{(1)\;-1}_{12}S^{(1)}_{21}\;\theta^{+}_{[x] 1}(K_{12}S^{(1)}_{12}S^{(1)\;-1}_{21})\;\varphi_{12}^{0}\;,\;\theta^{+}_{[x] 1}(v)\;\big]=0,\quad &
  &\forall v \in U_q({\mathfrak b}^{+}). 
                        \label{axiomGTDh}
\end{alignat}
\end{enumerate}
For $k\geq 1$, we will also define $S^{(k)}_{12}, W^{(k)}_{12}\in U_q(\h)^{\widehat{\otimes}2}$ as $S^{(k)}_{12} =(\theta^{+}_{[x] 1})^{k-1}(S^{(1)}_{12})$ and
$W^{(k)}_{12}=S^{(k)}_{12}(S^{(k+1)}_{12})^{-1}$. \\
A generalized translation quadruple is of effective dynamics $\l\subset\h$ if moreover 
\begin{equation}
  \forall u\in U_q(\b_\pm),\quad 
  \theta^{\pm}_{[xq^{h_2}]1}(u)\in U_q({\b}_\pm)\otimes U_q(\l)\otimes {\cal F}^{(0)}(\h). 
         \label{axiomGTDa}
\end{equation}
\end{definition}

\begin{definition}
A generalized translation quadruple  $(\theta^{+},\theta^{-},\varphi^{0}, S^{(1)})$ is non degenerate if \\
1. the restriction of $(\id-\theta^-)$ to $U_q^-(\g)$ is invertible.\\
2. the restriction of 
$(\id^{\otimes 3}-Ad_{(id\otimes \Delta)(\varphi^{(0)-1})}\circ \theta_1^+)$ to $ (U^+_q(\g)\otimes U_q(\g)^{\otimes 2}\otimes {\cal F}^{(0)}(\h))^{c(op)}$ is invertible.
\end{definition}

\begin{rem}
Let us remark that \eqref{axiomGTDh} implies that
\begin{alignat*}{2}
 0&=\big[\;K_{12}S^{(1)\;-1}_{12}S^{(1)}_{21}\;
          \theta^{+}_{[x] 1}(K_{12}S^{(1)}_{12}S^{(1)\;-1}_{21})\;\varphi_{12}^{0}\;,\;
          \theta^{+}_{[x] 1}(\widehat{R}_{13})\;\big] & & \\
 &=\big[\;K_{12}S^{(1)\;-1}_{12}S^{(1)}_{21}\;
          \theta^{+}_{[x] 1}(K_{12}S^{(1)}_{12}S^{(1)\;-1}_{21}\;\varphi_{12}^{0})\;,\;
          \theta^{+}_{[x] 1}(\widehat{R}_{13})\;\big] & &\text{(due to \eqref{axiomGTDe})}
\\
  &=\big[\;K_{12}S^{(1)\;-1}_{12}S^{(1)}_{21}\;(K_{32}S^{(1)}_{32}S^{(1)\;-1}_{23}\;
         \varphi_{32}^{0})^{-1}\;,\;\theta^{+}_{[x] 1}(\widehat{R}_{13})\;\big] &
         &\text{($\h$-invariance of $\widehat{R}$)}\\
  &=\big[\;K_{12}S^{(1)\;-1}_{12}S^{(1)}_{21}\;(K_{32}S^{(1)}_{32}S^{(1)\;-1}_{23}\;
         \varphi_{32}^{0})^{-1}\;,\;\theta^{-}_{[x] 3}(\widehat{R}_{13})\;\big] &
         &\text{(due to \eqref{axiomGTDd})}\\
  &=\big[\;\theta^{-}_{[x] 3}(K_{32}S^{(1)\;-1}_{32}S^{(1)}_{23})^{-1}\;
         (K_{32}S^{(1)}_{32}S^{(1)\;-1}_{23}\;\varphi_{32}^{0})^{-1}\;,\;
         \theta^{-}_{[x] 3}(\widehat{R}_{13})\;\big]\  & 
         &\text{($\h$-invariance of $\widehat{R}$)}.
\end{alignat*}
As a consequence,
\begin{equation}
 \big[\;\theta^{-}_{[x] 1}(K_{12}S^{(1)\;-1}_{12}S^{(1)}_{21})\;
    (K_{12}S^{(1)}_{12}S^{(1)\;-1}_{21})\;\varphi_{12}^{0}\;,\;
    \theta^{-}_{[x] 1}(v)\;\big]=0, \qquad
 \forall v \in U_q({\mathfrak b}_{-}).
\label{axiomGTDi}
\end{equation}
\end{rem}

\begin{rem}
If $\theta^{+}\vert_{U_q(\h)}$ is invertible with $\theta^{-}\vert_{U_q(\h)}$ as inverse, we will denote $S^{(0)}=\theta^-_{[x]1}(S^{(1)})$ and $W^{(0)}=S^{(0)}S^{(1)\;-1}.$ In this case (\ref{axiomGTDh}) can be rewritten as 
\begin{alignat}{2}
  &\big[\;K \theta^{+}_{[x] 1}(K)\varphi^{0}_{12}
\;(W_{12}^{(1)}\,W_{21}^{(0)})^{-1}\;,\;
\theta^{+}_{[x] 1}(v)
\;\big]=0,\qquad &%
 &\forall v \in U_q({\mathfrak b}_{+}). \label{axiomGTDh2}
\end{alignat}
\end{rem}

The following fundamental result holds:
 
\begin{theorem}{\label{th-cocycle}}\par\noindent
Let  $(\theta^{+},\theta^{-},\varphi^{0}, S^{(1)})$ be a given non degenerate generalized translation quadruple, the linear equation
\begin{equation}\label{ABRR2}
J(x)=W^{(1)}\;\theta^{-}_{[x] 2}\big(\widehat{R}\,J(x)\big)
\end{equation}
admits a unique solution $J(x)\in ( U_q(\n_+)\widetilde{\otimes} {\cal F}_{\h}^{(2)}\widetilde{\otimes}   U_q(\n_-))^c, $ under the assumption that
\begin{equation}\label{Jhyp}
(\iota^+_{\emptyset}\otimes \iota^-_{\emptyset})(\widehat{J}(x))=1\otimes 1,
\end{equation}
where $\widehat{J}_{12}(x)=S_{12}^{(1)}{}^{-1}J_{12}(x)$.
This solution satisfies the  QDCE \eqref{eq:s-cocycle}.\\
This solution can also be expressed as the infinite product
\begin{equation}\label{Jprod}
 \widehat{J}_{12}(x)=\prod_{k=1}^{+\infty} \widehat{J}_{12}^{(k)}(x)\quad \text{with}\quad
 \widehat{J}_{12}^{(k)}(x)=(\theta^{-}_{[x] 2})^k 
    \big( S_{12}^{(1)\; -1}\;\widehat{R}_{12}\;S_{12}^{(1)}\big).
\end{equation}
Moreover, such a cocycle $J(x)$ is of dynamics $ \l$ if the quadruple is of effective dynamics $\l$.
\end{theorem}

\begin{rem}
This solution $J(x)$ is of zero degree because $\widehat{R}$ is of degree zero and $\theta^-$ preserves the degree.
\end{rem}

\begin{rem}
Due to \eqref{axiomGTDd}, \eqref{axiomGTDe}, \eqref{axiomGTDg}, the expression \eqref{Jprod} of $\widehat{J}^{(k)}(x)$ can be rewritten as:
\begin{equation}\label{Jprod-bis}
    \widehat{J}_{12}^{(k)}(x)=
    \big[\Ad_{(\varphi_{12}^0)^{-k}}\circ(\theta^{+}_{[x] 1})^k \big]
    \big(S^{(1)\; -1}_{12}\;\widehat{R}_{12}\;S^{(1)}_{12}\big).
\end{equation}
It follows immediately that a solution $J(x)$ of \eqref{ABRR2}, \eqref{Jhyp} satisfies also the following linear equations:
\begin{align}
  &\theta^{+}_{[x] 1}(J_{12}(x))
          =\big[\Ad_{\varphi^0_{12}}\circ \theta^{-}_{[x] 2}\big](J_{12}(x)),\\
  &J_{12}(x)=W_{12}^{(1)}\;\big[\Ad_{(\varphi_{12}^0)^{-1}}\circ\theta^{+}_{[x] 1}\big]
                      \big(\widehat{R}_{12}\,J_{12}(x)\big).\label{ABRR1}
\end{align}
\end{rem}

\proof
Since $\theta^{\pm}$ preserves the degree, the proof of Proposition 3.1 of \cite{ESS} can be straightforwardly applied to our case. 
The linear  equation on $\widehat{J}(x)$ is
\begin{equation}
 \big[\id\otimes\id-\theta_{[x]2}^-(S_{12}^{(1)-1}\widehat{R}S_{12}^{(1)})\; \theta_{[x]2}^-\big]
 (\widehat{J}(x))=0.
 \label{linearJhat}
\end{equation}
This equation can be written as a system of linear  equations, triangular in term of the degree on the 
second space  as in  \cite{ESS}. If we write $\widehat{J}(x)=\sum_{n\geq 0}\widehat{J}_n$ where $\widehat{J}_n$ is of degree $-n$ on the second space, the linear equation can be written as:
\begin{equation}
(\id\otimes (\id-\theta^-))(\widehat{J}_n)=\sum_{i\atop 0\leq p<n}a_i \theta_{2}^-(\widehat{J}_p )
 \quad\text{with}\ a_i\in (U_q(\n_+)\otimes U_q(\n_-))U_q(\h)^{\widehat{\otimes}2}.
\end{equation}
The existence and uniqueness of the solution is a consequence of the initial condition $\widehat{J}_0=1^{\otimes 2}$ and the invertibility of the  linear operator 
$(\id-\theta^-)$ restricted to $U^{-}_q(\g)\otimes {\cal F}(\h)^{(0)}$.

It is easy to see, using \eqref{axiomGTDg}, that a solution of \eqref{ABRR2},\eqref{Jhyp} can be represented as an infinite product of the form \eqref{Jprod}.
The convergence of this product holds in the following sense:
for all  $\gamma, \gamma'\in P$, $a_{\gamma,\gamma'}=(\iota_{\gamma}^+\otimes \iota_{\gamma'}^-)(\prod_{k=1}^{+\infty} \widehat{J}_{12}^{(k)}(x))$ is a formal series in $x_0^2, \ldots, x_{r}^2, x_d^2$ with coefficients in $U_q(\h)^{{\widehat \otimes} 2}.$ The uniqueness of the solutions of the linear equation (\ref{ABRR2}) implies that   this formal series defines a unique element   of ${\cal F}^{(2)}(\h).$ 
We will use heavily the representation in term of infinite product because it highlights 
the computations.

Let now $J(x)$ be the solution  of \eqref{ABRR2}, \eqref{Jhyp}, we will show that it satisfies the QDCE.
Let us consider the element \begin{equation}
Y_{123}(x)=(\id \otimes \Delta)(J(x)^{-1})\,
(\Delta\otimes \id)(J(x))\,J_{12}(xq^{h_3}),
\end{equation} 
in order to prove that $Y_{123}(x)=J_{23}(x)$, {\it i.e.} that $J(x)$ is solution of the QDCE~\eqref{eq:s-cocycle}, we first notice that $Y_{123}(x)$ satisfies the following properties:
\begin{align}
  &\big(\varphi_{12}^{(0)}\,\varphi_{13}^{(0)}\big)^{-1}\;
   \theta^+_{[x] 1}(Y_{123}(x))\;
   \varphi_{12}^{(0)}\,\varphi_{13}^{(0)}
   = Y_{123}(x),
                    \label{property2}\\
  &W_{23}^{(1)}\; \theta^-_{[x] 3}\big(\widehat{R}_{23}\, Y_{123}(x)\big)
     = Y_{123}(x),
                    \label{property3}\\
 &(\iota_{\emptyset}^+\otimes \id\otimes \iota_{\emptyset}^-)(S_{23}^{(1)\; -1}\, Y_{123}(x))=1^{\otimes 3}, \label{property1a}\\
 & \forall\gamma'\not=\emptyset,\qquad \quad
  (\iota_{\emptyset}^+\otimes \id\otimes \iota_{\gamma'}^-)(S_{23}^{(1)\; -1}\, Y_{123}(x)) 
  \in (1\otimes  U_q^+({\mathfrak g})\otimes 1)(1\otimes {\cal F}^{(2)}(\h)), 
              \label{property1b}\\
&\forall \gamma\not=\emptyset,\ \forall \gamma',\quad 
 (\iota_{\gamma}^+\otimes \id\otimes \iota_{\gamma'}^-)(S_{23}^{(1)\; -1}\, Y_{123}(x)) 
 \in (1\otimes  U_q({\mathfrak g})\otimes 1)( {\cal F}^{(3)}(\h))\;\;. 
              \label{property1c}
\end{align}
The last three properties mean that the element
$\big[ S_{23}^{(1)\; -1}\, Y_{123}(x)-1^{\otimes 3}\big]$ lies  in an extension of  
$\big[\left( 1\otimes U_q^+({\mathfrak g})\otimes U_q^-({\mathfrak g}) \right)
\oplus \left(U_q^+({\mathfrak g}) \otimes U_q({\mathfrak g}) 
                                          \otimes U_q({\mathfrak b}^-)\right)
\big]\otimes {\cal F}^{(0)}(\h).$

The proof of these properties can be found in Appendix~\ref{append-cocycle} (cf. Lemma~\ref{lem1}, Lemma~\ref{lem2}, Lemma~\ref{lem3}).

Using then the fact that the kernel of   $(\id^{\otimes 3}-Ad_{(id\otimes \Delta)(\varphi^{(0)-1})}\circ \theta_1^+)$ is zero 
and property (\ref{property2}), we deduce  that
$Y_{123}(x)=1\otimes Z(x)$ for a certain $Z(x) \in 1^{\otimes 2} \oplus (U_q^{+}({\mathfrak g})\otimes U_q^{-}({\mathfrak g})){\cal F})^{(2)}(\h)^c.$ 
Using now the property (\ref{property3}) and the fact
that $\ker(\theta^{-}-\id) \cap (U_q^{-}({\mathfrak g})\otimes {\cal F}^{(0)}(h))^c=\{0\}$, we obtain that $Z(x)=J(x).$
\cqfd

\bigskip

We will emphasize special classes of generalized translation quadruples for which the analysis of the Quantum Dynamical Gauge Transformation can be described in a simple way.

\begin{definition}{\em Generalized Translation Datum}
 \par\noindent
A {\em generalized translation datum} is a collection $(\theta^+,\theta^-,\varphi^0,\varphi^+,\varphi^-,S^{(1)})$, where
$(\theta^+,\theta^-,\varphi^0,S^{(1)})$ is a generalized translation quadruple, and
$\varphi^{+}$, $\varphi^{-}$ are invertible elements of  $U_q({\h})^{\widehat{\otimes} 2}$ such that 
\begin{align}
  &\log_q(\varphi^{\pm})\in \h^{\otimes 2},\\
  &\Delta\circ\theta^{\pm}_{[x]}=\Ad_{\varphi^{\pm}}\circ 
   \big(\theta^{\pm}_{[x]}\otimes\theta^{\pm}_{[x]} \big)\circ \Delta.
              \label{axiomGTDc}
\end{align}
A generalized translation datum is said to be of effective dynamics $\l\subset\h$ if the corresponding generalized translation quadruple is of effective dynamics $\l$.
\end{definition}

\begin{definition}{\em Vertex and IRF types}
 \par\noindent
A generalized translation datum is said {\em of Vertex type} if 
it is of effective dynamics $\l\subset\C c$ in the affine case and $\l=0$ in the finite dimensional case.
\par\noindent
A generalized  translation datum is of {\em Restricted Vertex type} if it is of vertex type and moreover satisfies,  $\forall v\in U_q({\b}_+)$,
\begin{align}
   &\big[\,\varphi^{0}\, , \,v_2\,\big]=0,\label{axiomVertexa}\\
   &\big[\,\varphi^{+}\, , \,\theta_{[x]2}^+(v)\,\big]=0,\label{axiomVertexb}\\
   &\big[\,  \,K^{-1}\,
     (\theta^{+}_{[x]}\otimes\theta^{+}_{[x]})(K)\,\varphi_{21}^{+}\, , \,
     \theta_{[x]2}^+(v)\, \big]=0.\label{axiomVertexd}
\end{align}
\par\noindent
A generalized translation datum is said to be {\em of IRF type} if 
it is of effective dynamics $\l= \h$ and 
\begin{equation}\label{axiom-face}
  \theta^\pm\vert_{ U_q(\h)}=\id.
\end{equation}
A generalized  translation datum is of {\em Restricted IRF type} if it is of IRF type and moreover satisfies,  $\forall v\in U_q({\b}_-)$,
\begin{align}
 &\big[\,(\varphi^0)^{-1}\,\varphi^-\, ,\; \theta_{[x]1}^-(v)\,\big]=0,
                  \label{axiomPsi0Phi-}\\
 &\big[\, \varphi_{21}^{-}\,(\varphi^0_{12})^{-1}\, ,\;
      \theta_{[x]1}^-(v)\,\big]=0.
                   \label{axiomPhi-Psi0K}
\end{align}
\end{definition}

A restricted Vertex type generalized translation datum is said to be non degenerate if it is 	 associated to a non degenerate generalized quadruple and if moreover the operator $(\id-\Ad_{W_{12}^{(1)}(\varphi_{12}^{0})^{-1}}\circ \theta_1^+)$ defined on   $(U_q^+(\g)\otimes U_q(\g)\otimes {\cal F}^{(0)}(\h))^{c(op)}$ is invertible.

Although it would be very interesting to classify the generalized translation data we will adopt in this paper a more modest goal and show that the notion of generalized translation data  encompass the relevant Vertex and IRF examples.


\subsection{Basic examples of Quantum Dynamical coCycles}
\label{subsec-Examples}

\subsubsection{Quantization of Belavin-Drinfeld classical $r-$matrices}
\label{subsec-BelavinDrinfeld}

We show here that the formalism of Generalized Translation quadruple introduced in Definition~\ref{def-GTQ} gives an explicit quantization of  classical  $r-$matrices associated to any Belavin-Drinfeld triple. We think that this construction simplifies the work of  \cite{ESS} in the case of a nilpotent generalized Belavin-Drinfeld triple. 

Let ${\cal T}=(\Gamma_1,\Gamma_2,T)$ be a Belavin-Drinfeld triple, we define  $\varphi^0=1$  and set $\theta^{\pm}_{[x]}=\sigma^{\pm}.$ The effective dynamics is therefore $\l=\{0\}.$
The non degeneracy condition on $\theta^\pm_{[x]}$ is implied by the nilpotence condition (\ref{nilpotency}) of $T,$  equation (\ref{quasiInverse}) is implied from the definitions of $\sigma^{\pm}$ and  equation (\ref{axiomGTDd}) is   a consequence of the fact that $T$ preserves the scalar product.

It remains to compute $S^{(1)}$ such that all the axioms of a generalized Translation quadruple are satisfied. 

Having defined $S^{(1)}=q^{s+m}$, with $s\in \bigwedge^2 \h$, $m\in 
\text{Sym}({\mathfrak h}^{\otimes 2}),$  the relation (\ref{axiomGTDh}) is strictly equivalent to the requirement that $s$ is solution of (\ref{Sbelavindrinfeld}) ($m$ being not constrained by this relation).
Indeed, denoting $\Omega_{\overset{\circ}{\mathfrak h}}=\sum_{\alpha,\beta\in \Gamma}\Omega_{\alpha \beta}\,\overset{\circ}{\zeta}{}^{\alpha} \otimes \overset{\circ}{\zeta}{}^{\beta}$, and $s=\sum_{\alpha,\beta\in \Gamma} s_{\alpha \beta}\,\overset{\circ}{\zeta}{}^{\alpha} \otimes \overset{\circ}{\zeta}{}^{\beta}$, the equation (\ref{axiomGTDh}) is equivalent to 
\begin{equation*}
 \big[\, (\Omega_{\alpha \beta}-2s_{\alpha \beta}) \,
         \overset{\circ}{\zeta}{}^{\alpha} \otimes \overset{\circ}{\zeta}{}^{\beta} + 
  (\Omega_{\alpha \beta}+2s_{\alpha \beta})\, 
         \overset{\circ}{\zeta}{}^{T(\alpha)} \otimes \overset{\circ}{\zeta}{}^{\beta}\, ,\, 
  e_{T(\gamma)}\otimes 1\,\big]=0,\qquad \forall\gamma\in\Gamma_1,
\end{equation*}
{\it i.e.} 
\begin{equation*}
 (\Omega_{\alpha \beta}-2s_{\alpha \beta}) + (\Omega_{T^{-1}(\alpha) \beta}+2s_{T^{-1}(\alpha) \beta})=0,\qquad \forall\beta \in \Gamma, \ \forall\alpha \in \Gamma_2.
\end{equation*}
Changing $\alpha$ to $\alpha'=T^{-1}(\alpha)$ in the previous relation we obtain the relation (\ref{Sbelavindrinfeld}), {\it i.e.} 
\begin{equation*}
   (\Omega_{T(\alpha') \beta}-2s_{T(\alpha') \beta}) + (\Omega_{\alpha' \beta}+2s_{\alpha' \beta})=0,
   \qquad \forall \beta \in \Gamma, \ \forall\alpha' \in \Gamma_1.
\end{equation*}

It remains to show that, having fixed a particular Belavin-Drinfeld quadruple $({\cal T},s),$ there always exists $m=\sum_{\alpha,\beta\in \Gamma} m_{\alpha \beta}\,\overset{\circ}{\zeta}{}^{\alpha} \otimes \overset{\circ}{\zeta}{}^{\beta}$ solution of  (\ref{axiomGTDg}), {\it i.e.} such that
\begin{alignat}{2}
 &s_{\alpha \beta}+m_{\alpha \beta}=0,\qquad  & 
    &\alpha \in \Gamma\setminus\Gamma_2,\ \beta \in \Gamma_2,\label{contraintem1}\\
 &s_{\alpha \beta}+m_{\alpha \beta}=0,\qquad &
    &\beta \in \Gamma\setminus\Gamma_1,\ \alpha \in \Gamma_1,\label{contraintem2}\\
 &s_{\alpha \beta}+m_{\alpha \beta}=s_{T(\alpha) T(\beta)}+m_{T(\alpha) T(\beta)},\quad & 
    &\alpha \in \Gamma_1,\ \beta \in \Gamma_1.\label{contraintem3}
\end{alignat}
Let us describe the set of solutions of these equations and show that it is not empty. We define
\begin{alignat}{2}
  &A^{0}_1=\{ (\alpha,\beta), \beta \in \Gamma\setminus\Gamma_1, \alpha \in  \Gamma_1\},\quad 
   &&A^{0}_2=\{ (\alpha,\beta), \alpha \in \Gamma\setminus\Gamma_2, \beta \in  \Gamma_2\},\\
  &A^{0}=A^{0}_1 \cup A^{0}_2, \quad 
   &&B^0=A^{0}\cap (\Gamma_1\times \Gamma_1),\\
  &B^{k+1}=\{ (T(\alpha),T(\beta)),\mbox{ for } (\alpha,\beta) \in B^k \cap &&(\Gamma_1\times \Gamma_1)\}.
\end{alignat}
Note that, since $((\Gamma\setminus\Gamma_i) \times \Gamma_i) \cap (\Gamma_i \times (\Gamma\setminus\Gamma_i))=\emptyset$ for $i=1,2$, $(\beta,\alpha)\notin A^{0}$ if $(\alpha,\beta)\in A^{0}$. Therefore,  $\forall (\alpha,\beta)\in A^{0}$,  we have  $m_{\alpha \beta}=-s_{\alpha \beta}$, and 
$m_{\beta \alpha}=m_{\alpha \beta}$. 

The condition (\ref{Sbelavindrinfeld}) implies that,  $\forall\alpha,\beta \in \Gamma_1$,  $s_{\alpha \beta}=s_{T(\alpha) T(\beta)}.$ Indeed,
\begin{align} 
  s_{\alpha \beta}-s_{T(\alpha) T(\beta)}
  &=(s_{\alpha \beta}-s_{T(\alpha) \beta})-(s_{\beta T(\alpha)}-s_{T(\beta) T(\alpha)}) \nonumber\\
  &=\frac{1}{2}(\alpha+T(\alpha),\beta)-\frac{1}{2}(\beta+T(\beta),T(\alpha)))\nonumber \\
  &= \frac{1}{2}(\alpha,\beta)-\frac{1}{2}(T(\beta),T(\alpha))=0.
\end{align} 
Because of this property, we have  $m_{T(\gamma) T(\delta)}=m_{\gamma \delta},$ for $(\gamma,\delta) \in B^{k}\cap (\Gamma_1\times \Gamma_1)$. For any $(\gamma,\delta) \in B^{k},$ there exists $(\alpha,\beta) \in B^0$ such that $\gamma=T^k(\alpha),\ \delta=T^k(\beta)$; as a result, $m_{\gamma \delta}=m_{\alpha \beta}=-s_{\alpha \beta}.$ 
Compatibility of our definitions has to be checked as soon as there exist $k<l$ such that $B^k\cap B^l \not= \emptyset.$ Indeed, for such $k<l$, let $(\gamma,\delta)\in B^k\cap B^l$, and $(\alpha,\beta),\ (\alpha',\beta')$ two elements of $B^0$ such that $\gamma=T^{k}(\alpha)=T^{l}(\alpha'),\ \delta=T^{k}(\beta)=T^{l}(\beta').$ To be consistent, our definition requires that $s_{\alpha \beta}=s_{\alpha' \beta'}.$ This relation is satisfied since  $\alpha'=T^{k-l}(\alpha),\ \beta'=T^{k-l}(\beta)$ and   $s_{\alpha \beta}=s_{T^{k-l}(\alpha) T^{k-l}(\beta)}.$ 
Note finally that, due to the nilpotency of $T$, there exists an integer $N$ such that $B^N=\emptyset.$ Therefore  $m_{\alpha \beta}$ and $m_{\beta \alpha}$ are fixed for $(\alpha,\beta)$ in $A^0 \bigcup (\cup_{k=1}^N B^k)$, the other coefficients being unconstrained.

As a conclusion, the previous study shows that it is possible to build a generalized translation quadruplet associated to  any Belavin-Drinfeld Quadruple leadind to an explicit  quantization of the corresponding classical $r$-matrix.


\bigskip

Let us now give the explicit expression of $J$ in the cases of the  Belavin-Drinfeld triples already discussed.
\begin{itemize} 
 \item[-] {$\Gamma_1=\Gamma_2=\emptyset$}:

We define $\theta^{\pm}_{[x]}=i\circ \epsilon$, where $\epsilon$ is the counit and $i$ is the trivial embedding of $\C$ in $U_q(\g)$, 
and $\varphi^{0}=1$, $S^{(1)}=q^s, s\in \bigwedge^2(\h).$
We have $\widehat{J}(x)=1$ and  $\l=\{0\}.$

  \item[-] {Cremmer-Gervais's solution}:

Let $\g=A_r$ with $r\geq 2$ and  let us choose the following generalized translation datum:
\begin{alignat}{4}
&\theta^{+}_{[x]}(e_i)=e_{i-1},\quad &  &\theta^{+}_{[x]}(e_1)=0,\quad & &\theta^{-}_{[x]}(f_i)=f_{i+1},\quad &  &\theta^{-}_{[x]}(f_r)=0,\\
&\theta^{+}_{[x]}(\zeta^{\alpha_i})=\zeta^{\alpha_{i-1}},\quad & 
&\theta^+_{[x]}(\zeta^{\alpha_1})=0,\quad & 
&\theta^{-}_{[x]}(\zeta^{\alpha_i})=\zeta^{\alpha_{i+1}},\quad & 
&\theta_{[x]}^-(\zeta^{\alpha_r})=0,
\end{alignat}
and
\begin{equation}
  \varphi^{0}=1,\quad 
   S^{(1)}=q^{\sum_{i=1}^{r-1}\zeta^{\alpha_i}\otimes\zeta^{\alpha_{ i+1 }}},\quad
   \varphi^{+}=q^{-\zeta^{\alpha_{r-1}}\otimes\zeta^{\alpha_r}},\quad
   \varphi^{-}=q^{-\zeta^{\alpha_1}\otimes\zeta^{\alpha_2}}.
\end{equation}
The effective dynamics is  $\l=\{0\}$.
The expression of  $J(x)$  is the same as the one given in \cite{BRT}.
\end{itemize}

\subsubsection{ Standard IRF  solutions}
\label{subsec-StandardIRF}
We define $\A(\h)$ the commutative Hopf algebra generated by $x_i^{h}$, $h\in \h+\C 1$, with $i=1,\ldots,r$ in the finite case and $i=0,1,\ldots,r,d$ in the affine case, and with the relations $x_i^{h+h'}=x_i^h x_i^{h'}.$
We define $B(x)\in \A(\h)\otimes U_q(\h)$ by
\begin{equation}
  B(x)=\begin{cases}
      k^2\prod\limits_{j=1}^r x_j^{2\overset{\circ}{\zeta}{}^{\alpha_j}}
            &\text{if $\g$ is of finite type,}\\
      k^2\prod\limits_{j=0}^r x_j^{2{\zeta}{}^{\alpha_j}}x_d^{2\zeta^d} 
            &\text{if $\g$ is of affine type}.
       \end{cases}
\end{equation}
It  satisfies the relations:
\begin{equation}
   \Delta(B(x))=B_1(x)B_2(x)K^2,\qquad B_1(xq^{h_2})=B_1(x)K^2.
\end{equation}
The standard IRF type solution is obtained with the following non degenerate generalized translation datum:
\begin{equation}
\theta^{\pm}_{[x]}=\Ad_{B(x)}^{\pm 1},\quad \varphi^{+}=K^2,\quad \varphi^{-}=
\varphi^{0}=K^{-2},\quad S=1.\label{valeurStandardIRF}
\end{equation}

If $\g$ is a simple finite dimensional Lie algebra, then $\l=\h,$ and
the generalized linear difference equation is the ABRR equation has been written and solved in \cite{ABRR}.

In the case where ${\g}$ is an affine algebra, this solution does not depend on $x_d$ because $x_d^{2\zeta_d}$ commutes with 
$\widehat{R}.$ As a result this solution, which is {\it a priori} of effective dynamics $\h$, is also of effective dynamics 
$\l=\oplus_{i=0}^r\C h_i=\overset{\circ}{\h}\oplus\C c.$
The formula (\ref{Jprod}) for the twist $J(x)$ in the affine case has been obtained in \cite{JKOS}. 

\subsubsection{ Belavin-Baxter Elliptic Vertex solutions}
\label{subsec-BelavinBaxter}

In the work \cite{ESV}, the  quantization of $r$-matrices associated to a generalized Belavin-Drinfeld triplet of an affine Lie algebra when $T$ is an automorphism has been  constructed. It is possible to obtain a Vertex solution ({\it i.e.} with the dynamics reduced to $\l=\mathbb{C} c$) in the affine type case only if the corresponding Belavin-Drinfeld triple is such that $\Gamma_1=\Gamma_2=\Gamma$ and $T$ does not leave  invariant any proper sub-diagram of the Dynkin diagram. As a consequence, in this case, ${\g}=A_r^{(1)}$ and $T=T_{\aleph}$  has to be a rotation sending node $\lfloor i \rfloor$ on node $\lfloor i+\aleph \rfloor$, with 
$1\leq \aleph\leq r$ prime to $(r+1)$, where $\lfloor k \rfloor$ denotes the unique element of 
$\{0,\ldots,r\}$ congruent to $k \mod (r+1).$ 

We define such a datum in the following way.

We denote $p(x)\in {\cal D}^{-}({\h})$ the component of $x$ along $q^c$, {\it i.e.} $p(x)=\prod_{i=0}^r{x_i}$. The morphisms $\theta^{\pm}_{[x]}$ are defined as
\begin{equation}\label{datum-V1}
  \theta^{\pm}_{[x]}=\Ad_{D^{\pm}(p(x))}\circ \sigma^{\pm},\qquad \text{with}\quad
  D^{+}(p)=p^{\frac{2\varpi}{r+1}},\quad
  D^{-}(p)=p^{-\frac{2\varpi}{r+1}}q^{-\frac{2}{r+1}c\varpi},
\end{equation}
and $\sigma^\pm$ are defined as
\begin{alignat}{2}\label{datum-V2}
  &\sigma^{+}(e_i)=e_{\lfloor i-\aleph \rfloor}, &\qquad
  &\sigma^{-}(f_i)=f_{\lfloor i+\aleph \rfloor},\\
          \label{datum-V2bis}
  &\sigma^+(\zeta^{\alpha_j})=\zeta^{\alpha_{\lfloor j-\aleph \rfloor}} + v_j\, \zeta^d, & \quad
  &\sigma^-(\zeta^{\alpha_j})=\zeta^{\alpha_{\lfloor j+\aleph \rfloor}} - v_{\lfloor j+\aleph\rfloor}\, \zeta^d,\qquad
  \sigma^\pm(c)=c,
\end{alignat}
with
\begin{equation}\label{datum-V3bis}
  v_j=\frac{1}{2(r+1)}
    \big( j(r+1-j)- \lfloor j-\aleph \rfloor (r+1-\lfloor j-\aleph \rfloor)  \big).
\end{equation}
With this value of $v_j$ we have $(\sigma^{\pm}){}^{\otimes 2}(\Omega_{\h})=\Omega_{\h}.$
Note that \eqref{datum-V2bis} is equivalent to
\begin{equation}\label{datum-V4}
  \sigma^{\pm}(h_{\alpha_i})=h_{\alpha_{\lfloor i\mp \aleph \rfloor}},\qquad \sigma^{\pm}(\varpi)=\varpi. 
\end{equation}
Note that, because of the presence of $p^{\pm \frac{2\varpi}{r+1}}$ in the definition of $D^{\pm}$, $\theta^{\pm}$ actually maps $U_q(\b_\pm)$ to  $U_q(\b_\pm)\otimes \widetilde{\cal D}^{\pm}(\h)$, where 
$\widetilde{\cal D}^{\pm}(\h)=\C[\tilde{x}_0^{\mp 2},\ldots,\tilde{x}_r^{\mp 2}]$ with $\tilde{x}_i^{r+1}=x_i.$
We will therefore prove that the axioms of a generalized datum are satisfied after having replaced 
$ {\cal D}^{\pm}(\h)$ by 
$\widetilde{\cal D}^{\pm}(\h),$ which corresponds to a minor generalization of the axioms. 
The only non trivial property is the proof non degeneracy of this generalized datum. The restriction of  $(\theta^{-}-\id)$ to $U^{\pm}(\g)\otimes \widetilde{\cal F}^{(0)}(\h)$ is invertible as a  direct consequence of the identity
\begin{equation}
(1-\theta^{-})^{-1}=\sum_{n=0}^{+\infty}(\theta^{-})^n=
\sum_{n=0}^r\frac{\Ad^n_{D^{-}(p)}}{1-\Ad^{r+1}_{D^{-}(p)}}(\sigma^{-})^n.
\end{equation}

We will also choose:
\begin{align}
         &\varphi^{0}=q^{-\frac{2}{r+1}\varpi\otimes c},\quad
          \varphi^{+}=1,\quad
          \varphi^{-}=q^{-\frac{2}{r+1}(\varpi\otimes c+c\otimes \varpi)}.
                                 \label{datum-V5}
\end{align}
One can proceed analogously for the non degeneracy condition on $\theta^+.$

Concerning the expression of $S^{(1)}_{12}$, a precise analysis of the constraints leads to the solution  \eqref{ExpressionGenS1} as we will see.
In the particular case $\aleph=1$, it gives
\begin{multline}\label{ExpressionS1}
 \log_q(S_{12}^{(1)})=-\sum_{i=0}^r\zeta^{\alpha_i}\otimes 
                   (\zeta^{\alpha_i}-\zeta^{\alpha_{\lfloor i+1 \rfloor}})\\
  +\sum_{i=0}^r \frac{\lfloor i\rfloor}{r+1}(\zeta^{\alpha_i}\otimes c
            -c\otimes \zeta^{\alpha_{\lfloor i+1 \rfloor }})
  -\frac{1}{2(r+1)}(\varpi\otimes c+c\otimes \varpi).
\end{multline}

\bigskip

In order to explain the choice of such a datum, it is convenient to introduce the following notations.

Let $\varkappa_0,\ldots,\varkappa_r,\varkappa_d$ be the canonical basis vectors of $\mathbb{C}^{r+2},$ and $\varkappa^0,\ldots,\varkappa^r,\varkappa^d$ the dual basis vectors. 
Let $(E_{\mu}^{\nu})_{\mu,\nu\in \{0,\ldots,r,d\}}$ be the canonical basis of $(r+2)\times (r+2)$ matrices defined by $E_{\mu}^{\nu}=\varkappa_{\mu}{}^t\varkappa^{\nu},$  obeying $E_{\mu}^{\nu} E_{\rho}^{\sigma}=E_{\mu}^{\sigma}\delta_{\nu,\rho}$. 
We denote by $I_{r+2}$ the identity matrix and by $P_n$ the projectors
 $P_{n}=\sum_{i=0}^n E_i^i.$ 
The Cartan matrix of $A_{r}^{(1)}$ is extended to an $(r+2)\times (r+2)$ matrix $\underline{A}=\sum_{i,j=0}^{r}a_{ij}E^{j}_{i}=2 P_{r+1}-Y-{}^{t}Y$ where $
   Y=\sum_{i,j=0}^{r}\delta_{i,\lfloor j-1 \rfloor}E^{j}_{i}.$
We define 
\begin{equation}
{\cal   A}=\underline{A}+E^{0}_{d}+E^{d}_{0}.
\label{DefAffineInitiale}
\end{equation} 

Let $\Theta^{\pm}$ be the matrix of $\sigma^{\pm}$ in the basis $\zeta^{\alpha_0},\ldots, \zeta^{\alpha_r}, \zeta^d$.
Let $\sigma^+$ defined on  $\h$ as in \eqref{datum-V2bis},
we therefore have: $\Theta^{+}=Y^\aleph+E^d_d+\sum^r_{i=0}v_i E^i_d= Y^\aleph+E^d_d+ \varkappa_d .{}^t v,$ where  ${}^{t}v=\sum_{i=0}^r v_i \varkappa^i$.
 We now prove that the corresponding choice of  $v_i$ is equivalent to  $(\sigma^+)^{\otimes 2}(K)=K$.
This last condition is equivalent to
\begin{equation}
 \Theta^{+}{\cal  A}\; {}^{t}\Theta^{+}={\cal A}, \label{sigmaK=K}
\end{equation}
which  reduces to the following equations on $v$:
\begin{equation}\label{sigmaK=Kbis}
 \underline{A}v= ({}^{t}Y^\aleph-I_{r+2})\varkappa_0\qquad 
  \text{and} \qquad 
  {}^{t}v({}^{t}Y^\aleph+I_{r+2})\varkappa_0=0. 
\end{equation}
In order to solve these equations, let us introduce
\begin{equation}
  T=-\frac{1}{2(r+1)}\sum_{l=1}^{r}l(r+1-l)Y^l. 
\end{equation}
We have 
\begin{equation}
   \underline{A}T = T\underline{A}=\Pi,\qquad
   \text{with}\quad \Pi=P_{r+1} -\frac{1}{r+1}w.{}^{t}w,
\end{equation}
where  $w=\sum_{i=0}^{r}\varkappa_i$
and $\Pi$
is an orthogonal  projector, commuting with $Y$, of kernel $\C w\oplus\C\varkappa_d.$ 
As a consequence, we obtain that the solution of the relations \eqref{sigmaK=Kbis} is given by 
\begin{equation}
  v=T({}^{t}Y^\aleph-I_{r+2})\varkappa_0=-\frac{1}{2(r+1)}\sum_{j=0}^{r}
    \big( \lfloor j-\aleph \rfloor (r+1-\lfloor j-\aleph \rfloor) - j(r+1-j) \big)\varkappa_j,
\end{equation}
which justifies the choice \eqref{datum-V3bis} for $v_j$.
The action \eqref{datum-V4} of $\sigma^+$ on $h_{\alpha_j}$, $0\le j\le r$, and on $\varpi$ follows then directly from the fact that
\begin{align}
 &\big(\Theta^+ {\cal A}\big)_\mu^i=\sum_{j=0}^r\big(Y^\aleph\big)_j^i\,\big(\underline{A}+E_0^d\big)_\mu^j,
  \qquad \mu\in\{0,\ldots,r,d\},\ i\in\{0,\ldots,r\},\\
 &(\Theta^+-I_{r+2}) w =0.
\end{align}
Note that the kernel of $\Theta^+I_{r+2}$ is generated by $\varkappa_d$ and $ w.$
\bigskip

Let $\varphi^0$ be given as in \eqref{datum-V5}. We now show that  $S^{(1)}_{12}$ exits and give an  explicit expression of it.
Let us denote 
\begin{equation}
\log_q(S^{(k)})=\sum_{\mu,\nu=0,\ldots,r,d}\varsigma_{\mu\,\nu}^{(k)} \zeta^{\alpha_{\mu}}\otimes \zeta^{\alpha_{\nu}}  ,\qquad 
  {\cal S}^{(k)}=\sum_{\mu,\nu=0,\ldots,r,d}\varsigma_{\mu\,\nu}^{(k)} E^{\nu}_{\mu},\quad \forall k\in \N.
\end{equation}
The equation  (\ref{axiomGTDh2}) can be solved as
\begin{equation}
  W_{12}^{(0)}=K_{12}(\varphi_{12}^{0}\varphi_{21}^{0})^{1/2},
\end{equation}
which can be rewritten as:
\begin{equation}\label{jaugeS}
  (1-\Theta^+){\cal S}^{(0)}={\cal A}+\Phi, \qquad\text{with}\quad
  \Phi=-\frac{1}{(r+1)}( \varkappa_d .{}^{t}w + w.\varkappa^d ). 
\end{equation} 
Because ${}^{t}w(1-\Theta^+)=0$, we must have  ${}^{t}w({\cal A}+\Phi)=0$, which is the case for the choice  $\varphi^{0}= q^{-\frac{2}{r+1}\varpi\otimes c}.$
In order to solve \eqref{jaugeS}, let us define 
\begin{equation}
  \Omega^{(\aleph)}= - \frac{1}{r+1}\sum^r_{l=1} l Y^{\aleph l},
\end{equation}
then
\begin{equation}
   \Omega^{(\aleph)} (1-Y^\aleph) =(1-Y^\aleph)\Omega^{(\aleph)}
   =\frac{r}{r+1} P_{r+1} -\frac{1}{r+1}\sum^r_{l=1}Y^{\aleph l}= \Pi.
\end{equation}
$(1-\Theta^+)$ admits the following quasiinverse
\begin{equation}  
 (\Omega^{(\aleph)}-\varkappa_d.{}^{t}v\,(\Omega^{(\aleph)})^2)(1-\Theta^+)
 = (1-\Theta^+)(\Omega^{(\aleph)}-\varkappa_d.{}^{t}v\,(\Omega^{(\aleph)})^2)
 =\Pi^{(\aleph)}, 
\end{equation}
where 
\begin{equation}
  \Pi^{(\aleph)}=\Pi-\varkappa_d.{}^{t}v\,\Omega^{(\aleph)}
\end{equation}
is a projector commuting with $\Theta^+$ with kernel reducing to  $\C w\oplus\C\varkappa_d.$ 
As a consequence, we have 
\begin{equation}
  \Pi^{(\aleph)} {\cal S}^{(0)}=(\Omega^{(\aleph)}-\varkappa_d.{}^{t}v\,(\Omega^{(\aleph)})^2)({\cal A}+\Phi).
\end{equation}
A direct computation shows that the right handside of the previous equation satisfies   
\begin{equation}
  \Theta^+ \big( \Pi^{(\aleph)}(\Omega^{(\aleph)}-\varkappa_d.{}^{t}v\,(\Omega^{(\aleph)})^2)({\cal A}+\Phi) \big) {}^{t}\Theta^+ = \big( \Pi^{(\aleph)}(\Omega^{(\aleph)}-\varkappa_d.{}^{t}v\,(\Omega^{(\aleph)})^2)({\cal A}+\Phi) \big),
\end{equation}
which is equivalent to 
$\theta^{+}_{[x]1}(S_{12}^{(0)})=\theta^{-}_{[x]2}(S_{12}^{(0)}).$

As a result we obtain that a solution  of (\ref{axiomGTDg}) and (\ref{axiomGTDh}) is 
\begin{multline}
 {\cal S}^{(0)}=\Omega^{(\aleph)} \underline{A} 
  + \varkappa_d.\Big(\varkappa^0-\frac{1}{r+1}{}^{t}w\Big)\,\Omega^{(\aleph)} \\
  + \Omega^{(\aleph)}\, \Big(\varkappa_0-\frac{1}{r+1}w\Big).\varkappa^d
  + \Big(\varkappa^0\, T\, \Omega^{(\aleph)}\, \varkappa_0\Big)\, E^d_d.
\end{multline}
However, any element of the form 
$Z=(x\, w.{}^{t}w+y \,\varkappa_d.{}^{t}w+ z\, w.\varkappa^d+t\, \varkappa_d.\varkappa^d)$, for $x,y,z,t\in {\mathbb C}$, is such that $\Pi^{(\aleph)} Z=0$ and $\Theta^+ Z \; {}^{t}\Theta^+ = Z,$ and therefore can be freely added to the previous  choice of  ${\cal S}^{(0)}.$ 
Using the basic property stating that $\Omega^{(\aleph)} w ={}^{t}\Omega^{(\aleph)} w= -\frac{r}{2}w$ we can  choose a simpler expressions for ${\cal S}^{(0)},$ for example 
\begin{equation}
  {\cal S}^{(0)}=\Omega^{(\aleph)} \underline{A} + \varkappa_d.\varkappa^0\,\Omega^{(\aleph)} 
  + \Omega^{(\aleph)} \varkappa_0.\varkappa^d.
\end{equation}
However, we will choose $x=y=z=0$ and $t=-(\varkappa^0\, T\, \Omega^{(\aleph)}\, Y^\aleph\, \varkappa_0)$ in order to obtain the following expression for  $ {\cal S}^{(1)}$
\begin{equation}
 {\cal S}^{(1)}= \Theta^+ {\cal S}^{(0)}=Y^\aleph \, \Omega^{(\aleph)}\, \underline{A} 
  + \varkappa_d.\varkappa^0 \,\Pi\, Y^\aleph\, \Omega^{(\aleph)} 
  + Y^\aleph\, \Omega^{(\aleph)} \,\Pi\, \varkappa_0 .\varkappa^d.
  \label{ExpressionGenS1}
\end{equation}
This special choice of ${\cal S}^{(1)}$ satisfies 
\begin{equation}
  {\cal A} +{\cal S}^{(1)}+  {}^{t}{\cal S}^{(1)}=0,\quad\text{{\it i.e.}}\quad
 K\,S^{(1)}_{12}\,S^{(1)}_{21}=1\otimes 1.
\end{equation}

It will be convenient to decompose $S^{(1)}_{12}$  as follows:
\begin{equation}\label{decompS}
  S{}^{(1)}_{12}= \breve{S}{}^{(1)}_{12}\; O^{(1)}_{12}\; P^{(1)}_{12}\; Q^{(1)}_{12},
\end{equation}
with
\begin{alignat}{2}
 &\breve{S}{}^{(1)}_{12}
   =q^{\sum_{i,j=0}^r(Y^\aleph \Omega^{(\aleph)} \underline{A})^i_j\; \zeta^{\alpha_i} \otimes \zeta^{\alpha_j}},\qquad &
 &O^{(1)}_{12}=q^{\sum_{i=0}^r(\Pi Y^\aleph \Omega^{(\aleph)})^0_i \;\zeta^{d} \otimes \zeta^{\alpha_i}},\label{decompS1} \\
 &P^{(1)}_{12}= q^{\sum_{i=0}^r(\Pi Y^\aleph \Omega^{(\aleph)})^i_0 \; \zeta^{\alpha_i} \otimes \zeta^{d}},\qquad &
 &Q^{(1)}_{12}=1.\label{decompS2}
\end{alignat}
It is possible to compute explicitely $S{}^{(1)}_{12},$ but we only give the expression of 
$\breve{S}{}^{(1)}_{12}$ in the general case:
let $\aleph'$ be the  positive integer in $[0, r]$ such that $\aleph \aleph'=1 \mod (r+1),$
\begin{align}
Y^\aleph \Omega^{(\aleph)} \underline{A}
 &=-\frac{1}{r+1}Y^{\aleph}\sum_{l=0}^{r}l Y^{\aleph l}(2-Y-{}^t Y)\nonumber\\
 &=-\frac{1}{r+1}Y^{\aleph}\sum_{u=0}^{r}\lfloor u \aleph'\rfloor Y^{u}(2-Y-{}^t Y)\nonumber\\
 &=\sum_{u=0}^r \chi_{\aleph}(u)Y^{u+\aleph},
\end{align}
with $\chi_{\aleph}:\Z\rightarrow \{0,1,-1\},$ the function defined by
\begin{equation}
\chi_{\aleph}(u)=-\frac{1}{r+1}(2\lfloor u\aleph' \rfloor- \lfloor (u+1)\aleph'\ \rfloor-\lfloor (u-1)\aleph'\rfloor).
\end{equation}
In the particular case where $\aleph=1$ we can simplify the  expression of $S^{(1)}$ because  $\Omega^{(1)} \underline{A}=P_{r+1} -{}^{t}Y,$ and  we find \eqref{ExpressionS1}.

\bigskip

The universal formula for the twist $J(x)$ has been first obtained by \cite{JKOS} in the case where $\aleph=1$ and for any $\aleph$ in \cite{ESV}.

If $V$ is an irreducible finite dimensional representation of $U_q(\g)$ then by a theorem of V.~Chari and A. Pressley \cite{CP} this representation is such that 
$q^c$ is represented by $1$ (there is no twisting by an outer automorphism sending $q^c$ to $-q^c$ because the presentation  of $U_q(\g)$ we are using contains the whole  $q^{\h})$.
This theorem explains the terminology  we are using:
indeed in a finite dimensional irreducible representation of $U_q(\g)$, $c$ is represented by  $0,$ we therefore obtain that the expression of the dynamical $R$-matrix of vertex type in irreducible finite dimensional representations  has no dynamics and therefore is a  matrix solution of the Yang-Baxter equation, which is called Vertex solution in the mathematical physics litterature.

\subsection{Vertex-IRF transformation and Quantum Dynamical coBoundary Problem}
\label{subsec-QDBE}

Let $R(x)$ (resp. $\underline{R}(x)$) be  $\End(V^{\otimes 2})$  solutions of  the Quantum Dynamical Yang-Baxter Equation. 
$R(x)$ and $\underline{R}(x)$ are said to be related by a dynamical gauge transformation if there exists a meromorphic map 
$M: ({\mathbb C}^{\times})^{\dim \h} \rightarrow GL(V)$ such that
\begin{align} 
  \uR(x)\, M_1(xq^{h_2})\, M_2(x)= M_2(xq^{h_1})\, M_1(x)\, {R}(x).
\end{align}
In the case where ${R}(x)$ is of Vertex type and $\uR(x)$ is of IRF type,  this   dynamical gauge transformation, if it exists, is called Vertex-IRF transformation.

We have the following proposition \cite{EN}

\begin{proposition}{}
\par\noindent
If ${R}$ is a solution of the QDYBE of effective dynamics ${\l}$, and if 
$M:({\mathbb C}^{\times})^{\dim \h} \rightarrow GL(V)$ 
is such that $[M(x), h]=0$,  $\forall h\in {\l},$ 
then  a sufficient condition for 
\begin{equation}
 \uR(x)= M_2(xq^{h_1})\, M_1(x)\,{R}(x)\, M_2(x)^{-1}\,M_1(xq^{h_2})^{-1}
\end{equation}
to satisfy the QDYBE of effective dynamics $\underline{\l}\supset{\l}$ is that 
\begin{equation}
[h_1+h_2, \uR_{12}(x)]=0,\quad \forall h\in \underline\l.
\end{equation}
\end{proposition}

Little is known  on  the existence of dynamical gauge transformation between universal solutions of the QDYBE.

This problem is of course simpler to address in the classical setting: 

Let $r,\underline{r}: ({\mathbb C}^\times)^{\dim \h}\rightarrow \g^{\otimes 2}$
be given solutions of the CDYBE. These two 
solutions are said to be dynamically gauge equivalent if there exists 
$m: ({\mathbb C}^{\times })^{\dim \h} \rightarrow G$, with $\mathrm{Lie}(G)=\g$, such that
\begin{equation}
  {r}(x)=m_1(x)^{-1}\,m_2(x)^{-1}\,
  \Big(\underline{r}(x)+\frac{1}{2}\sum_{\alpha\in\Gamma} A_\alpha(x)\wedge h_{\alpha}\Big)\,
  m_1(x)\, m_2(x),
\end {equation}
where 
$A=\sum_{\alpha\in \Gamma} A_\alpha dx^\alpha$ is a flat connection defined as 
$A_\alpha=x_\alpha (\partial_{\alpha}m)m^{-1}\in \mathfrak{g}$.

But even in the classical case little is known. We can only state two important results related to this problem: 

1) In the case where $\g$ is of finite type, a  
classification of CDYBE of effective dynamics $\l,$ where $\l$ contains a regular semi simple element, up   to dynamical gauge transformation subject to the constraint that $m(x)\in L$ has been obtained in \cite{Sch} and is in one to one correspondence with generalized Belavin-Drinfeld Triple.
 
2) The theorem of \cite{BDF}:  if $\underline{r}$ is the standard solution of the CDYBE associated to $\g$ ($\underline\l=\h$),  then there exists a dynamical gauge transformation connecting $\underline{r}$ to  ${r}$ with ${\l}=\{0\}$ if and only if  $\g=A_r$ and ${r}$ is the Cremmer-Gervais solution.

\bigskip

A partial solution to the quantum version of the  above results is known:

The explicit quantization of the solutions of the CDYBE  associated to any generalized Belavin-Drinfeld triple of a finite dimensional simple Lie algebra has been done in \cite{ESS}.

The construction of the universal Vertex-IRF transform between the universal standard solution of the QDYBE and the universal Cremmer-Gervais solution has been done in our previous work 
\cite{BRT}.

\bigskip

Instead of working at the level of solutions of the QDYBE, we can also formulate the notion of dynamical gauge  transformation at the more fundamental level of solutions of the QDCE. 

Let $F(x)$, $\underline{F}(x)$  be Universal Quantum Dynamical Cocycles. 
$F(x)$ and $\underline{F}(x)$ are said to be related by a dynamical gauge transformation if there exists  $M(x)$ in a completion of  $(A\otimes  U_q(\mathfrak g))$, with $A$ a commutative algebra containing ${\cal F}^{(0)}(\h)$ such that %
\begin{align}
  &M(x)\ \text{is invertible},\\
  &\underline{F}(x)= \Delta(M(x))\,{F}(x)\,M_2(x)^{-1}\, M_1(xq^{h_2})^{-1}.
\end{align}
This definition is formal at the present time because the product of the right handside has to be defined. Nonetheless this formal definition implies that if $\pi$ is a finite dimensional $\h$-semisimple representation such that $\pi(M(x))$ is meromorphic then 
  the associated solutions $R(x)$, $\underline{R}(x)$ of the QDYBE are related by this gauge transformation.

In our previous work \cite{BRT}, we studied the case where $\g$ is finite dimensional and ${F}$ is constant (${l}=\{0\}$). In that case, such a dynamical gauge transformation $M$ was called a Quantum Dynamical coBoundary.

By extension, if $\g$ is of affine type, a dynamical gauge transformation relating an IRF type solution $\underline{F}(x)$ of the QDCE and a solution  ${F}(x)$ of the QDCE with effective dynamics ${\l}=\C c$ will be called a {\em generalized Quantum Dynamical Coboundary}. A generalized Dynamical coBoundary is of course a Vertex-IRF transform.

We now precise the meaning of the 
{\em Generalized Quantum Dynamical coBoundary}.

\begin{definition}{}
Let $F(x), \underline{F}(x) \in 
(U_q(\n_+)\widetilde{\otimes} {\cal F}^{(2)}(\h)\widetilde{\otimes} U_q(\n_-))^{c},$ zero degree solutions  of the quantum dynamical coCycle equation,  an  element $M(x)\in ({\cal F}^{(0)}(\h)\otimes U_q(\g))^{c(op)}$ such that $(M(x)_{\emptyset,\emptyset})$ is invertible,  is said to be a generalized Quantum Dynamical coBoundary, if the following equation hold:
\begin{equation}
 \Delta(M(x))\,{F}(x)=   \underline{F}(x)  M_1(xq^{h_2}) M_2(x),
\end{equation}
in the sense that:
\begin{equation}
\forall \gamma,\gamma'\in P, (\iota^+_{\gamma}\otimes \iota^-_{\gamma'})\Big( \Delta(M(x))\,{F}(x)\Big)=(\iota^+_{\gamma}\otimes \iota^-_{\gamma'})\Big( \underline{F}(x)  M_1(xq^{h_2}) M_2(x)\Big).
\end{equation}
\end{definition}

Due the zero degree property of $F(x)$ and $\underline{F}(x),$ 
$(\id\otimes \iota^-_{\gamma'})(\Delta(M(x))\,{F}(x))$ is a well defined element of $({\cal F}^{(0)}(\h)\otimes U_q(\g))^c \otimes ({\cal F}^{(0)}(\h)\otimes U_q(\b_+))^c$, whereas $(\iota^+_{\gamma}\otimes \id)( \underline{F}(x)  M_1(xq^{h_2}) M_2(x))$ is a well defined element of $({\cal F}^{(0)}(\h)\otimes U_q(\b_-))^c \otimes ({\cal F}^{(0)}(\h)\otimes U_q(\g))^c.$


\section{Gauss Decomposition of Generalized  Quantum Dynamical coBoundary}
\label{subsec-GaussQDC}

Let $\underline{F}(x)$ and $J(x)$ be quantum dynamical cocycles associated respectively with the generalized translation data 
$(\utheta^+, \utheta^-,  \uphi^0,\uphi^+,\uphi^-,\uS {}^{(1)})$ of Restricted IRF type, 
and 
$(\theta^+, \theta^-, \varphi^0, \varphi^+,\varphi^-,S^{(1)})$ of Restricted Vertex type. We will prove in this section a  generalization of our previous  result  \cite{BRT} giving  sufficient conditions for the existence of a Generalized  Quantum Dynamical Coboundary between these quantum Dynamical Cocycles.

\bigskip

Let ${   M}^{(0)}(x)\in \A(\h)\otimes  U_q({\mathfrak h})$ and 
${\mathfrak C}^{[\pm 1]}(x)\in 1\oplus \left({\cal F}^{0}(\h)
\otimes U^{\pm}_q({\mathfrak g})\right)^c$, we define 
 ${   M}^{(\pm)}(x)\in 1\oplus \left({\cal F}^{0}(\h)
\otimes U^{\pm}_q({\mathfrak g})\right)^c$ 
 by
\begin{align}
  {  M}^{(\pm)}(x)=\prod_{k=1}^{+\infty} {\mathfrak C}^{[\pm k]}(x)^{\pm 1},
  \quad \text{with}\quad
  &{\mathfrak C}^{[+ k]}(x)
       = \big(\theta^+_{[x]}\big)^{k-1}\big({\mathfrak C}^{[+1]}(x)\big),
          \nonumber\\
  &{\mathfrak C}^{[- k]}(x)
       =\big(\utheta^-_{[x]})^{k-1}\big({\mathfrak C}^{[-1]}(x)\big) .
   \label{MdeXprod2}
\end{align}
We define  ${M}(x)\in \A(\h)\otimes\left({\cal F}^{(0)}(\h)\otimes U_q({\mathfrak g})\right)^c$  as  
\begin{equation}\label{MdeXprod}
  {   M}(x)= {   M}^{(0)}(x)\, {  M}^{(-)}(x)^{-1} 
               {   M}^{(+)}(x).
\end{equation}

Let  $\Gamma^{(\pm)}$ such that 
$\log_q(\Gamma^{(\pm)})\in \h^{\otimes 2}$ and define
\begin{align}
   &\Lambda^{(+)}_{12}=W_{12}^{(1)}\,
        \theta_{[x]1}^+\big(K^{-1}\,\Gamma_{21}^{(+)}\big),
                 \label{choix+}\\
   &\Lambda^{(-)}_{12}=K\,\Gamma_{21}^{(-)}.
                 \label{choix-}
\end{align}
\begin{theorem}{\label{maintheorem}}
\par\noindent 
We assume that the previous  $\underline{F}(x)$ and $J(x)$ are associated to non degenerate generalized translation datum.
The following algebraic relations on ${  M}^{(0)}(x)$ and 
${\mathfrak C}^{[\pm 1]}(x) $ are sufficient conditions to ensure that 
${  M}(x)$ is a solution of the generalized Quantum Dynamical coBoundary Equation:
\begin{align}
   &\Delta\big({  M}^{(0)}(x)\big)=\uS^{(1)}_{12}\, S^{(1)\; -1}_{12}\,
        {  M}^{(0)}_1(xq^{h_2})\,  {  M}^{(0)}_2(x),
    \label{axiomABRR0}\\
   &\Delta\big({\mathfrak C}^{[\pm 1]}(x)\big) =
    \Ad_{\Gamma^{(\pm)}_{12}}\big({\mathfrak C}^{[\pm 1]}_{1}(x)\big)\;
    \Ad_{\Lambda^{(\pm)}_{12}}\big({\mathfrak C}^{[\pm 1]}_{2}(x)\big),
     \label{axiomABRR1d}\\
   &{\mathfrak C}^{[\pm 1]}_{1}(xq^{h_2})=
     \Ad_{[S_{12}^{(1)\; -1}\,\Gamma^{(\pm)}_{12}]}\big(
     {\mathfrak C}^{[\pm 1]}_{1}(x)\big),
     \label{axiomABRR1}\\
   &\Ad_{\Lambda^{(-)}_{12}}\big({\mathfrak C}^{[- 1]}_{2}(x)\big)\ 
    \Ad_{\Gamma^{(+)}_{12}}\big({\mathfrak C}^{[+ 1]}_{1}(x)\big)\ 
    W_{12}^{(1)}\;\Ad_{\varphi^0_{12}}^{-1}\circ\theta^+_{[x]1}\big(\widehat{R}\big)
                    \nonumber\\
   &\hspace{3.5cm}
    =\utheta^-_{[x] 2}\big(\widehat{R}\big)\ 
     \Ad_{\Gamma^{(+)}_{12}} \big({\mathfrak C}^{[+ 1]}_{1}(x)\big)\ 
     W^{(1)}_{12}\ 
    \Ad_{\theta_{[x]1}^+(\Lambda^{(-)}_{12})}
                \big({\mathfrak C}^{[- 1]}_{2}(x)\big).
     \label{axiomABRR2prime}
\end{align}

\end{theorem}

\proof
{}From the property \eqref{axiomABRR0} of $M^{(0)}(x) \in U_q({\mathfrak h})$
and the zero $\h$-weight property of $\underline{F}(x)$, it results  that $M(x)$ satisfies the generalized dynamical coBoundary equation if and only if the  equality $U(x)=V(x)$ holds  with
\begin{align}
  &U(x)=\Delta(M^{(+)}(x))\, J_{12}(x)\, M^{(+)}_2(x)^{-1}\,S^{(1)\; -1}_{12},
                  \label{def-U}\\
  &V(x)=\Delta(M^{(-)}(x))\, \underline{F}_{12}(x)\;
   \uS_{12}^{(1)\; -1} S^{(1)}_{12}\,
   M^{(-)}_1(xq^{h_2})^{-1}\, M^{(+)}_1(xq^{h_2})\, M_2^{(-)}(x)^{-1}\,
   S^{(1)\; -1}_{12}. \!\!
                  \label{def-V}
\end{align}
Note that because of the degree zero property of $J(x)$ and $\underline{F}(x)$, $U(x)$ (resp. $V(x)$) are well defined elements of $ (U_q(\b_+)\otimes U_q(\g)\otimes {\cal F}^{(0)}(\h))^{c(op)}$ (resp.  $(U_q(\g)\otimes U_q(\b_-)\otimes {\cal F}^{(0)}(\h))^{c(op)}$.
Because of their expressions, it is clear that
\begin{equation}
U(x)\in 1^{\otimes 2}\oplus (U^+_q(\g)\otimes U_q(\g)\otimes {\cal F}^{(0)}(\h))^{c(op)}.
\end{equation}
Note however that we do not  have 
\begin{equation}
V(x)\in 1^{\otimes 2}\oplus (U_q(\g)\otimes U^-_q(\g)\otimes {\cal F}^{(0)}(\h))^{c(op)}.
\end{equation}


Note that if the theorem holds, {\it i.e.}  $U(x)=V(x)$, it will constrain $U(x)$ and $V(x)$ to belong to $(U_q(\n_+)\widetilde{\otimes} {\cal F}^{(2)}(\h)\widetilde{\otimes} U_q(\n_-))^c.$ 
The  proof of the theorem is  decomposed in several lemmas. 

We first show that $U(x)$ satisfies a linear equation and deduce from it the triangular property  of  $U(x).$

\begin{lemma}{\label{lemma1coboundary}}
\par\noindent
${U}(x)$  obeys the following linear relation:
\begin{equation}
  {U}_{12}(x)=\Ad_{\Gamma^{(+)}_{12}}\big({\mathfrak C}^{[+1]}_1(x)\big)\
    \Ad_{[W_{12}^{(1)}\,(\varphi^0_{12})^{-1}]}\circ\theta^+_{[x]1}\big(\widehat{R}\;
    {U}_{12}(x)\big).
  \label{ABRR1U}
\end{equation}
\end{lemma}

\proof
{}From the definition~\eqref{def-U} of $U(x)$, one has,
\begin{align*}
    \Ad_{\varphi^0_{12}}^{-1}\circ\theta^+_{[x]1}\big(\widehat{R}_{12}\; {U}_{12}(x)\big)
  &=\Ad_{\varphi^0_{12}}^{-1}\circ\theta^+_{[x]1}\big(
    \widehat{R}_{12}\,\Delta(M^{(+)}(x))\,J_{12}(x)\, M^{(+)}_2(x)^{-1}\,
        S^{(1) -1}_{12}\big)\\
  &= \Ad_{\varphi^0_{12}}^{-1}\circ\theta^+_{[x]1}\big(K^{-1}\,
    \Delta'(M^{(+)}(x))\,K \big)\; W_{12}^{(1)\; -1}\\
  &\hspace{3cm}\times 
   W_{12}^{(1)}\;\Ad_{\varphi^0_{12}}^{-1}\circ\theta^+_{[x]1}\big(\widehat{R}_{12}\,J_{12}(x)\big)\;\\
  &\hspace{3cm}\times 
   \Ad_{\varphi^0_{12}}^{-1}\circ\theta^+_{[x]1}\big(M^{(+)}_2(x)^{-1}\big)\, S^{(2) -1}_{12}\\
  &= \Ad_{\varphi^0_{12}}^{-1}\circ\theta^+_{[x]1}\big(K^{-1}\;
    \Delta'(M^{(+)}(x))\,K\big)\;W_{12}^{(1)\; -1}\,\\
  &\hspace{3cm}\times 
         J_{12}(x)\, M^{(+)}_2(x)^{-1}\, S^{(1) -1}_{12}\, W_{12}^{(1)},
\end{align*}
in which the last equality follows from \eqref{axiomVertexa} and from the fact that $J(x)$ satisfies the linear equation \eqref{ABRR1}.
Therefore, it remains to prove that
\begin{equation}\label{eq-to-prove-lem1}
  \Delta(M^{(+)}(x))
  =\Ad_{\Gamma^{(+)}_{12}}\big( {\mathfrak C}^{[+1]}_1(x)\big)\ 
  \Ad_{[W^{(1)}_{12}\,(\varphi^0_{12})^{-1}]}\circ\theta^+_{[x]1}\big(K^{-1}\,
    \Delta'(M^{(+)}(x))\,K\big).
\end{equation}
Using~\eqref{axiomABRR1d}, and defining
\begin{equation}\label{defphik}
  \varphi^{[\pm k]}=\prod_{i=1}^{k}
  \big(\theta^\pm_{[x]}\otimes\theta^\pm_{[x]}\big)^{i-1}(\varphi^\pm),
\end{equation}
we have (from eq (\ref{axiomGTDc}))
\begin{align*}
  \Delta(M^{(+)}(x))
     &=\prod_{k=0}^{+\infty} \Big[ \Ad_{\varphi_{12}^{[+k]}}\circ
        \big(\theta^+_{[x]}\otimes\theta^+_{[x]}\big)^{k}\Big]
    \Big( \Delta\big( {\mathfrak C}^{[+1]}(x) \big)\Big)
                          \\
     &=\prod_{k=0}^{+\infty} 
     \Big\{ P_{12}^{[+k]}(x)\;\widetilde{P}_{12}^{[+k]}(x)\Big\},
\end{align*}
with, for $k\ge 0$,
\begin{align}
   &P_{12}^{[+k]}(x)= \Big[ \Ad_{\varphi_{12}^{[+k]}}\circ
        \big(\theta^+_{[x]}\otimes\theta^+_{[x]}\big)^{k}\Big]
    \Big( \Ad_{\Gamma_{12}^{(+)}}\big( \mathfrak{C}_1^{[+1]}(x)\big)\Big),
                   \\
   &\widetilde{P}_{12}^{[+k]}(x)= \Big[ \Ad_{\varphi_{12}^{[+k]}}\circ
        \big(\theta^+_{[x]}\otimes\theta^+_{[x]}\big)^{k}\Big]
    \Big( \Ad_{\Lambda_{12}^{(+)}}\big( \mathfrak{C}_2^{[+1]}(x)\big)\Big).
\end{align}
The infinite product expression of $\Delta'(M^{(+)}(x))$ can be written in a similar way and can be reorganized as
\begin{align*}
  \Delta'(M^{(+)}(x))
  &=\prod_{k=0}^{+\infty}
    \Big\{ P_{21}^{[+k]}(x)\;\widetilde{P}_{21}^{[+k]}(x)\Big\}\\
  &=P_{21}^{[+0]}(x)\;
    \prod_{k=1}^{+\infty} 
    \Big\{\widetilde{P}_{21}^{[+(k-1)]}(x)\;P_{21}^{[+k]}(x)\Big\}.
\end{align*}
Using this expression, one can express the right hand side of \eqref{eq-to-prove-lem1} as
\begin{align*}
 \text{r.h.s.}
  &={P}_{12}^{[+0]}(x)\;
     \Big[\Ad_{[(\varphi^{0})^{-1}\, W^{(1)}_{12}\,
                \theta^{+}_{[x] 1}(K^{-1})]}
     \circ\theta^{+}_{[x] 1}\Big]
     \Big( P_{21}^{[+0]}(x)\Big)\;    \\
  &\hspace{1cm}
     \qquad\times
     \prod_{k=1}^{+\infty}\Big\{
     \Big[\Ad_{[(\varphi^{0})^{-1}\, W^{(1)}_{12}\,
                \theta^{+}_{[x] 1}(K^{-1})]}\circ\theta^{+}_{[x] 1}\Big]
     \Big(\widetilde{P}_{21}^{[+(k-1)]}(x)\Big)\  \\
 &\hspace{2.1cm}
     \qquad\times
     \Big[\Ad_{[(\varphi^{0})^{-1}\, W^{(1)}_{12}\,
                \theta^{+}_{[x] 1}(K^{-1})]}\circ\theta^{+}_{[x] 1}\Big]
     \Big({P}_{21}^{[+(k)]}(x)\Big) \Big\}    \\
  &=P_{12}^{[+0]}(x)\;
    \Ad_{\widetilde{Q}_{12}^{[+0]}(x)}\Big(\widetilde{P}_{12}^{[+0]}(x)\Big)  \\
  &\hspace{1cm}
     \qquad\times
     \prod_{k=1}^{+\infty} \Big\{
     \Ad_{{Q}_{12}^{[+k]}(x)}\Big({P}_{12}^{[+k]}(x)\Big)\ 
    \Ad_{\widetilde{Q}_{12}^{[+k]}(x)}\Big(\widetilde{P}_{12}^{[+k]}(x)\Big)
                   \Big\}.
\end{align*}
Here
\begin{equation*}
   \widetilde{Q}_{12}^{[+0]}(x)=(\varphi^{0})^{-1}\, W_{12}^{(1)}\,
    \Lambda_{12}^{(+)\; -1}\, \theta^+_{[x] 1}\big(K^{-1}\,\Gamma_{21}^{(+)}\big),
\end{equation*}
and, for $k\ge 1$,
\begin{align*}
   &{Q}_{12}^{[+k]}(x)=\big\{ \varphi^0\,K\,S_{12}^{(1)\;-1}\,S_{21}^{(1)}\;
         \theta^+_{[x] 1}(K\, S_{12}^{(1)}\,S_{21}^{(1)\;-1})\big\}^{-1}\;
                \widehat{Q}_{21}^{[+k]}(x)^{-1},\\
   &\widetilde{Q}_{12}^{[+k]}(x)=(\varphi^{0}\;\varphi^{+})^{-1}\;
       S_{12}^{(1)}\, (\theta^+_{[x]}\otimes\theta^+_{[x]})(S_{12}^{(1)\; -1})\;
       \theta^+_{[x] 1}\big( \widehat{Q}_{12}^{[+k]}(x) \big),
\end{align*}
with
\begin{multline*}
   \widehat{Q}_{12}^{[+k]}(x)
    =\varphi_{21}^{[+k]}\,
     \theta^+_{[x] 2\!}\big(\varphi_{12}^{[+(k-1)]\; -1}\big)\;\varphi^0\;
     (K\,S_{12}^{(1)}) ^{-1}\,
     \big(\theta^+_{[x]}\otimes\theta^+_{[x]}\big)^{k}(K\,S_{12}^{(1)})\;\\
   \times
     \theta^+_{[x] 2}\big(S_{12}^{(1)}\,
     (\theta^+_{[x]}\otimes\theta^+_{[x]})^{k-1}(S_{12}^{(1)\;-1})\big)\;\
     \Big[\big(\theta^+_{[x]}\big)^{k-1}\otimes\big(\theta^+_{[x]}\big)^{k}\Big]
     \Big(\widetilde{Q}_{12}^{[+0]}(x) \Big).
\end{multline*}
{}From \eqref{quasiInverse} and \eqref{axiomGTDg}, one has
\begin{equation}\label{axiomS}
    \big[\, S_{12}^{(1)}\, 
    \big(\theta^+_{[x]}\otimes\theta^+_{[x]}\big)(S_{12}^{(1)\; -1})\; ,\;
        \theta^+_{[x]2}(v)\, \big]=0,\quad \forall v\in U_q(\b_+).
\end{equation}
Therefore, taking also into account \eqref{axiomGTDh} and \eqref{axiomVertexb}, the relation \eqref{eq-to-prove-lem1} is satisfied if
\begin{align}
  &\big[\, \widetilde{Q}_{12}^{[+0]}(x)\, ,\, {\mathfrak C}^{[+1]}_{2}(x)\, \big]=0,
               \label{cond1-lem1}\\
  &\big[\, \widehat{Q}_{12}^{[+k]}(x)\, ,\, {\mathfrak C}^{[+(k+1)]}_{2}(x)\, \big]=0,
   \qquad \forall k\ge 1.
               \label{cond2-lem1}
\end{align}
\eqref{cond1-lem1} is ensured by the choice \eqref{choix+} and \eqref{axiomVertexa}.
Using \eqref{cond1-lem1}, \eqref{axiomS} and \eqref{axiomVertexb}, the condition \eqref{cond2-lem1} reduces to
\begin{equation}\label{cond2prime-lem1}
  \big[\, \varphi_{21}^{[+k]}\,(\varphi^0)^k\;
          K^{-1}\,\big(\theta^+_{[x]}\otimes\theta^+_{[x]}\big)^{k}(K)\; ,\;
        {\mathfrak C}^{[+(k+1)]}_{2}(x)\, \big]=0,
\end{equation}
which follows from \eqref{axiomVertexd}.
\cqfd

\begin{lemma}{\label{lemma1primecoboundary}}
 \par\noindent
The linear equation \eqref{ABRR1U} has a unique solution in  
$1^{\otimes 2} \oplus (U^+_q(\g)\otimes U_q(\g)\otimes {\cal F}^{(0)}(\h))^{c(op)} $ and this solution  necessarily belongs to  
$1^{\otimes 2} \oplus (U^+_q({\mathfrak g})\otimes U_q(\b_-)\otimes
 {\cal F}^{(0)}(\h))^c$. Therefore $U(x)$ belongs to this space 
 and satisfies 
\begin{equation}
{U}_{12}(x)=S^{(1)}_{12}\ \prod_{k=0}^{+\infty}\Big\{
  \big( \Ad^{-1}_{\varphi_{12}^0}\circ\theta^+_{[x] 1}\big)^k\left( 
  \Ad_{[S^{(1)\; -1}_{12}\,\Gamma^{(+)}_{12}]} \big({\mathfrak C}^{[+]}_1(x)\big) \ 
  \widehat{J}^{(1)}_{12}(x)
   \right)\Big\}\ S^{(1)\; -1}_{12}.
   \label{Uinfprod}
\end{equation}
\end{lemma}

\proof
One writes the linear equation \eqref{ABRR1U}as a triangular system with respect to the degree in the first space. The  existence and uniqueness of the solution is a consequence of the non degenerate assumption. Indeed this  linear equation can be rewritten as:
\begin{equation}
S^{(1)-1}{U}_{12}(x)S^{(1)}=\Ad_{\Gamma^{(+)}S^{(1)-1}}({\mathfrak C}^{[+]}_1(x)). \Ad_{\varphi^{(0)-1}}\circ \theta^+_{[x] 1}\big( S^{(1)-1}\widehat{R}_{12} S^{(1)} S^{(1)-1}{U}_{12}(x)S^{(1)}\big).
\end{equation}
The non degeneracy condition implies that $id^{\otimes 2}-\Ad(\varphi^{(0)-1})\circ \theta^+_{[x] 1}$ is an invertible operator on $(U_q^{+}(\g)\otimes U_q(\g)\otimes {\cal F}^{(0)(\h)})^c,$ as a result the existence and unicity of the linear equation follows.
Since $\widehat{R}\in 1^{\otimes 2}\oplus(U^+_q(\g)\otimes U^-(\g))^c$, we obtain that $U(x)$  belongs to  $1^{\otimes 2} \oplus (U^+_q({\mathfrak g})\otimes U_q(\b_-)\otimes {\cal F}^{(0)}(\h))^c$.

The infinite product formula follows by iteration of the linear equation.
\cqfd

We now show that $V(x)$ satisfies a linear equation.

\begin{lemma}{\label{lemma3coboundary}}
 \par\noindent
$V(x)$  defined by \eqref{def-V}
is an element of  $(U_q(\g)\otimes U_q(\b_-)\otimes {\cal F}^{(0)}(\h))^{c(op)}$ and 
 obeys the following linear relation:
\begin{equation}
  \Ad_{\Lambda_{12}^{(-)}}\big( \mathfrak{C}_2^{[-1]}(x)\big)\; V(x)=
    \utheta_{[x]2}^-\big(\widehat{R}\, V(x)\big)\;
        \Ad_{S^{(1)}_{12}}\big(\mathfrak{C}_2^{[-1]}(x)\big).
   \label{ABRR2V}
\end{equation}
\end{lemma}

\proof
Using the fact that, the corresponding generalized datum being of face type, $\utheta^-_{[x]}$ acts as the identity on $U_q(\h)$, one has
\begin{align}\label{theta-M1}
  &\utheta_{[x]2}^-\big( M_1^{(\pm)}(xq^{h_2})\big) = M_1^{(\pm)}(xq^{h_2}) \;\;\text{and}\;\; \uW_{12}^{(1)}=1^{\otimes 2}.
\end{align}
Moreover,
\begin{align}\label{theta-M2}
  &\utheta_{[x]2}^-\big( M_2^{(-)}(x)^{-1}\big) =  M_2^{(-)}(x)^{-1}\,
     \mC^{[-1]}_2(x)^{-1}.
\end{align}
Therefore, using these properties as well as   the linear equation \eqref{ABRR2} for $F(x),$
one can rewrite $\utheta_{[x]2}^-\big(\widehat{R}\, V(x)\big)$ as
\begin{align*}
  \utheta_{[x]2}^-\big(\widehat{R}\, V(x)\big)
  &=\utheta_{[x]2}^-\big( K^{-1}\;\Delta'(M^{(-)}(x))\; K\big)\; 
    \utheta_{[x]2}^-\big(\widehat{R}\, \underline{F}_{12}(x)\big)
              \\
  &\hspace{1.3cm}\times
    \utheta_{[x]2}^-\big(S_{12}^{(1)}\,\uS_{12}^{(1)\; -1}\;
    M_1^{(-)}(xq^{h_2})^{-1}\, M_1^{(+)}(xq^{h_2})\;
    M_2^{(-)}(x)^{-1}\, S_{12}^{(1)\; -1}\big)
              \\ 
  &=\utheta_{[x]2}^-\big( K^{-1}\;\Delta'(M^{(-)}(x))\; K\big)\; 
    \underline{F}_{12}(x)\; S_{12}^{(1)}\,\uS_{12}^{(1)\; -1}\;
    M_1^{(-)}(xq^{h_2})^{-1}\, 
              \\
  &\hspace{1.3cm}\times
    M_1^{(+)}(xq^{h_2})\;
    M_2^{(-)}(x)^{-1}\,\mC^{[-1]}_2(x)^{-1}\;
    S_{12}^{(1)\; -1}
              \\ 
  &=\utheta_{[x]2}^-\big( K^{-1}\;\Delta'(M^{(-)}(x))\; K\big)\; 
      \Delta(M^{(-)}(x)^{-1})\;
    V(x)\;
    \Ad_{S_{12}^{(1)}}\big(\mC^{[-1]}_2(x)^{-1}\big).
\end{align*}
Hence, \eqref{ABRR2V} is satisfied if and only if 
\begin{equation}\label{eq-to-prove-lem3}
   \Ad_{\Lambda_{12}^{(-)}}\big( \mathfrak{C}_2^{[-1]}(x)\big)\;\Delta(M^{(-)}(x))
    =
     \utheta_{[x]2}^-\big(K^{-1}\;\Delta'(M^{(-)}(x))\; K\
      \big).
\end{equation}
This last equality is obtained using the definition of $M^{(-)}(x)$ as an infinite product, like in the proof of Lemma~\ref{lemma1coboundary}. Indeed, using \eqref{axiomABRR1d}, and defining $\uphi_{12}^{[-k]}$ similarly as in \eqref{defphik} (note that in the IRF case $\uphi_{12}^{[-k]}=(\uphi_{12}^-)^k$), we have
\begin{align*}
  \Delta(M^{(-)}(x))
     &=\prod_{k=0}^{+\infty} 
     \Big\{ \widetilde{\uP}_{12}^{[-k]}(x)^{-1}\;\uP_{12}^{[-k]}(x)^{-1}\Big\},
\end{align*}
with
\begin{align*}
   &\uP_{12}^{[-k]}(x)= \Big[ \Ad_{\uphi_{12}^{[-k]}}\circ
        \big(\utheta^-_{[x]}\otimes\utheta^-_{[x]}\big)^{k}\Big]
    \Big( \Ad_{\Gamma_{12}^{(-)}}\big( \mathfrak{C}_1^{[-1]}(x)\big)\Big),
                   \\
   &\widetilde{\uP}_{12}^{[-k]}(x)= \Big[ \Ad_{\uphi_{12}^{[-k]}}\circ
        \big(\utheta^-_{[x]}\otimes\utheta^-_{[x]}\big)^{k}\Big]
    \Big( \Ad_{\Lambda_{12}^{(-)}}\big( \mathfrak{C}_2^{[-1]}(x)\big)\Big),
\end{align*}
and we reorganize the infinite product expression of $\Delta'(M^{(-)}(x))$  as
\begin{align*}
  \Delta'(M^{(-)}(x))
  &=\widetilde{\uP}_{21}^{[-0]}(x)^{-1}\;
    \prod_{k=1}^{+\infty} 
    \Big\{{\uP}_{21}^{[-(k-1)]}(x)^{-1}\;\widetilde{\uP}_{21}^{[-k]}(x)^{-1}\Big\}.
\end{align*}
The  right hand side of \eqref{eq-to-prove-lem3} is equal to
\begin{align*}
 \text{r.h.s.}
  &=
    \Ad_{ {\uQ}_{12}^{[-0]}(x)}\Big({\uP}_{12}^{[-0]}(x)^{-1}\Big)
     \prod_{k=1}^{+\infty} \Big\{
     \Ad_{ \widetilde{\uQ}_{12}^{[-k]}(x)}\Big(\widetilde{\uP}_{12}^{[-k]}(x)^{-1}\Big)\ 
    \Ad_{ {\uQ}_{12}^{[-k]}(x)}\Big({\uP}_{12}^{[-k]}(x)^{-1}\Big)
                   \Big\},
\end{align*}
where
\begin{align*}
   & {\uQ}_{12}^{[-0]}(x)=K^{-1}\,\Lambda_{21}^{(-)}\,
        \Gamma_{12}^{(-)\; -1} ,
                                      \\
   & {\uQ}_{12}^{[-k]}(x)=\uphi_{12}^{[-k]\; -1}\,
     \utheta^-_{[x] 2\!}\big(\uphi_{21}^{[-k]}\big)\;
     \utheta^-_{[x] 2}(K^{-1})\,
     \big(\utheta^-_{[x]}\otimes\utheta^-_{[x]}\big)^{k}
     \big(\utheta^-_{[x] 2}(\Lambda_{21}^{(-)})\,
     \Gamma_{12}^{(-)\; -1}\big)
                                            \\
   &\phantom{ {\uQ}_{12}^{[-k]}(x)}
      = \uphi_{12}^{[-k]\; -1}\,\uphi_{21}^{[-k]}\,  {\uQ}_{12}^{[-0]}(x),
                                            \\
   & \widetilde{\uQ}_{12}^{[-k]}(x)
     =\uphi_{12}^{[-k]\; -1}\,
     \utheta^-_{[x] 2\!}\big(\uphi_{21}^{[-(k-1)]}\,K^{-1}
     \big(\utheta^-_{[x]}\otimes\utheta^-_{[x]}\big)^{k-1}
     (\Gamma_{21}^{(-)})\big)\,
     (\utheta^-_{[x]}\otimes\utheta^-_{[x]})^k(\Lambda_{12}^{(-)\; -1})
                                           \\
   &\phantom{ \widetilde{\uQ}_{12}^{[-k]}(x)}
     =(\uphi_{12}^{-})^{-1}\,K^{-2}\, {\uQ}_{21}^{[-(k-1)]}(x)^{-1}.
\end{align*}
\eqref{eq-to-prove-lem3} is satisfied if
\begin{align}
  &\big[\, \widetilde{\uQ}_{12}^{[-k]}(x)\, ,\, \mC^{[-(k+1)]}_{2}(x)\, \big]=0,
           \qquad\forall k\ge 1,\\
  &\big[\, {\uQ}_{12}^{[-k]}(x)\, ,\, \mC^{[-(k+1)]}_{1}(x)\, \big]=0,
           \qquad\forall k\ge 0,
\end{align}
which follows from \eqref{choix-}, \eqref{axiomGTDi}, \eqref{axiomPsi0Phi-} and \eqref{axiomPhi-Psi0K}.
\cqfd


Let us now prove that $U(x)$ satisfies also the same linear equation \eqref{ABRR2V} as $V(x)$. It is at this point that we will use the hexagonal relation (\ref{axiomABRR2prime}).

\begin{lemma}{\label{lemma2coboundary}}
U(x) satisfies the linear equation 
\begin{equation}
   U(x)=\Ad_{\Lambda_{12}^{(-)}}\big( \mathfrak{C}_2^{[-1]}(x)^{-1}\big)\ 
   \utheta_{[x]2}^-\big(\widehat{R}\, U(x)\big)\ 
        \Ad_{S^{(1)}_{12}}\big(\mathfrak{C}_2^{[-1]}(x)\big).\label{ABRR2U}
\end{equation}
\end{lemma}

\proof
 \par\noindent
Let us define
\begin{equation}
  \widetilde{U}(x)
  =\Ad_{\Lambda_{12}^{(-)}}\big( \mathfrak{C}_2^{[-1]}(x)^{-1}\big)\ 
   \utheta_{[x]2}^-\big(\widehat{R}\, U(x)\big)\ 
        \Ad_{S^{(1)}_{12}}\big(\mathfrak{C}_2^{[-1]}(x)\big).
\end{equation} 
{}From its definition and the fact that $U(x)$ belongs to $1^{\otimes 2} \oplus (U_q^+({\mathfrak g})\otimes U_q({\mathfrak b}_-)\otimes {\cal F}^{(0)}(\h))^c$, it is obvious that $\widetilde{U}(x)\in 1^{\otimes 2} \oplus (U_q^+({\mathfrak g})\otimes U_q({\mathfrak b}_-)\otimes {\cal F}^{(0)}(\h))^c$. Let us now prove that it satisfies the same linear equation \eqref{ABRR1U} as $U(x)$.

Indeed, one has,
\begin{align*}
  \Ad^{-1}_{\varphi_{12}^0}\circ\theta^+_{[x] 1}\big(\widetilde{U}_{12}(x)\big)
  &=\big[\Ad_{[\theta^{+}_{[x] 1}(\Lambda_{12}^{(-)})\,(\varphi_{12}^0)^{-1}]}\circ\theta^+_{[x] 1}\big]\big( \mathfrak{C}_2^{[-1]}(x)^{-1}\big)\ \\
  &\quad\times
    \big[
          \Ad^{-1}_{\varphi_{12}^0}\circ\theta^+_{[x] 1}\circ
                  \utheta_{[x]2}^-\big]\big(\widehat{R}\, U(x)\big)\ 
    \big[\Ad_{[S^{(2)}_{12}\,(\varphi_{12}^0)^{-1}]}\circ\theta^+_{[x] 1}\big]
     \big(\mathfrak{C}_2^{[-1]}(x)\big)
                \\ 
  &=\Ad_{\theta^{+}_{[x] 1}(\Lambda_{12}^{(-)})}
    \big( \mathfrak{C}_2^{[-1]}(x)^{-1}\big)\ \\
  &\quad\times
    \big[
         \Ad_{W^{(1)\; -1}_{12}}
         \circ\utheta_{[x]2}^-\big]
    \Big(\Ad_{[W^{(1)}_{12}\,(\varphi_{12}^0)^{-1}]}\circ\theta^+_{[x] 1}
    \big(\widehat{R}\, U(x)\big)\Big)\
\\
  &\quad\times   
    \Ad_{S^{(2)}_{12}}\big(\mathfrak{C}_2^{[-1]}(x)\big),
\end{align*}
in which we have used \eqref{axiomVertexa} and \eqref{axiom-face}.
{}From the fact that $U(x)$ satisfies \eqref{ABRR1U}, one gets,
\begin{align*}
 \Ad_{[W^{(1)}_{12}\,(\varphi_{12}^0)^{-1}]}\circ\theta^+_{[x] 1}\big(\widetilde{U}_{12}(x)\big)
  &=\Ad_{[W^{(1)}_{12}\,\theta^{+}_{[x] 1}(\Lambda_{12}^{(-)})]}
    \big( \mathfrak{C}_2^{[-1]}(x)^{-1}\big)\ \\
  &\qquad\times
       \utheta_{[x]2}^-
    \Big(
    \Ad_{\Gamma_{12}^{(+)}}\big(\mC_1^{[+1]}(x)^{-1} \big)\; U(x)\Big)\ 
\\
  &\qquad\times
    \Ad_{S^{(1)}_{12}}\big(\mathfrak{C}_2^{[-1]}(x)\big)
                     \\ 
  &=\Ad_{[W^{(1)}_{12}\,\theta^{+}_{[x] 1}(\Lambda_{12}^{(-)})]}
    \big( \mathfrak{C}_2^{[-1]}(x)^{-1}\big)\ 
\\
  &\qquad\times
    \Ad_{\Gamma_{12}^{(+)} } \big(\mC_1^{[+1]}(x)^{-1} \big)\ 
     \utheta_{[x]2}^-\big(\widehat{R}^{-1}\big)\;\\
  &\qquad\times
    \Ad_{
       \Lambda_{12}^{(-)}}\big(\mathfrak{C}_2^{[-1]}(x)\big)\;
    \widetilde{U}_{12}(x).
\end{align*}
As a result, $\widetilde{U}(x) $ satisfies (\ref{ABRR1U}) as soon as
\begin{multline}
  \Ad_{\Gamma_{12}^{(+)}}\big(\mC_1^{[+1]}(x) \big)\ 
  \Ad_{[W^{(1)}_{12}\,(\varphi_{12}^0)^{-1}]}\circ\theta^+_{[x] 1}\big(\widehat{R}\big)\
  \Ad_{[W^{(1)}_{12}\,\theta^{+}_{[x] 1}(\Lambda_{12}^{(-)})]}
    \big( \mathfrak{C}_2^{[-1]}(x)^{-1}\big)\ 
                     \\
 \times
  \Ad_{\Gamma_{12}^{(+)} } \big(\mC_1^{[+1]}(x)^{-1} \big)\
     \utheta_{[x]2}^-\big(\widehat{R}^{-1}\big)\;
    \Ad_{\Lambda_{12}^{(-)}}\big(\mathfrak{C}_2^{[-1]}(x)\big)=1,
\end{multline}
which is equivalent to the hexagonal relation (\ref{axiomABRR2prime}). 

Therefore, $\widetilde{U}(x) $ and  $U(x)$ are both in $1^{\otimes 2} \oplus (U_q^+({\mathfrak g})\otimes U_q(\b)_-\otimes {\cal F}^{(0)}(\h))^c$ and obey the same linear relation (\ref{ABRR1U}). {}From Lemma~\ref{lemma1primecoboundary}, this equation admits a unique solution, which means that $\widetilde{U}(x)=U(x).$ 
\cqfd

\begin{lemma}
$U(x)^{-1}V(x)$ is an element of $1^{\otimes 2} \oplus (U_q({\mathfrak g})\otimes U^-_q({\mathfrak g})\otimes {\cal F}^{(0)}(\h))^c$ which obeys the linear relation
\begin{equation}
   U_{12}(x)^{-1}V_{12}(x)
   = \Ad_{S_{12}^{(1)}}\big( {\mathfrak C}^{[-1]}_2(x)^{-1}\big)\
     \utheta_{[x]2}^-\big( U_{12}(x)^{-1}V_{12}(x) \big)\
     \Ad_{S_{12}^{(1)}}\big( {\mathfrak C}^{[-1]}_2(x)\big).
  \label{ABRRUV}
\end{equation}
\end{lemma}

\proof
In order to show that $U(x)^{-1}V(x)\in 1^{\otimes 2} \oplus (U_q({\mathfrak g})\otimes U^-_q({\mathfrak g})\otimes {\cal F}^{(0)}(\h))^c)$, let us compute $(\iota_-)_2(U(x))$ and $(\iota_-)_2(V(x))$.

Using (\ref{Uinfprod}), the fact that $\widehat{J}_{12}^{(1)}(x)\in 1^{\otimes 2}\oplus (U_q^+({\mathfrak g})\otimes U^-_q({\mathfrak g}))^c$, and \eqref{axiomABRR1}, we have,
\begin{align*}
(\iota_-)_2\big(U_{12}(x)\big) 
  &=S^{(1)}_{12}\;
    \prod_{k=1}^{+\infty}\left\{
    \big( \Ad^{-1}_{\varphi_{12}^0}\circ\theta^+_{[x] 1}\big)^k
    \circ \Ad_{[S^{(1)\; -1}_{12}\,\Gamma_{12}^{(+)}]} 
          \big({\mathfrak C}^{[+]}_1(x)\big) 
    \right\}\ S^{(1)\; -1}_{12}
                \\
  &=S^{(1)}_{12}\; \prod_{k=1}^{+\infty}\left\{
    \big(\theta^{+}_{[xq^{h_2}]1}\big)^k \big({\mathfrak C}^{[+]}_1(xq^{h_2})\big)
   \right\}\ S^{(1)\; -1}_{12}\\
  &=S^{(1)}_{12}\; M_1^{(+)}(xq^{h_2})\; S^{(1)\; -1}_{12}.
\end{align*}

On the other hand, with the same notations as in the proof of Lemma~\ref{lemma3coboundary},
\begin{align*}
 (\iota_-)_2 \big( \Delta(M^{(-)}(x)) \big) 
 &= \prod_{k=0}^{+\infty}
    \left\{\Ad_{\uphi_{12}^{[-k]}} \circ\big(\utheta^{-}_{[x]1}\big)^k
            \Big(\Ad_{\Gamma_{12}^{(-)}}\big( {\mathfrak C}^{[-]}_1(x) \big) 
            \Big)\right\}
                       \\
 &= \prod_{k=0}^{+\infty}
    \left\{\Ad_{[(\uphi_{12}^{0\; -1}\,\uphi_{12}^{-})^k\, S^{(1)}_{12}]}
           \circ\big(\utheta^{-}_{[xq^{h_2}]1}\big)^k
            \big( {\mathfrak C}^{[-]}_1(xq^{h_2}) \big)\right\}
                       \\
 &= S^{(1)}_{12}\; M_1^{(-)}(xq^{h_2}) \;S^{(1)\; -1}_{12},
\end{align*}
in which we have used \eqref{axiomGTDb}, \eqref{axiomPsi0Phi-} and \eqref{axiomABRR1}.
As  $\uS_{12}^{(1)\; -1}{F}_{12}(x)\in 1^{\otimes 2}\oplus (U_q^+({\mathfrak g})\otimes U^-_q({\mathfrak g})\otimes {\cal F}^{(0)}(\h))^c)$, it follows that $(\iota_-)_2\left( V(x) \right) =S^{(1)}_{12}\;M^{(+)}_1(xq^{h_2})\;S^{(1)\; -1}_{12}.$

Combining these results we obtain that $U(x)^{-1}V(x) \in 1^{\otimes 2} \oplus (U_q({\mathfrak g})\otimes U^-_q({\mathfrak g})\otimes {\cal F}^{(0)}(\h))^c).$ 
The relation (\ref{ABRRUV}) is a direct consequence of (\ref{ABRR2U}) and (\ref{ABRR2V}).
\cqfd

Finally, since the  solution for $U^{-1}(x)V(x)   \in 1\otimes 1 \oplus (U_q({\mathfrak g})\otimes U^-_q({\mathfrak g})\otimes {\cal F}^{(0)}(\h))^c)$ of  (\ref{ABRRUV}) is unique (because of the non degeneracy condition),  we must have $U^{-1}(x)V(x)=1\otimes 1.$ This ends the proof of the main theorem.
\cqfd

It remains to find interesting examples of this theorem, which is the purpose of the next section.

\section{Universal Vertex-IRF transformation in the standard case}
\label{section-UVIRF}

In this section, $(\utheta^+, \utheta^-,  \uphi^0, \uS^{(1)},\uphi^+,\uphi^-)$ denotes the standard IRF generalized translation datum defined in the basic example \ref{subsec-StandardIRF}. Our purpose is to construct explicitely, in this case, a solution $M(x)$ of the Quantum Dynamical Coboundary Equation along the line proposed in Theorem~\ref{maintheorem}. 
We therefore have to determine $\mathfrak{C}^{[\pm 1]}(x)$ and $M^{(0)}(x)$ such that the set of axioms \eqref{axiomABRR0}- \eqref{axiomABRR2prime} are satisfied.

{}From now on, we will denote, similarly as in our previous article \cite{BRT}, $\mathfrak{C}^{[+]}(x):=\mathfrak{C}^{[+1]}(x)$, $\mathfrak{C}^{[-]}(x):=(\utheta^{-}_{[x]})^{-1}(\mathfrak{C}^{[-1]}(x))=\Ad_{B(x)}(\mathfrak{C}^{[-1]}(x))$, and we will choose
\begin{alignat}{2}
  &\Lambda^{(\pm)}_{12}=S_{12}^{(1)}, &\qquad& \Gamma_{12}^{(\pm)}=K^{\pm 1}\,S_{21}^{(1)},
\end{alignat}
as a special solution of \eqref{choix+}-\eqref{choix-}. The conditions \eqref{axiomABRR0}-\eqref{axiomABRR2prime} become
\begin{align}
   &\Delta({M}^{(0)}(x))=S^{(1)\; -1}_{12}\,{ M}^{(0)}_1(xq^{h_2})\,{ M}^{(0)}_2(x),
    \label{OLDaxiomABRR0}\\
   &K_{12}^{-1}\,\Delta({\mathfrak C}^{[\pm]}(x))\,K_{12}=
    \Ad_{S^{(1)}_{21}}\big({\mathfrak C}^{[\pm]}_{1}(x)\big)\;
    \Ad_{[K^{\mp 1}_{12}\,S^{(1)}_{12}]}\big({\mathfrak C}^{[\pm]}_{2}(x)\big),
     \label{OLDaxiomABRR1d}\\
   &{\mathfrak C}^{[\pm]}_{1}(xq^{h_2})=
     \Ad_{[S^{(1)\; -1}_{12}\,S^{(1)}_{21}\,K_{12}]}\big(
     {\mathfrak C}^{[\pm]}_{1}(x)\big),
     \label{OLDaxiomABRR1}\\
   &{\mathfrak C}^{[-]}_2(x)\; {\mathfrak C}^{[+]}_1(xq^{h_2}) \ 
      \Ad_{  B_2(x)}\circ \theta^-_{[x] 2} 
                       \big(S^{(1)\; -1}_{12}\,\widehat{R}_{12}\, S^{(1)}_{12}\big)
                          \nonumber\\ 
   &\hspace{6cm}
     =\big\{ S^{(1)\; -1}_{12}\, \widehat{R}_{12}\,  S^{(1)}_{12}\big\}\; 
     {\mathfrak C}^{[+]}_1(xq^{h_2})\; {\mathfrak C}^{[-]}_2(x).
     \label{OLDaxiomABRR2prime}
\end{align}
Our aim is now to solve this set of equations in the case where ${\mathfrak g}=A_r^{(1)},$ the $A_r$ case was already done in \cite{BRT}.

\subsection{Gauss decomposition of Quantum Dynamical Coboundaries and Sevostyanov's characters in the $U_q(A_r^{(1)})$ standard case}
\label{Sevostyanov}

In the following, we develop a method enabling us to find solutions to  the system of equations \eqref{OLDaxiomABRR0}-\eqref{OLDaxiomABRR2prime},  in order to obtain a universal Vertex-IRF transformation. We will show that, with the use of characters of a twisted version of $U_q(\n_\pm)$, which first appeared in  \cite{Sev}, we can effectively  construct 
${\mathfrak C}^{[\pm]}(x)$ satisfying automatically the equations \eqref{OLDaxiomABRR1d},  \eqref{OLDaxiomABRR1}. This is  a serious  improvement of our previous work \cite{BRT} where the formulas for ${\mathfrak C}^{[\pm]}(x)$ were simply guessed.
The hexagonal relation is then proved, when it holds, by verifying it in tensor products of extended evaluation representations.

\begin{definition}
Let  $\omega\in \h^{\otimes 2}$, we define the algebra $U_q(\n_\pm)^{\omega}$ as follows:
\begin{itemize}
\item $U_q(\n_+)^{\omega}$ is the subalgebra of $U_q(\b_+)$ generated by 
$\{\tilde{e}_i=q^{-( \alpha_i\otimes\id )(\omega)}e_i,\ i=0,\ldots,r\}.$
\item $U_q(\n_-)^{\omega}$ is the subalgebra of $U_q(\b_-)$ generated by 
$\{\tilde{f}_i=f_iq^{( \alpha_i\otimes\id)(\omega)},\ i=0,\ldots,r\}.$
\end{itemize}
\end{definition}

The introduction of these algebras aims at fixing a bad property of $U_q(\n_+)$ compared to $U(\n_+)$: $U_q(\n_+)$ does not admit any non trivial one dimensional representations. This is in far contrast with the fact that  any $(a_i)\in (\C^\times)^{r+1}$ defines a character of $U(\n_+)$ by assigning $e_i\mapsto a_i.$
We will see that there exists $\omega$ such that the space of characters of 
$U_q(\n_\pm)^{\omega}$ is $\C^{r+1}.$

A property, which is trivial to show but will be important in the sequel, is:
\begin{align}
\forall\gamma\in P,\
&q^{\omega}(e_\gamma\otimes f_{\gamma})q^{-\omega}\in U_q(\b_+)\otimes  U_q(\n_-)^{\omega},\\
&q^{\omega}(f_\gamma\otimes e_{\gamma})q^{-\omega}\in U_q(\b_-)\otimes  U_q(\n_+)^{\omega}.
\end{align}

\begin{lemma}{\label{lem-defCpm}}
Let us suppose that there exists $M^{(0)}(x)\in\A(\h)\otimes U_q(\h)$ invertible satisfying \eqref{OLDaxiomABRR0} and the additional condition 
\begin{equation}
 {  M}^{(0)}_1(xq^{h_2})^2 = {  M}^{(0)}_1(x)^2\, S^{(1)}_{12} \, S^{(1)\; -1}_{21} \, K^{-1},
    \label{OLDaxiomABRR0b}
\end{equation} 
and assume that $\epsilon^+$ is a character of $U_q(\n_-)^{\omega_-}$, with $\omega_-=\log_q(K_{12}S_{21}^{(1)})$, and that $\epsilon^-$ is a character of $U_q(\n_+)^{\omega_+}$, with $\omega_+=\log_q(S_{21}^{(1)})$. Then 
 the following elements 
\begin{align}
  {\mathfrak C}^{[+]}(x) 
  &= M^{(0)}(x)^{-2}\; (\id \otimes \epsilon^+)\big(K_{12}S^{(1)}_{21}\, \widehat{R}_{12}\,   
        K^{-1}_{12}\, S^{(1)\; -1}_{21} \big)  \; M^{(0)}(x)^{2}
                   \label{epsCplus2}\\
 &= (\id \otimes \epsilon^+)\big(M^{(0)}_2(x)^{2}\; S^{(1)\; -1}_{12}\, \widehat{R}_{12}\, 
        S^{(1)}_{12}\; M^{(0)}_2(x)^{-2}\big),
                   \label{epsCplus1} \\
{\mathfrak C}^{[-]}(x) 
&= M^{(0)}(x)^{-2}\; (\id \otimes \epsilon^-)\big(S^{(1)}_{21}\, \widehat{R}^{-1}_{21}\,
        S^{(1)\; -1}_{21} \big)  \; M^{(0)}(x)^{2}
                   \label{epsCmoins1} \\
 &=  (\id \otimes \epsilon^-)\big(M^{(0)}_2(x)^{2} \; K^{-1}_{12}\, S^{(1)\; -1}_{12} 
        \widehat{R}^{-1}_{21} \,S^{(1)}_{12}\, K_{12}\; M^{(0)}_2(x)^{-2}),  
                   \label{epsCmoins2}
\end{align}
obey the properties (\ref{OLDaxiomABRR1d}), (\ref{OLDaxiomABRR1}).\\
We will denote in the following $a_i^+=\epsilon^+(\tilde{f}_i)$ and 
$a_i^-= \epsilon^-(\tilde{e}_i).$
\end{lemma}

\proof
Let us first remark that the expressions (\ref{epsCplus1}) and (\ref{epsCplus2}) (resp. (\ref{epsCmoins1}) and (\ref{epsCmoins2})) are equivalent due to the weight zero property of the $R-$matrix and to the formulae (\ref{OLDaxiomABRR0}) and (\ref{OLDaxiomABRR0b}). 

Using the quasitriangularity and the weight zero property of the $R-$matrix, it is  easy to show that
\begin{align}
 &(\Delta \otimes \id)\big(S^{(1)\; -1}_{12}\, \widehat{R}_{12}\, S^{(1)}_{12}\big)
  =\Ad_{[K_{12}\, S^{(1)}_{21}]} \big(S^{(1)\; -1}_{13}\, \widehat{R}_{13}\, S^{(1)}_{13}\big)\;
   \Ad_{S^{(1)}_{12}}\big(S^{(1)\; -1}_{23}\,\widehat{R}_{23}\,S^{(1)}_{23}\big),
                           \label{deltasrs1}\\
 &(\Delta \otimes \id)\big(S^{(1)}_{21}\,\widehat{R}^{-1}_{21}\,S^{(1)\; -1}_{21}\big)
  =\Ad_{S^{(1)\; -1}_{12}}\big(S^{(1)}_{31}\,\widehat{R}^{-1}_{31}\,S^{(1)\; -1}_{31}\big)\;
   \Ad_{[K_{12}\,S^{(1)\; -1}_{21}]}\big(S^{(1)}_{32}\,\widehat{R}^{-1}_{23}\,S^{(1)\; -1}_{32}\big).
                           \label{deltasrs2}
\end{align}
Applying $(\id^{\otimes 2}\otimes(\epsilon^+\circ \Ad_{M^{(0)}(x)^{-2}}))$ to (\ref{deltasrs1}) and using (\ref{epsCplus1}) one obtains the equation 
(\ref{OLDaxiomABRR1d}) on ${\mathfrak C}^+(x).$
Analogously  using  (\ref{deltasrs2}) with (\ref{OLDaxiomABRR0}) and (\ref{epsCmoins1}) we obtain  the equation  (\ref{OLDaxiomABRR1d})  on ${\mathfrak C}^-(x).$
Using (\ref{epsCmoins1}), (\ref{epsCplus2}) and (\ref{OLDaxiomABRR0b}), we deduce  (\ref{OLDaxiomABRR1}).
\cqfd

\begin{lemma}\label{lem-M0}
An invertible solution $M^{(0)}(x)\in \A[\h]\otimes U_q(\h)$ solution of the equations  (\ref{OLDaxiomABRR0}), (\ref{OLDaxiomABRR0b}) is given by:
\begin{equation}
M^{(0)}(x){}^2=k^{-1}\; m(S^{(1)-1})\prod_{\mu=0,\ldots,r,d}
x_\mu^{-\zeta^{\alpha_\mu}+[\id\otimes \nu(\zeta^{\alpha_\mu})-\nu(\zeta^{\alpha_\mu})\otimes \id](\log_q(S^{(1)}))}.
\end{equation}
This expression can be simplified when the relation $KS_{12}^{(1)}S_{21}^{(1)}=
1^{\otimes 2}$ holds and we have 
\begin{equation}
M^{(0)}(x){}^2=\prod_{\mu=0,\ldots,r,d}
x_\mu^{-\zeta^{\alpha_\mu}+[\id\otimes \nu(\zeta^{\alpha_\mu})-\nu(\zeta^{\alpha_\mu})\otimes \id](\log_q(S^{(1)}))}.
\end{equation}
\end{lemma}

We now prove 
\begin{lemma}
Let $\omega=\sum_{\mu,\nu=0,\ldots,r,d} \omega_{\mu,\nu} \zeta^{\alpha_\mu}\otimes\zeta^{\alpha_\nu}\in \h^{\otimes 2}$.
The space  of characters of $U_q(\n_\pm)^{\omega}$  is  in bijection with $\C^{r+1}$ when the following condition is satisfied
\begin{equation}
\forall i,j\in \Gamma,i\not=j,\quad \sum_{k=0}^{1-a_{ij}}q^{k(\omega_{ji}-\omega_{ij})}(-1)^k\bmatrix 1-a_{ij} \\ k \endbmatrix_{q} =0,
\end{equation}
which is equivalent to $\omega_{ij}-\omega_{ji}\in \{a_{ij},a_{ij}+2,\ldots,-a_{ij}\}. $
\end{lemma}

\proof
Trivial computation from the Serre's relations.
\cqfd

\begin{proposition}Let us assume that $\g=A_r^{(1)}$, $\aleph=1$,
 and let $S^{(1)}$ be given by \eqref{ExpressionS1}.
Let us define  $\omega_-=\log_q(K_{12}S_{21}^{(1)})=-\log_q(S_{12}^{(1)})$ and $\omega_+=\log_q(S_{21}^{(1)}),$ then the space  of characters of $U_q(\n_-)^{\omega_-}$ and  $U_q(\n_+)^{\omega_+}$  are both in bijection with $\C^{r+1}.$ 
\end{proposition}

\proof
In particular, when $a_{ij}=0$, the condition of the previous lemma is satisfied if $(\omega_{\pm})_{ ij}=0.$
{}From the explicit expression of $S^{(1)}$ this condition  holds. 
\cqfd

In particular when $\aleph=1$ we obtain
\begin{equation}
\tilde{e}_i=
q^{(\zeta^{\alpha_i}-\zeta^{\alpha_{\lfloor i-1\rfloor}}+\frac{\lfloor i-1\rfloor}{r+1}c+\frac{1}{2(r+1)}c)}e_i,\qquad
\tilde{f}_i=f_i
q^{(\zeta^{\alpha_i}-\zeta^{\alpha_{\lfloor i+1\rfloor}}-\frac{\lfloor i\rfloor}{r+1}c+\frac{1}{2(r+1)}c )}.
\end{equation}

It is not clear at all whether the previous construction now allows us to solve the hexagonal relation \eqref{OLDaxiomABRR2prime}.
The meaning of this relation is still mysterious for us  and it is clear, at least in the finite dimensional situation \cite{BRT}, that the hexagonal relation can only be verified for very specific vertex Generalized Translation Data and for specific $\g$ (we remind the reader that in the finite dimensional case $\g$ has to be $A_r$ whereas the vertex Generalized Translation Data is necessarily of Cremmer-Gervais type).

We will show that the hexagonal relation can be proved in the case $\g=A_r^{(1)}$ and when $J(x)$ is the Belavin-Baxter solution with $\aleph=1.$ 
It would be interesting to extend  the theorem of Balog-Dabrowski-Feher to the affine case. We conjecture the following result:

\begin{conjecture} Let $\g$ be of affine type  and $\underline{F}(x)$ be the standard IRF solution. There exists a generalized quantum dynamical coboundary between  $\underline{F}(x)$ and $J(x)$ of vertex type if and only if $\g=A_r^{(1)}$ and $J(x)$ is the Belavin-Drinfeld solution associated to $\aleph=1.$
\end{conjecture}

In the rest of this article, we will restrict ourselves to the study of the case $\g=A_r^{(1)}$ and of the vertex Generalized Translation Datum $(\theta^+, \theta^-, \varphi^0, \varphi^+,\varphi^-,S^{(1)})$ associated to the Belavin-Baxter solution with $\aleph=1.$

\begin{rem}
$\mathfrak{C}^{[\pm]}(x)$, defined as in Lemma~\ref{lem-defCpm}, belongs to 
$({\cal F}^{(0)}(\h')\otimes U^{\pm}(\g'))^c.$ Indeed, since  for $a\in U_q(\g)$, $\Ad_{a}$ leaves $U_q(\g')$ invariant, it is sufficient to show that  
$(\id \otimes \epsilon^-)\big(S^{(1)}_{21}\, \widehat{R}^{-1}_{21}\,
        S^{(1)\; -1}_{21} \big)$ and 
$(\id \otimes \epsilon^+)\big(K_{12}S^{(1)}_{21}\, \widehat{R}_{12}\,   
        K^{-1}_{12}\, S^{(1)\; -1}_{21} \big)$ belong to $U_q(\g').$ 
Indeed one has
\begin{align*}
 &\Ad_{S_{21}^{(1)}}(f_i\otimes e_i) =
 \big(f_i\otimes q^{\zeta^{\alpha_i}-\zeta^{\alpha_{\lfloor i-1\rfloor }}
                +\frac{\lfloor i-1\rfloor}{r+1}c+\frac{1}{2(r+1)}c} \big)
 \big(q^{-\zeta^{\alpha_i}+\zeta^{\alpha_{\lfloor i+1\rfloor}}
                +\frac{\lfloor i\rfloor}{r+1}c-\frac{1}{2(r+1)}c}\otimes e_i \big),\\
 &\Ad_{KS_{21}^{(1)}}(e_i\otimes f_i) =
\big( q^{-\zeta^{\alpha_i}+\zeta^{\alpha_{\lfloor i-1\rfloor}}
      -\frac{\lfloor i-1\rfloor}{r+1}c-\frac{1}{2(r+1)}c} \otimes f_i \big)
 \big(e_i\otimes q^{\zeta^{\alpha_i}-\zeta^{\alpha_{\lfloor i+1\rfloor}}
      -\frac{\lfloor i\rfloor}{r+1}c+\frac{1}{2(r+1)}c } \big),
\end{align*}
and the elements $-\zeta^{\alpha_i}+\zeta^{\alpha_{\lfloor i-1\rfloor}}\in \h'.$
\end{rem}

We define $U_q(\b_\pm')=U_q(\b_\pm)\cap U_q(\g').$

\begin{definition}{}
 \par\noindent
For $z\in \C^\times$, $\tilde{c}\in \C$, we define the extended evaluation representations 
$Ev^{+}_{(z,\tilde{c})} $  of $U_q(\b_+')$ and $Ev^{-}_{(z,\tilde{c})} $ of $U_q(\b_-')$ by
\begin{alignat*}{3}
 &Ev^{+}_{(z,\tilde{c})}(e_i)=\ev_z(e_i), \quad  &
 &Ev^{-}_{(z,\tilde{c})}(f_i)=\ev_z(f_i), \quad  &
 &Ev^{\pm}_{(z,\tilde{c})}(h_{\alpha_i})=\ev_z(h_{\alpha_i}),\quad
                      i=1,\ldots, r,\\
 &Ev^{+}_{(z,\tilde{c})}(e_0)=\ev_z(e_0), \quad  &
 &Ev^{-}_{(z,\tilde{c})}(f_0)=\ev_z(f_0), \quad  &
 &Ev^{\pm}_{(z,\tilde{c})}(c)= \tilde{c}.
\end{alignat*}
For a $n-$ tuple ${\cal Z}=\big(z_i,\tilde{c}_i)\in (\C^{\times}\times \C)^n$ we will denote 
$\overset{n}{E}v^{\pm}_{\cal Z}=\bigotimes_{i=1}^n Ev_{(z_i,\tilde{c}_i)}^{\pm}$.
\end{definition}

Let ${\cal F}_{\h{}'}^{(n)}$ be the subfield of ${\cal F}_{\h}^{(n)}$ generated by 
$U_q(\h')^{\widehat{\otimes}n}$ and $D^{-}(\h')=\C[x_0^2,\ldots, x_r^2].$
An element $y^\pm(x)\in  (U_q({\mathfrak n}^{\pm})\otimes {\cal F}_{\h{}'}^{(1)})^c$ is said to be meromorphic if the series 
$\overset{n}{E}v^\pm_{\cal Z}(y^\pm(x))$
is  a meromorphic function in the variables ${\cal Z}\in (\C^{\times}\times \C)^n$, $x\in ({\C}^{\times})^r$ for every $n\geq 1.$
An element $y(x)\in (U_q(\n_+)\widetilde{\otimes}{\cal F}(\h')^{\widehat{\otimes} 2}\widetilde{\otimes} U_q(\n_-))^c$ is said to be meromorphic if the series $(\overset{n}{E}v^+_{\cal Z}\otimes 
\overset{m}{E}v_{\cal Z'}^-)(y(x))$ is a meromorphic function in the variables
 ${\cal Z}\in (\C^{\times}\times \C)^n, {\cal Z}'\in (\C^{\times}\times \C)^m, x\in  ({\C}^{\times})^r$ for every $m,n\geq 0.$

We now show  that it is sufficient to prove the hexagonal relation in any extended evaluation representation in order to prove it at the universal level.

\begin{proposition}
Let $\mathfrak{C}^{[\pm]}(x)\in 1\oplus ({\cal F}^{(0)}(\h')\otimes  U_q^\pm(\g))^c$ satisfying \eqref{OLDaxiomABRR1d} and \eqref{OLDaxiomABRR1} and assume that $Ev^\pm(\mathfrak{C}^{[\pm]}(x))$ are meromorphic functions, we define
\begin{align*}
{\cal W}_{12}(x)&= {\mathfrak C}^{[+]}_1(xq^{h_2}) \;
     \Ad_{B_2(x)}\circ\theta^-_{[x] 2}\big(S^{(1)\; -1}_{12}\,\widehat{R}_{12}\,
             S^{(1)}_{12}\big)\; {\mathfrak C}^{[-]}_2(x)^{-1},
                                \\
\widetilde{\cal W}_{12}(x)&= {\mathfrak C}^{[-]}_2(x)^{-1}\, 
      \big\{ S^{(1)\; -1}_{12}\, \widehat{R}_{12}\, S^{(1)}_{12}\big\}\,  
     {\mathfrak C}^{[+]}_1(xq^{h_2}),\, 
\end{align*}
${\cal W}_{12}(x)$ and $\widetilde {\cal W}_{12}(x)$ are meromorphic elements of $(U_q(\n_+)\widetilde{\otimes}{\cal F}(\h')^{\widehat{\otimes} 2}\widetilde{\otimes} U_q(\n_-))^c$.

If the meromorphic functions  $(Ev^+\otimes Ev^-) ({\cal W}_{12}(x))$ and   $(Ev^+\otimes Ev^-) (\widetilde{\cal W}_{12}(x))$ are equal then ${\cal W}_{12}(x)=\widetilde {\cal W}_{12}(x).$
\end{proposition}

This proposition is divided in three lemmas:

\begin{lemma}{}
Let $\mathfrak{C}^{[\pm]}\in 1\oplus ({\cal F}^{(0)}(\h')\otimes  U_q^\pm(\g'))^c$ satisfying \eqref{OLDaxiomABRR1d} and \eqref{OLDaxiomABRR1} and assume that 
$(Ev^+\otimes Ev^-) ({\cal W}_{12}(x))$ and   $(Ev^+\otimes Ev^-)(\widetilde{\cal W}_{12}(x))$ are meromorphic functions and equal, then the functions  $(\overset{n}{E}v^+\otimes \overset{m}{E}v^-) ({\cal W}_{12}(x))$ and   $(\overset{n}{E}v^+\otimes \overset{m}{E}v^-) (\widetilde{\cal W}_{12}(x))$ are meromorphic and  equal for all $n,m\geq 0.$

\end{lemma}

\proof
Using \eqref{OLDaxiomABRR1d} and \eqref{OLDaxiomABRR1}, we have
\begin{align*}
   &(\id \otimes \Delta)({\cal W}_{12} (x))
        =\Ad_{[K_{23}^{\phantom{1}}\, S^{(1)}_{32}]}\big( {\cal W}_{13}(x q^{h_2})\big)\;
         {\mathfrak C}^{[+]}_1(xq^{h_2+h_3})^{-1}\;
         \Ad_{S^{(1)}_{23}}\big( {\cal W}_{12}(x q^{h_3})\big),\\
  &(\id \otimes \Delta)(\widetilde{\cal W}_{12}(x))
       =\Ad_{[K_{23}^{\phantom{1}}\, S^{(1)}_{32}]}\big(\widetilde{\cal W}_{13}(x q^{h_2})\big)\;
         {\mathfrak C}^{[+]}_1(xq^{h_2+h_3})^{-1}\;
         \Ad_{S^{(1)}_{23}}\big(\widetilde{\cal W}_{12}(x q^{h_3})\big).
 \end{align*}
{}From these relations, we deduce the following result: 
if $(\overset{}{E}v^+\otimes \overset{m}{E}v^-) ({\cal W}_{12}(x))$  (resp.  
$(Ev^+\otimes \overset{m}{E}v^-) (\widetilde{\cal W}_{12}(x))$) is a meromorphic function then  
$(Ev_1^+\otimes \overset{m+1}{E}\!\!v^-) ({\cal W}_{12}(x))$ is  a meromorphic function (resp.  $(Ev^+\otimes \overset{m+1}{E}\!\!v^-) (\widetilde{\cal W}_{12}(x))$) is a meromorphic function).   
Moreover if  $(Ev^+\otimes \overset{m}{E}v^-) ({\cal W}_{12}(x)-\widetilde{\cal W}_{12}(x))=0$ then 
 $(Ev^+\otimes \overset{m+1}{E}\!\!v^-) ({\cal W}_{12}(x)-\widetilde{\cal W}_{12}(x))=0.$
Similarly, from the relations
\begin{align*}
  &(\Delta  \otimes \id)({\cal W}_{12} (x))
       =\Ad_{[K_{12}^{\phantom{1}}\, S^{(1)}_{21}]}\big( {\cal W}_{13}(x)\big)\;
         {\mathfrak C}^{[-]}_3(x)\;
         \Ad_{S^{(1)}_{12}}\big( {\cal W}_{23}(x)\big),\\
  &( \Delta  \otimes \id)(\widetilde{\cal W}_{12}(x))
       =\Ad_{[K_{12}^{\phantom{1}}\, S^{(1)}_{21}]}\big(\widetilde{\cal W}_{13} (x)\big)\;
	{\mathfrak C}_3^{[-]}(x)\;
         \Ad_{S^{(1)}_{12}}\big( \widetilde{\cal W}_{23} (x)\big),
 \end{align*}
we deduce that
if $(\overset{n}{E}v^+\otimes Ev^-) ({\cal W}_{12}(x))$  (resp.  $(\overset{n}{E}v^+\otimes Ev^-) (\widetilde{\cal W}_{12}(x))$) is a meromorphic function then  
$(\overset{n+1}{E}\!\!v^+\otimes Ev^-) ({\cal W}_{12}(x))$ is  a meromorphic function (resp.  $(\overset{n+1}{E}\!\!v^+\otimes Ev^-) (\widetilde{\cal W}_{12}(x))$) is a meromorphic function).   
Moreover if  $(\overset{n}{E}v^+\otimes Ev^-) ({\cal W}_{12}(x)-\widetilde{\cal W}_{12}(x))=0$ then 
 $(\overset{n+1}{E}\!\!v^+\otimes Ev^-) ({\cal W}_{12}(x)-\widetilde{\cal W}_{12}(x))=0.$
As a result by induction on $m,n$, we obtain the result of the lemma.
\cqfd

\begin{lemma}{}Let $\omega_z,\ z\in (\C)^{\times}$ be the matrix 
\begin{equation}
\omega_z=\begin{pmatrix}
   0& 1 & 0& \hdotsfor{3}\\
   0 &0 &1&0&\hdotsfor{2}\\
   \hdotsfor{6}\\
    \hdotsfor{6}\\
\hdotsfor{4}&0&1\\
  z&0& \hdotsfor{2}& 0&0
\end{pmatrix},
\end{equation}
we have the following property, which is proved by a direct verification on the generators of 
 $U_q(\b_-'):$
\begin{equation}
\forall u\in U_q(\b_\pm'),\ \forall z\in \C^{\times},\quad
\ev_z\circ \sigma^\pm(u)=\Ad_{\omega_z}^{\mp 1}\circ \ev_z(u).
\end{equation}
\end{lemma}

\begin{lemma}{}
Let $y^{\pm}\in (U_q({\mathfrak n}^{\pm})\otimes {\cal F}^{(1)}({\h{}'}))^c$ be meromorphic elements, if the meromorphic functions 
$\overset{n}{E}v^\pm(y^{\pm})$ are zero for every $n$  then $y^{\pm}=0.$
Let $y\in (U_q(\n_+)\widetilde{\otimes}{\cal F}(\h')^{\widehat{\otimes} 2}\widetilde{\otimes} U_q(\n_-))^c$ a  meromorphic element such that $(\overset{n}{E}v^+\otimes \overset{m}{E}v^-)(y)=0$ for all $n,m\geq 0$, then $y=0.$
\end{lemma}

\proof
Let $J$ be the set of  meromorphic elements $y^+$ of   $(U_q({\mathfrak n}^{+})\otimes {\cal F}^{(1)}({\h{}'}))^c$  such that $\overset{n}{E}v^\pm(y^{\pm})$ are zero for every $n.$
We first show that $J\cap  (U_q({\mathfrak n}^{+})\otimes {\cal F}^{(1)}({\h{}'}))=\{0\}$ implies that $J=0.$
Indeed let $y^+=\sum_{\gamma\in P}a_\gamma e_\gamma\in J$ with $a_\gamma\in {\cal F}^{(1)}({\h{}'}),$ we therefore have 
\begin{equation}
\sum_{\gamma\in P} \bigotimes_{i=1}^n Ev^{+}_{(z_i, \tilde{c}_i)} (a_\gamma)
 \bigotimes_{i=1}^n Ev^{+}_{(z_i, \tilde{c}_i)} (e_\gamma)=0.
\end{equation}
If we develop this meromorphic function in power series in $z_1,\ldots, z_n$ and consider the term of total degree $n_0$, because of the expression of the evaluation representation,  it selects the elements $\gamma$ such that $\langle \Lambda_0, \gamma\rangle=n_0,$ therefore we obtain 
\begin{equation}
\sum_{\gamma\in P,\langle \Lambda_0, \gamma\rangle=n_0 } \bigotimes_{i=1}^n Ev^{+}_{(z_i, \tilde{c}_i)} (a_\gamma)
 \bigotimes_{i=1}^n Ev^{+}_{(z_i, \tilde{c}_i)} (e_\gamma)=0,
\end{equation}
{\it i.e.}
\begin{equation}
\sum_{\gamma\in P,\langle \Lambda_0, \gamma\rangle=n_0 } \bigotimes_{i=1}^n Ev^{+}_{(z_i, \tilde{c}_i)} (a_\gamma)
 \bigotimes_{i=1}^n Ev^{+}_{(z_i,0)} (e_\gamma)=0.
\end{equation}

Applying the previous lemma,  we obtain 
\begin{equation}
\sum_{\gamma\in P,\langle \Lambda_0, \gamma\rangle=n_0 }
 \bigotimes_{i=1}^n \Ad_{ \omega_{z_i}}^{-1} ( Ev^{+}_{(z_i, \tilde{c}_i)} (a_\gamma))
 \bigotimes_{i=1}^n  \Ad_{ \omega_{z_i}}^{-1}( Ev^{+}_{(z_i, 0)} (e_\gamma))=0,
\end{equation}
{\it i.e.}
\begin{equation}
\sum_{\gamma\in P,\langle \Lambda_0, \gamma\rangle=n_0 }
 \bigotimes_{i=1}^n  \Ad_{ \omega_{z_i}}^{-1} ( Ev^{+}_{(z_i, \tilde{c}_i)} (a_\gamma))
 \bigotimes_{i=1}^n  Ev^{+}_{(z_i,0)} (\sigma^+(e_\gamma))=0.
\end{equation}
Let $n_1$ be a fixed non negative integer, the term of total degree $n_1$ in $z_1,\ldots, z_r$ in this series  is given by   $\gamma$ such that 
$\langle \Lambda_0, \sigma^+(\underline{\gamma}) \rangle=n_1.$
An immediate recursion implies that for any $n_0, n_1,\ldots, n_r\in \N$ we must have 
\begin{equation}
\sum_{\gamma\in P(n_0,\ldots, n_r) }
 \bigotimes_{i=1}^n (\Ad_{ \omega_{z_i}})^{-r} ( Ev^{+}_{(z_i, \tilde{c}_i)} (a_\gamma))
 \bigotimes_{i=1}^n Ev^{+}_{(z_i, 0)} ((\sigma^{+})^r(e_\gamma))=0, 
\end{equation}
where $P(n_0,\ldots, n_r)=\{\gamma\in P,\langle \lambda_0,( \sigma^+)^i( \underline{\gamma})\rangle=n_i\}.$ Note that  $P(n_0,\ldots, n_r)$ is finite.
As a result, by applying once more $\bigotimes_{i=1}^n \Ad_{ \omega_{z_i}}^{-1}$ we obtain 
\begin{equation}
\forall n_0,\ldots, n_r\in \N, \sum_{\gamma\in P(n_0,\ldots, n_r) }
 \bigotimes_{i=1}^n  Ev^{+}_{(z_i, \tilde{c}_i)} (a_\gamma) 
 \bigotimes_{i=1}^n  Ev^{+}_{(z_i, 0)} (e_\gamma)=0. 
\end{equation}

Let $y_{n_0,\ldots, n_r}=\sum_{\gamma\in P(n_0,\ldots, n_r) }a_{\gamma}e_\gamma,$ we have shown that $y_{n_0,\ldots, n_r}\in J\cap  (U_q({\mathfrak n}^{+})\otimes {\cal F}^{(1)}({\h{}'})).$
We therefore obtain that $y_{n_0,\ldots, n_r}=0$ {\it i.e.} $a_\gamma=0.$

The proof that $J\cap ( U_q({\mathfrak n}^{+})\otimes {\cal F}^{(1)}({\h{}'}))=\{0\}$ is implied by  the result Proposition 3.15 of Etingof-Kazhdan \cite{EK}. Indeed by applying this result (which has been proved in the formal case but holds also in the evaluated $q$ case as well)  one obtains if $y=\sum_{\gamma\in P}a_\gamma e_\gamma \in  J\cap  (U_q({\mathfrak n}^{+})\otimes {\cal F}^{(1)}({\h{}'}))$  that 
$ \bigotimes_{i=1}^n Ev^{+}_{(z_i, \tilde{c}_i)} (a_\gamma)=0.$
As a result one obtains $a_\gamma=0.$ \\
The other statements of the lemma are proved analogously.
\cqfd

In the following, we fix $M^{(0)}(x)\in \A[\h]\otimes U_q(\h)$ satisfying \eqref{OLDaxiomABRR0} and \eqref{OLDaxiomABRR0b}, for example as the solution given in Lemma~\ref{lem-M0}, and we define, as in Lemma~\ref{lem-defCpm}, $\mathfrak{C}^{[\pm]}(x)=\Ad_{M^{(0)}(x)}^{-2}(C^{\pm}),$ 
with
\begin{align}
 &C^+=(\id\otimes\epsilon^+)  
  \big({S}^{(1)}_{21}\, {K}_{12}\, \widehat{R}_{12}\,
       {K}_{12}^{-1}\,{S}^{(1)\; -1}_{21}\big),\label{def-C+}\\
 &C^{-}=(\id \otimes\epsilon^-)
  \big({S}^{(1)}_{21}\,\widehat{R}^{-1}_{21}\, {S}^{(1)\; -1}_{21} \big).\label{def-C-}
\end{align}

It will be convenient to  use the  decomposition (\ref{decompS}) of $S^{(1)}$,  and a similar one for $K$:
\begin{equation}
  S{}^{(1)}_{12}= \breve{S}{}^{(1)}_{12}\; O^{(1)}_{12}\; P^{(1)}_{12},\qquad
 K_{12}= \breve{K}_{12}\; G_{12}\; H_{12},
\end{equation}
with $\breve{S}{}^{(1)}_{12}$, $O^{(1)}_{12}$, $P^{(1)}_{12}$ given by \eqref{decompS1}, \eqref{decompS2}, and 
\begin{equation}
 \breve{K}_{12}=q^{\sum_{i=0}^r\zeta^{\alpha_i}\otimes
           (2\zeta^{\alpha_{\lfloor i\rfloor}}
           -\zeta^{\alpha_{\lfloor i-1\rfloor}}-\zeta^{\alpha_{\lfloor i+1\rfloor}})},\qquad  
 G_{12}=q^{\zeta^d \otimes \zeta^{\alpha_0}},\qquad 
 H_{12}=q^{\zeta^{\alpha_0} \otimes\zeta^d}.\label{DefAffineFinale}
\end{equation}
With this decomposition, we introduce 
the following notations:
\begin{align}
  &\breve{C}^{[+]}=(\id \otimes\epsilon^+)  
  \big(\breve{S}^{(1)}_{21}\, \breve{K}_{12}\, \widehat{R}_{12}\,
       \breve{K}_{12}^{-1}\,\breve{S}^{(1)\; -1}_{21}\big),\\
  &\breve{C}^{[-]}=(\id \otimes\epsilon^-)
  \big(\breve{S}^{(1)}_{21}\,\widehat{R}^{-1}_{21}\, \breve{S}^{(1)\; -1}_{21} \big).
\end{align}
We will also use the following elements 
\begin{equation}
  u^+=m(O{}^{(1)}_{12}), \quad u^-=m(P^{(1)}_{12}),\quad \breve{k}=m(\breve{K}),\quad
  g=m(G_{12})=m(H_{12}).
\end{equation}

\begin{lemma}{}
Let $(z,\tilde{c}),(z',\tilde{c}')\in \C^{\times}\times\C$.
The relation
\begin{equation}
  (\overset{}{E}v^+_{(z,\tilde{c})}\otimes {E}v^-_{(z',\tilde{c}')}) ({\cal W}_{12}(x))= 
(Ev^+_{(z,\tilde{c})}\otimes {E}v^-_{(z',\tilde{c}')}) (\widetilde{\cal W}_{12}(x))
\end{equation}
is equivalent to 
\begin{multline}
(\overset{}{E}v^+_{(z,\tilde{c})}\otimes {E}v^-_{(z',\tilde{c}')})  \big( \breve{C}^{[-]}_2\,\breve{k}_2\; \Ad_{\breve{S}^{(1)\; -2}_{12}}({\breve{C}}^{[+]}_1) \;
 \Ad_{\breve{S}^{(2)\; -1}_{12}} \circ \sigma^-_{2}(\widehat{R}_{12} ) \big)\\
  =
(\overset{}{E}v^+_{(z,\tilde{c})}\otimes {E}v^-_{(z',\tilde{c}')}) \big(   \Ad_{\breve{S}^{(1)\; -1}_{12}} (\widehat{R}_{12})  \;
   \Ad_{\breve{S}^{(1)\; -2}_{12}}({\breve{C}}^{[+]}_1 ) \; {\breve{C}}^{[-]}_2\,\breve{k}_2 \big).\label{reformhexag3}
\end{multline}
Both handside of this equation being  constant functions of  $\tilde{c},\tilde{c}',$ it is sufficient to show this equality in the case where $ (\tilde{c},\tilde{c}')=(0,0)$ for proving the hexagonal relation.
\end{lemma}

\proof
Let us first notice that
\begin{alignat}{2}
 &\Ad_{O^{(1)}_{12}}(\widehat{R})=\Ad_{u^+_1}^{-1}(\widehat{R}),  & \qquad
 &\Ad_{P^{(1)}_{12}}(\widehat{R})=\Ad_{u^-_2}^{-1}(\widehat{R}),  \qquad \\
 &\Ad_{G_{12}}(\widehat{R})=\Ad_{{g}_1}^{-1}(\widehat{R}), & \qquad 
 &\Ad_{H_{12}}(\widehat{R})=\Ad_{{g}_2}^{-1}(\widehat{R}).
\end{alignat}
In particular, using the explicit expressions \eqref{epsCplus2} and \eqref{epsCmoins1} of ${\mathfrak C}^{[\pm]},$ we have the following property:
\begin{align}
  &{\mathfrak C}^{[+]}(x)=\Ad_{[M^{(0)}(x)^{-2}(u^{-}g)^{-1}]}\big(\breve{C}^{[+]}\big),
  \qquad 
  {\mathfrak C}^{[-]}(x)=\Ad_{[M^{(0)}(x)^{-2}(u^{-})^{-1}]}\big(\breve{C}^{[-]}\big),
\end{align}
Using also the fact that, from \eqref{OLDaxiomABRR0} and \eqref{OLDaxiomABRR0b},
\begin{align*}
 &\Ad_{[M^{(0)}_1(x)^{2}M^{(0)}_2(x)^{2}]}(\widehat{R})
  =\Ad_{[S^{(1)}_{12}S^{(1)}_{21}K_{12}]}(\widehat{R})=\widehat{R},
\end{align*}
the hexagonal relation (\ref{OLDaxiomABRR2prime}), written using (\ref{OLDaxiomABRR1}),
\begin{multline}
  {\mathfrak C}^{[-]}_2(x)\; 
  \Ad_{[S^{(1)\; -1}_{12} S^{(1)}_{21} K_{12}^{\phantom{1}} ]} ({\mathfrak C}^{[+]}_1(x)) \;
  \Ad_{[B_2^{\phantom{1}}(x)\, \theta^-_{[x] 2}( S^{(1)\; -1}_{12})]}\circ 
  \theta^-_{[x] 2}(\widehat{R}_{12})
                     \\
   =
  \Ad_{S^{(1)\; -1}_{12}} (\widehat{R}_{12}) \;     
  \Ad_{[S^{(1)\; -1}_{12} S^{(1)}_{21} K_{12}^{\phantom{1}}]}( {\mathfrak C}^{[+]}_1(x) ) \; 
  {\mathfrak C}^{[-]}_2(x)
  ,
  \label{reformhexag1}
\end{multline}
can therefore be transformed into
\begin{multline}
  \breve{C}^{[-]}_2\; 
  \Ad_{[\breve{S}^{(1)\; -1}_{12} \breve{S}^{(1)}_{21} \breve{K}_{12}^{\phantom{1}}]}
       ({\breve{C}}^{[+]}_1) \;
  \Ad_{[X_2(x)\,
      u_2^- \, \theta^-_{[x] 2}( u_2^-)^{-1}\breve{S}^{(1)}_{21} \breve{K}_{12}^{\phantom{1}})] } 
    \circ \theta^-_{[x] 2}(\widehat{R}_{12} )
                    \\
 =
   \Ad_{[\breve{S}^{(1)}_{21} \breve{K}_{12}^{\phantom{1}}]} (\widehat{R}_{12})  \;
    \Ad_{[\breve{S}^{(1)\; -1}_{12}\breve{S}^{(1)}_{21}\breve{K}_{12}^{\phantom{1}}]}
    ({\breve{C}}^{[+]}_1 ) \; {\breve{C}}^{[-]}_2
  .
  \label{reformhexag2}
\end{multline}
with $X(x)=B(x)\, M^{(0)}(x)^2\,\theta^-_{[x]}(M^{(0)}(x)^{-2})$.

Let us now compute the adjoint action of $[X_2(x)\,u_2^- \, \theta^-_{[x] 2}( u_2^-)^{-1}\breve{S}^{(1)}_{21} \breve{K}_{12}^{\phantom{1}})]$ on 
$\theta^-_{[x] 2}(\widehat{R}_{12} )$.

On the one hand, we have, 
\begin{equation*}
 \Delta \big(X(x)  \big)  = K_{12}^{2}\, (S^{(1)}_{12}S^{(1)}_{21}K_{12})^{-1}\,
         (\theta^-_{[x]})^{\otimes 2} (S^{(1)}_{12}S^{(1)}_{21}K_{12})\;
         X_1(x)\, X_2(x),
\end{equation*}
and on the other hand, using \eqref{axiomGTDi},
\begin{align*}
  \Ad_{X_2(xq^{h_3}) }  \circ\theta^-_{[x] 2}(\widehat{R}_{12})
  &= \Ad_{[ X_2(x)\; S^{(1)}_{23} S^{(1)\; -1}_{32} K_{23}\, 
         \theta^-_{[x] 2}(S^{(1)\; -1}_{23} S^{(1)}_{32} K_{23}) ] } 
    \circ\theta^-_{[x] 2}(\widehat{R}_{12}) \\
  &= \Ad_{[(\varphi^0_{23})^{-1} X_2(x)]} \circ \theta^-_{[x] 2}(\widehat{R}_{12}).
\end{align*}
{}From these two equations, we can deduce that
\begin{align*}
\Ad_{X_2(x)} \circ \theta^-_{[x] 2}(\widehat{R}_{12} ) 
  &= \Ad_{[p^{\frac{2}{r+1}\varpi} k]_2 } \circ \theta^-_{[x] 2}(\widehat{R}_{12} )  \\
  &= \Ad_{[p^{\frac{2}{r+1}\varpi}q^{(\underline{A} )^i_j \zeta^{\alpha_i}\zeta^{\alpha_j}+2\zeta^{\alpha_0}\zeta^{d}}]_2 } \circ \theta^-_{[x]2}(\widehat{R}_{12} ).
\end{align*}
Then, using the explicit expression 
$u^-=q^{\sum_{i=0}^r(\Pi Y^\aleph \Omega^{(\aleph)})^i_0 \, \zeta^{\alpha_i}\zeta^{d}}$, we obtain
\begin{align*}
 \Ad_{[u_2^- \theta^-_{[x] 2}((u_2^-)^{-1})] } \circ \theta^-_{[x] 2}(\widehat{R}_{12} ) 
  &=\Ad_{q^{\sum_{i=0}^r(\Pi (Y^\aleph-1) \Omega^{(\aleph)})^i_0 \, \zeta^{\alpha_i}\zeta^{d}}} \circ 
      \theta^-_{[x] 2}(\widehat{R}_{12} ) \\
  &=\Ad_{q^{-\sum_{i=0}^r \Pi^i_0 \, \zeta^{\alpha_i}\zeta^{d}}} \circ 
      \theta^-_{[x] 2}(\widehat{R}_{12} ).
\end{align*}
Finally, due to the fact that $\breve{S}^{(1)}_{21}\breve{K}_{12}=\breve{S}^{(1)\; -1}_{12}$, we have
\begin{align*}
  \Ad_{\theta^-_{[x] 2}(\breve{S}^{(1)}_{21}\breve{K}_{12}^{\phantom{1}}) } \circ 
      \theta^-_{[x] 2}(\widehat{R}_{12} ) 
 &=\Ad_{\theta^-_{[x] 2}(\breve{S}^{(1)\; -1}_{12}) } \circ \theta^-_{[x] 2}(\widehat{R}_{12} )\\
 &=\Ad_{q^{-\sum_{i,j=0}^r (\Omega^{(\aleph)} \underline{A} )^i_j\, 
                                         \zeta^{\alpha_i} \otimes \zeta^{\alpha_j}
           -\sum_{i=0}^r(Y^\aleph \Omega^{(\aleph)}\underline{A} v)^i\, 
                                         \zeta^{\alpha_i} \otimes \zeta^d } } 
    \circ \theta^-_{[x] 2}(\widehat{R}_{12} )\\
 &=\Ad_{[q^{-\sum_{i,j=0}^r (\Omega^{(\aleph)} \underline{A} )^i_j\, 
                                         \zeta^{\alpha_i} \otimes \zeta^{\alpha_j}}
       (q^{\sum_{i=0}^r(\Omega^{(\aleph)} \underline{A} v)^i\, \zeta^{\alpha_i}\zeta^d })_2] } 
    \circ \theta^-_{[x] 2}(\widehat{R}_{12} ) \\
&=\Ad_{[q^{-\sum_{i,j=0}^r (\Omega^{(\aleph)} \underline{A} )^i_j\, 
                                         \zeta^{\alpha_i} \otimes \zeta^{\alpha_j}}
       (q^{-\sum_{i=0}^r(\Pi \varkappa_0)^i\, \zeta^{\alpha_i}\zeta^d})_2 ] } 
    \circ \theta^-_{[x] 2}(\widehat{R}_{12} ).
\end{align*}
Therefore, combining these results, we obtain
\begin{align*}
  &\Ad_{[X_2(x)^2 u_2^-
         \theta^-_{[x] 2}(u_2^-)^{-1}\breve{S}^{(1)}_{21} \breve{K}_{12} )] } 
   \circ \theta^-_{[x] 2}(\widehat{R}_{12} ) \\
  &\hskip 2cm 
    =\Ad_{[q^{-(\Omega^{(\aleph)} \underline{A} )^i_j\, 
                     \zeta^{\alpha_i} \otimes \zeta^{\alpha_j}} 
          ( q^{(\underline{A} )^i_j\, \zeta^{\alpha_i}\zeta^{\alpha_j}
               +2\, \zeta^{\alpha_0}\zeta^{d}
               -2 (\Pi \varkappa_0)^i\,\zeta^{\alpha_i}\zeta^d  
               - \frac{2}{r+1} w^i \zeta^{\alpha_i}\zeta^d})_2 ]} 
     \circ \sigma^-_{2}(\widehat{R}_{12} )\\
  &\hskip 2cm 
    =\Ad_{[q^{-(\Omega^{(\aleph)} \underline{A} )^i_j \zeta^{\alpha_i} \otimes \zeta^{\alpha_j}}  
         (q^{(\underline{A} )^i_j \zeta^{\alpha_i}\zeta^{\alpha_j}})_2 ]} 
     \circ \sigma^-_{2}(\widehat{R}_{12} ) \\
  &\hskip 2cm 
    =\Ad_{[\breve{S}^{(2)\; -1}_{12}\breve{k}_2]} \circ \sigma^-_{2}(\widehat{R}_{12} ).
\end{align*}
Hence, the hexagonal  $(\ref{reformhexag2})$ can equivalently  be written as 
\begin{multline}
 \breve{C}^{[-]}_2\,\breve{k}_2\; \Ad_{\breve{S}^{(1)\; -2}_{12}}({\breve{C}}^{[+]}_1) \;
 \Ad_{\breve{S}^{(2)\; -1}_{12}} \circ \sigma^-_{2}(\widehat{R}_{12} ) \\
  =
 \Ad_{\breve{S}^{(1)\; -1}_{12}} (\widehat{R}_{12})  \;
   \Ad_{\breve{S}^{(1)\; -2}_{12}}({\breve{C}}^{[+]}_1 ) \; {\breve{C}}^{[-]}_2\,\breve{k}_2.
\end{multline}
Note that both sides of this equation belong to 
$(U_q(\n_+)\widetilde{\otimes}U_q(\overset{\circ}{\h})^{\widehat{\otimes }2} \widetilde{\otimes}U_q(\n_-))^c.$ 
As a result, the evaluation of both  sides of this equation under $(Ev^+_{(z,\tilde{c})} \otimes Ev^-_{(z',\tilde{c}')})$ do not depend on $(\tilde{ c},\tilde {c}'),$ which is the statement of the lemma.
\cqfd

\begin{lemma}{}
Let us define $\mho=\breve{C}^{[+]\;-1}\,\breve{C}^{[-]}\,\breve{k}.$ The relation 
(\ref{reformhexag3}) is satisfied if and only if 
\begin{equation}
\ev_z(\Ad_{\mho^{-1}}(u))=\ev_z(\sigma^-(u)),\quad
\forall u\in U_q({\mathfrak b}_-'),\;\;\ \forall z\in\mathbb{C}^{\times}.
\end{equation}
\end{lemma}

\proof
Using the braiding relation deduced from (\ref{OLDaxiomABRR1d}), we have immediatly
\begin{multline}
 \widehat{R}_{12}\, (K_{12}S^{(1)}_{21})\,{\mathfrak C}^{[+]}_1(x)\, (K_{12}S^{(1)}_{21})^{-1}
    S^{(1)}_{12} \\
  = S^{(1)}_{12}\, {\mathfrak C}^{[+]}_2(x)\, S^{(1)\; -1}_{12} K^{-1}_{12} S^{(1)}_{21}\, 
    {\mathfrak C}^{[+]}_1(x)\, S^{(1)\; -1}_{21} K_{12}\, \widehat{R}_{12}\, S^{(1)}_{12}\,
    {\mathfrak C}^{[+]\; -1}_2(x),\label{reformfusion1}
\end{multline}
which can be rewritten equivalently as
\begin{align}
&\Ad_{\breve{S}^{(1)\; -1}_{12}}(\widehat{R}_{12})\; 
 \Ad_{H^{2}_{12}\breve{S}^{(1)\;-2}_{12}}(\breve{C}^{[+]}_1) 
  = \breve{C}^{[+]}_2 \; \Ad_{(\breve{S}^{(1)}_{21})^2}(\breve{C}^{[+]}_1)\;
 \Ad_{\breve{S}^{(1)\; -1}_{12}}(\widehat{R}_{12})\;
 \breve{C}^{[+]\; -1}_2.\label{reformfusion2}
\end{align}
As a consequence, the relation (\ref{reformhexag3}) for zero central charge can be equivalently rewritten as
\begin{multline}\label{reformhexag4}
 (\ev_z \otimes \ev_{z'})\Big( \breve{C}^{[+]\; -1}_2\, \breve{C}^{[-]}_2 \, \breve{k}_2\; 
 \Ad_{(\breve{S}^{(1)\; -2}_{12})}({\breve{C}}^{[+]}_1) \;
 \Ad_{\breve{S}^{(2)\; -1}_{12}} \circ \sigma^-_{2}(\widehat{R}_{12} )  \Big)\\
 =(\ev_z \otimes \ev_{z'})\Big(   \Ad_{(\breve{S}^{(1)}_{21})^2}(\breve{C}^{[+]}_1)\;
 \Ad_{\breve{S}^{(1)\; -1}_{12}}(\widehat{R}_{12})\;
 \breve{C}^{[+]\; -1}_2\, {\breve{C}}^{[-]}_2\, \breve{k}_2 \Big).
\end{multline}
Using the fact that $(\ev_z \otimes \ev_{z'})\big(\breve{S}^{(1)\; -2}_{12} \sigma^-_2(\breve{S}^{(1)\; -2}_{21}) -1\big)=0$,
we deduce that (\ref{reformhexag4}) is equivalent to
\begin{multline}\label{reformhexag5}
 (\ev_z \otimes \ev_{z'})\Big( \mho_2 \; 
 \sigma^-_{2}\big(\Ad_{(\breve{S}^{(1)}_{21})^2}({\breve{C}}^{[+]}_1) \;
 \Ad_{\breve{S}^{(1)\; -1}_{12}}(\widehat{R}_{12}) \big) \Big)\\
 =(\ev_z \otimes \ev_{z'})\Big(   \Ad_{(\breve{S}^{(1)}_{21})^2}(\breve{C}^{[+]}_1)\;
 \Ad_{\breve{S}^{(1)\; -1}_{12}}(\widehat{R}_{12})\; \mho_2 \Big).
\end{multline}
This concludes the proof of the lemma.
\cqfd

This remark together with the previous lemma imply the following proposition:

\begin{proposition}\label{hexagonal=ad}
The hexagonal relation  \eqref{OLDaxiomABRR2prime} is satisfied if and only if

1) $Ev^{\pm}(\mathfrak{C}^{[\pm]}(x))$ are meromorphic functions;

2) the following simple relation holds in $M_{r+1}(\C)$: 
\begin{equation}
\forall z\in \C^{\times},\quad \Ad_{\ev_z(\mho)}=\Ad_{\omega_z}\label{relationmhoomega}.
\end{equation} 
\end{proposition}

\proof
The sequence of the previous lemmas is designed to prove this result.
\cqfd

In the following section we give explicit universal formulas for  $C^+, C^-$ and verify the relation (\ref{relationmhoomega}).


\subsection{Explicit Solutions  of Quantum Dynamical Coboundaries
in the $U_q(A_r^{(1)})$ standard case}
\label{subsec-Constr}

\begin{proposition}{\label{prop-C-}}
\par\noindent The explicit expression of 
$C^-=(\id\otimes \epsilon^-)(S_{21}^{(1)}\,\widehat{R}_{21}^{-1}\, S_{21}^{(1)\; -1})$ reads:
\begin{align}
 &C^-= X^-\, Y^-\, Z^-, 
\end{align}
with
\begin{align}
 &X^-=\prod_{i=1}^r 
  \exp_q \bigg[-(q-q^{-1})\, f_{\delta-\alpha_{i,r+1}} \,
               q^{-\zeta^{\alpha_0}+\zeta^{\alpha_i}+\frac{i(i-2)}{2(r+1)}c}\,
                (1-q^{-2})^{i-1}\prod_{l=0}^{i-1}a_l^{-} \bigg],\\
 &Y^-=\exp \bigg[\sum_{k>0,j=1,\ldots,r} (-1)^{kr} 
      \frac{(1-q^{-2})^{k(r+1)}}{[k(r+1)]_q} \frac{[kj]_q}{[k]_q}\,
      f_{k\delta}^{(\alpha_j)} \, q^{\frac{k(r-1)}{2}c} \prod_{l=0}^r (a_l^-)^k \bigg],\\
 &Z^-=\prod_{i=r}^1
 \exp_q \Big[ -(q-q^{-1})\, f_{\alpha_i} \, 
     q^{-\zeta^{\alpha_i}+\zeta^{\alpha_{\lfloor i+1\rfloor}}+\frac{\lfloor i \rfloor}{r+1}c-\frac{1}{2(r+1)}c}\,
         a_i^- \Big].
\end{align}
\end{proposition}

\proof
{}From $\epsilon^- (\tilde{e}_i)=a_i^-,$ we obtain 
\begin{equation}
\epsilon^-(\tilde{e}_{\delta-\alpha_{i,j+1}})=
\begin{cases}0& \text{if $j<r$ },\\
(1-q^{-2})^{i-1}\prod_{k=0}^{i-1}a_k^- & \text{if $j=r$ }.
\end{cases}
\end{equation}
We therefore obtain 
\begin{equation}
\epsilon^-(e_{\delta}^{(\alpha_i)}{}')=
\begin{cases}0& \text{if $i<r$ },\\
q^{-1}(-1)^{r-1}(1-q^{-2})^{r}\prod_{k=0}^{r}a_k^- & \text{if $i=r$ }.
\end{cases} 
\end{equation}
For $ \alpha\in\overset{\circ}{\Delta}_{+},\ n\geq 1$, we have   $\epsilon^-(\tilde{e}_{\alpha+n\delta})=0,\ \epsilon^-(\tilde{e}_{\delta-\alpha+n\delta})=0.$
As a result we obtain:
\begin{align}
 &X^{-}=(\id\otimes\epsilon^-)
  \Big(S_{21}^{(1)}\; \prod_{i=1}^r\exp_{q}\big[-(q-q^{-1})\, 
       f_{\delta-\alpha_{i,r+1}}\otimes e_{\delta-\alpha_{i,r+1}}\big]\;
       S_{21}^{(1)\; -1}\Big),\\
 &Y^{-}=(\id\otimes\epsilon^-)\Big(S_{21}^{(1)}\; 
    \exp\Big[-(q-q^{-1})\sum_{k>0}\sum_{i,j=1}^r c_{ij}^{(k)}\, 
             f_{k\delta}^{(\alpha_j)}\otimes e_{k\delta}^{(\alpha_i)}\Big]\;
       S_{21}^{(1)\; -1}\Big),\\
 &Z^{-}=(\id\otimes\epsilon^-)\Big(S_{21}^{(1)}\;
 \prod_{\alpha\in\overset{\circ}{\Delta}_+}^{>}
     \exp_{q}\big[-(q-q^{-1})\, f_{\alpha}\otimes e_{\alpha}\big]\;
       S_{21}^{(1)\; -1}\Big).
\end{align}

The expression of $X^-$ is trivially evaluated.

In order to obtain an explicit expression for  $Z^-$, we remark that only the $\widehat{R}$ matrix of $U_q(\overset{\circ}{\g})$ appears in   $Z^-$. We can therefore express  this element as:
\begin{equation}
 \prod_{\alpha\in\overset{\circ}{\Delta}_+}^{>}\exp_{q}[-(q-q^{-1})\, f_{\alpha}\otimes e_{\alpha}]
 = \prod_{\alpha\in\overset{\circ}{\Delta}_+}^{<}\exp_{q}[-(q-q^{-1})\,
                 f_{\alpha}'\otimes e_{\alpha}'],\label{prod-Rmat}
\end{equation} 
in which the root vectors $e'_\alpha$, $f'_\alpha$ ($\alpha$ not simple) are constructed in accordance with the normal order corresponding to the order of the products in the right hand side of \eqref{prod-Rmat}.
It is easy to check that
$\epsilon^-(\tilde{e}'_{\alpha})=0,\ \alpha\in\overset{\circ}{\Delta}_{+}$,
$\alpha$ {\em not} simple. 
As a result we get  the formula for $Z^-$ given in the statement of the proposition.

The only contributing term in $Y^-$ are those coming from 
$e_{k\delta}^{(\alpha_r)}.$
Using $\epsilon^-(e_{k\delta}^{(\alpha_r)}{}')=0$ for $k>1$ and the functional relation between $e_{k\delta}^{(\alpha_r)}$ and  
$e_{k\delta}^{(\alpha_r)}{}'$ we obtain:
\begin{equation}
 \epsilon^-\big(e_{k\delta}^{(\alpha_r)}{}\big)
 =
  \frac{(q^{-1}-q)^{k-1}}{k}\;
  \epsilon^-\big(e_{\delta}^{(\alpha_r)}{}'\big)^k
 =\frac{q^{-1}(-1)^{kr-1}}{k}\; (1-q^{-2})^{k(r+1)-1}\;\Big(\prod_{l=0}^r a_l^{-}\Big)^k.
\end{equation}
Using the explicit expression (\ref{cijk}) of $c_{ij}^{(k)}$ we finally get the exact expression for $Y^-.$
\cqfd

The computation of $C^+$ is completely analogous.

\begin{proposition}{\label{prop-C+}}
 \par\noindent
The explicit expression of $C^+=(\id\otimes \epsilon^+)\big(KS_{21}^{(1)}\,\widehat{R}_{12}\,(KS_{21}^{(1)})^{-1}\big)$ reads:
\begin{align*}
&C^+=Z^+\, Y^+\, X^+,
\end{align*}
with
\begin{align*}
 &Z^+=\prod_{i=1}^r\exp_{q^{-1}}\Big[(q-q^{-1})\,  
      q^{-\zeta^{\alpha_i}+\zeta^{\alpha_{\lfloor i-1\rfloor}}
         -\frac{\lfloor i \rfloor}{r+1}c+\frac{c}{2(r+1)}}\, e_i\,  a_i^+ \Big],
\\
 &Y^+=\exp\bigg[ \sum_{k>0,j=1,\ldots,r} (-1)^{kr} q^{-k(r+1)} \,
      \frac{(1-q^{2})^{k(r+1)}}{[k(r+1)]_q}\,\frac{[kj]_q}{[k]_q}\,
      e_{k\delta}^{(\alpha_j)}\, q^{-\frac{k(r+1)}{2}c} \,\prod_{l=0}^r (a_l^+)^k \bigg],
\\
 &X^+=\prod_{i=r}^1 \exp_{q^{-1}}\bigg[(q-q^{-1})\, 
      q^{-\zeta^{\alpha_{\lfloor i-1\rfloor}}+\zeta^{\alpha_r}-(1+\frac{i(i-2)}{2(r+1)})c-i+1}\,
      e_{\delta-\alpha_{i,r+1}}\, (1-q^{2})^{i-1}\,
      \prod_{l=0}^{i-1}a_l^{+} \bigg].
\end{align*}
\end{proposition}

\begin{rem}
The proof is completely similar to the one of Proposition~\ref{prop-C-}.
But, in order to bypass this  computation and link the expression of $C^+$ and $C^-$,  we can provide another proof using the following lemma:
\begin{lemma}{}
Let  $\epsilon^+$ be a character of  $U_q(\n_-)^{\omega_-}$  (resp. $\epsilon^-$ a character of 
$U_q(\n_+)^{\omega_+}$) with values in $\C(q^{\frac{1}{r+1}})$, and such that $(\epsilon^+(\tilde f_i))^*= q^{-1}\epsilon^-(\tilde e_i)$ where $*$ is defined in \eqref{defstar}. Then
 \begin{equation}
(C^{+})^*=\Ad_{k^{-1}}(C^{-}).
\end{equation}
\end{lemma}

\proof
The assumption implies the relation
\begin{equation}
\big(\Ad_{k^{-1}}\otimes \epsilon^-\big)\big(S_{21}^{(1)}(f_i\otimes e_i)(S_{21}^{(1)})^{-1}\big)=(*\otimes (*\circ\epsilon^+))\big(KS_{21}^{(1)}(e_i\otimes f_i)(KS_{21}^{(1)})^{-1}\big).
\end{equation}
As a result one obtains the relation
\begin{equation}
\big(\Ad_{k^{-1}}\otimes \epsilon^-\big)\big(
S_{21}^{(1)}\widehat{R}_{21}^{-1}(S_{21}^{(1)})^{-1}\big)=
(*\otimes (*\circ\epsilon ^+))\big(KS_{21}^{(1)}\widehat{R}_{12}(KS_{21}^{(1)})^{-1}\big).
\end{equation}
Application of the previous lemma immediately gives the announced formula for $C^+$ once the exact expression for $C^-$ is known.
\cqfd
\end{rem}

We have reduced in the previous section the proof of the hexagonal relation to the verification of the relation $\Ad_{\ev_z(\mho)}=\Ad_{\omega_z}$. This relation is satisfied for specific choices of $\epsilon^\pm.$

Let us define $a_i^+$ and $a_i^-$ by $(q-q^{-1})\, a_i^{-}\, q^{-\frac{1}{r+1}}=1$, $a_i^+=(q^{-1}a_i^-)^*,$ we thefore have 
\begin{equation*}
 a_i^+a_i^-=-\frac{q}{(q-q^{-1})^2}.
\end{equation*}

\begin{proposition}{\label{prop-evCpm}}
\par\noindent
With these specific values for $a_i^+, a_i^-$ we have 
\begin{align}
&\ev_z(C^{+})^{-1}=f_{q}(z)\,(1+\omega_{z}), \qquad  
\ev_z(C^{-})^{-1}=f_{q^{-1}}(z^{-1})\,(1+\omega_{z}^{-1}), \\
&\text{with}\;\; f_q(z)=\exp\bigg[ \sum_{k>0}
     \frac{(-1)^{k(r+1)}q^{kr}}{[k(r+1)]_q }\,\frac{[k]_q}{k}\,z^{k}\bigg].
\end{align}
As a result 
\begin{equation}
\ev_z(C^{+})^{-1}\ev_z(C^{-})=\frac{f_{q}(z)}{f_{q^{-1}}(z^{-1})}\omega_z,
\end{equation} which implies the assumption 2) of Proposition \ref{hexagonal=ad}. 
Moreover,  $Ev^{\pm}(\mathfrak{C}^{[\pm]}(x))$ are meromorphic functions.
As a result, both assumptions of Proposition \ref{hexagonal=ad}  are satisfied and the hexagonal relation is satisfied. 
\end{proposition}

\proof
Because $\ev_z(\overset{\circ}{\zeta}_i)=-\frac{i}{r+1}1+\sum_{j=1}^iE_{j,j}$, we obtain:
\begin{equation} 
 \ev_z((Z^-)^{-1})= 1+(q-q^{-1})\sum_{i=1}^r a_i^- E_{i+1,i}\, q^{-\frac{1}{r+1}}
 = 1+\sum_{i=1}^rE_{i+1,i}.
\end{equation}
We also have
\begin{align*}
 \ev_z((X^-)^{-1}) &=1+\sum_{i=1}^r (-1)^{i-1}z^{-1}E_{i,r+1}
       \prod_{l=0}^{i-1}\big[a_l^-(q-q^{-1})q^{-\frac{1}{r+1}}\big]\\
  &=1+z^{-1}\sum_{i=1}^r(-1)^{i-1}E_{i,r+1},
\end{align*}
as well as:
\begin{align*}
 \ev_z((Y^-)^{-1})&=\exp\bigg[-\sum_{k>0}\sum_{j=1}^{r}
        \frac{(-1)^{k(r+1)-1}\,q^{-k(r+1-j)}\,[kj]_q}{k\;[k(r+1)]_q}\,
        z^{-k}\, \big(E_{j,j}-q^{2k}E_{j+1,j+1}\big)\bigg]\\
     &=\sum_{j=1}^{r+1}y_q^j(z)E_{j,j}
\end{align*}
with 
\begin{equation*}
 y_q^1(z)=\cdots= y_q^{r}(z)=\exp\bigg[\sum_{k>0}\frac{(-1)^{k(r+1)}\,q^{-kr}}{[k(r+1)]_q }\,
                                 \frac{[k]_q}{k}\, z^{-k}\bigg]=f_{q^{-1}}(z^{-1}).
\end{equation*}
Therefore we obtain the announced formula for $\ev_z(C^{-})^{-1}.$
The computation of $\ev_z(C^{+})^{-1}$ is straightforward from 
 $\ev_z(C^{+})=\ev_z((C^{-})^{*}).$

{}From the explicit expression of $C^+$ and $C^-$, it is now straightforward to compute  $Ev^{\pm}(\mathfrak{C}^{[\pm]}(x))$ and show that they are meromorphic functions.
\cqfd

\subsection{Expression of the Vertex-IRF transformation in the evaluation representation of  $U_q(A^{(1)}_1)$}
\label{subsec-evVIRF}

Here we explain how, from the explicit universal solution for
$U_q(A_1^{(1)})$, one can obtain the well-known solution of Baxter of the Vertex-IRF 
transformation in the two-dimensional evaluation representation.

Let $\g=A_1^{(1)}$, and $p=x_0 x_1$, $w=x_1$.
{}From Proposition~\ref{prop-evCpm}, we have
\begin{align}
 &\ev_z(C^+)   =\frac{(q^2  z;q^4)_{\infty}}{( z;q^4)_{\infty}}
     \begin{pmatrix}
       1 & -1 \\
       -z & 1
     \end{pmatrix},
            \\
  &\ev_z(C^-)
    = \frac{(q^4  z^{-1};q^4)_{\infty}}{(q^2 z^{-1};q^4)_{\infty}}
     \begin{pmatrix}
       1 & -z^{-1} \\
       -1 & 1
     \end{pmatrix}.
\end{align}
On the other hand, choosing $S^{(1)}$ as in \eqref{ExpressionS1}, {\it i.e.} $\log_q(S^{(1)})=-\frac{1}{4} h_1\otimes h_1 +\frac{1}{8} h_1\otimes c -\frac{1}{8} c\otimes h_1 -c\otimes d$, and $M^{(0)}$ as in Lemma~\ref{lem-M0},  we obtain
\begin{equation}
 \ev_z\big(M^{(0)}(x)^2\big)
    = \begin{pmatrix}
       p^{1/4} w^{-1/2} & 0 \\
       0 & p^{-1/4}w^{1/2}
     \end{pmatrix}.
\end{equation}

Therefore, if we denote 
$\mathbf{C}^{[\pm]}(z;p,w) =\ev_z \big(\mathfrak{C}^{[\pm]}(p,w)\big)$, 
$\mathbf{C}^{[\pm k]}(z;p,w) =\ev_z \big(\mathfrak{C}^{[\pm k]}(p,w)\big)$,
we have
\begin{align}
 &{\mathbf C}^{[+]}(z;p,w)
    =\frac{(q^2  z;q^4)_{\infty}}{( z;q^4)_{\infty}}
     \begin{pmatrix}
       1 & -p^{-1/2} w \\
       -p^{1/2} w^{-1}z & 1
     \end{pmatrix},
            \\
  &{\mathbf C}^{[-]}(z;p,w)
    = \frac{(q^4  z^{-1};q^4)_{\infty}}{(q^2  z^{-1};q^4)_{\infty}}
     \begin{pmatrix}
       1 & -p^{-1/2}wz^{-1} \\
       -p^{1/2}w^{-1} & 1
     \end{pmatrix}.
\end{align}
The expressions of $\mathbf{C}^{[\pm k]}(z;p,w)$ can be easily obtained from the fact 
that
\begin{equation}
   \ev_z\circ\sigma^+=\Ad_{\omega_z}\circ\ev_z, \qquad
   \text{with}\quad \omega_z =\begin{pmatrix}
                           0 & 1\\
			   z & 0\end{pmatrix},			   
\end{equation}
and that
\begin{equation}			   
   \ev_z\big( D^+(p)\big) = p^{2z\frac{\dd }{\dd z}}
                                \begin{pmatrix}
				p^{1/2} & 0 \\
				0 & p^{-1/2}
				\end{pmatrix}, \qquad 
   \ev_z\big( B(p,w)\big)= q^{1/2}	 p^{2z\frac{\dd }{\dd z}}
                                \begin{pmatrix}
				w & 0 \\
				0 & w^{-1}
				\end{pmatrix}.		   
\end{equation}
In particular, we have
\begin{align}
   &{\mathbf C}^{[- 1]}(z;p,w) =
  \frac{(q^4 p^{2} z^{-1};q^4)_{\infty}}{(q^2 p^{2} z^{-1};q^4)_{\infty}}
      \begin{pmatrix}
         1 & -p^{3/2} w^{-1} z^{-1}\\
	 -p^{1/2} w  & 1
      \end{pmatrix},
                     \\
   &{\mathbf C}^{[+ 2]}(z;p,w) =
        \frac{(q^2 p^{2} z;q^4)_{\infty}}{( p^{2} z;q^4)_{\infty}}
      \begin{pmatrix}
         1 & -p^{3/2} w^{-1}\\
	 -p^{1/2} w z & 1
      \end{pmatrix}.
\end{align}

In order to compute
$\mathbf{M}^{(\pm)}(z;p,w) =\ev_z \big({M}^{(\pm)}(p,w)\big)$, 
we use the following $p$-difference linear equations, which are direct consequences of the expressions of $\mathbf{M}^{(+)}(z;p,w)$ and $\mathbf{M}^{(-)}(z;p,w)^{-1}$ as infinite products:
\begin{equation}  \label{eq-M+}
   \mathbf{M}^{(+)}(z;p,w)={\mathbf C}^{[+1]}(z;p,w)\,
              {\mathbf C}^{[+2]}(z;p,w)\ 
	      \begin{pmatrix}
	         p & 0\\
		 0 & p^{-1}
	      \end{pmatrix}
	      \mathbf{M}^{(+)}(p^4 z;p,w)
	      \begin{pmatrix}
	         p^{-1} & 0\\
		 0 & p
	      \end{pmatrix},
\end{equation}
\begin{equation}  \label{eq-M-}
   \mathbf{M}^{(-)}(z;p,w)^{-1}=
              \begin{pmatrix}
	         w^{-1} & 0\\
		 0 & w
	      \end{pmatrix}
	         \mathbf{M}^{(-)}(p^{-2} z;p,w)^{-1}
	      \begin{pmatrix}
	         w & 0\\
		 0 & w^{-1}
	      \end{pmatrix}\ 
              {\mathbf C}^{[-1]}(z;p,w).
\end{equation}
\eqref{eq-M+}, \eqref{eq-M-} turn into linear $p$-difference equations for the matrix coefficients of $\mathbf{M}^{(\pm)}(z;p,w)^{\pm 1}$.

More precisely, if we express $\mathbf{M}^{(+)}(z;p,w)$ in the form,
\begin{align}
  & \mathbf{M}^{(+)}(z;p,w) = 
   \frac{(q^2 z ;p^2,q^4)_\infty}{( z ;p^2,q^4)_\infty}(p^2 z;p^4)_\infty
           \begin{pmatrix}
                         a^{(+)}(z;p,w) & p^{-1/2} b^{(+)}(z;p,w)\\
                         p^{1/2} c^{(+)}(z;p,w) & d^{(+)}(z;p,w)
                   \end{pmatrix},
		        \label{ev-M+}
\end{align}
we have
\begin{multline}
 (1-p^2 z)\, a^{(+)}(z;p,w)= \big[1+p^{-2} + (p^{-2} w^2+p^2 w^{-2}) p^2 z\big]\, a^{(+)}(p^4 z;p,w)
                          \\
                      -p^{-2}(1-p^4 z)\, a^{(+)}(p^8 z ;p,w),\label{eq-a+}
\end{multline}
with
\begin{equation}
   a^{(+)}(0;p,w)=1,
\end{equation}
and 
the other matrix coefficients $b^{(+)}(z;p,w)$, $c^{(+)}(z;p,w)$ and $d^{(+)}(z;p,w)$ can be obtained from $a^{(+)}(z;p,w)$ as
\begin{align}
   &c^{(+)}(z;p,w)=\frac{p^2 w}{p^2+w^2}\big[ (1+p^{-4} w^2 z)\, a^{(+)}(z;p,w)
                        -(1-p^{-2}z)a^{(+)}(p^{-4}z;p,w)\big],\\
   &b^{(+)}(z;p,w)=p z^{-1} c^{(+)}\big(z;p,p w^{-1}\big),\qquad
    d^{(+)}(z;p,w)=a^{(+)}\big(z;p,p w^{-1}\big).
\end{align}
Equation \eqref{eq-a+} can be solved using the fact that
\begin{equation}
 (1-z)\, f(z) + \big[ (a+b)z-q^{-1}c-1\big]\, f(qz) + (q^{-1}c-abz)\, f(q^2 z)=0,
\end{equation}
in which $a$, $b$, $c$, $q$ are complex parameters,
admits the solution $f(z)=\hyp{a}{b}{c}{q}{z}$.
Here $\hyprs$ denotes the basic hypergeometric series
\begin{equation}\label{def-hyp}
   \hyprs=\sum_{k=0}^\infty 
        \frac{(a_1;q)_k\ldots (a_r;q)_k}{(q;q)_k\,(b_1;q)_k \ldots (b_s;q)_k}\, 
         \big[(-1)^k q^{\frac{n(n-1)}{2}}\big]^{1+s-r} z^k,
\end{equation}
with
\begin{equation}
   (a;q)_k =\prod_{l=0}^{k-1}(1-a\, q^l)=\frac{(a;q)_\infty}{(a\, q^k;q)_\infty}.
\end{equation}
This leads to 
%
\begin{align}
   &a^{(+)}(z;p,w)=
         \hyp{-p^{-2} w^2 }{-p^2 w^{-2}}{p^2}{p^4}{p^2 z},
   		  \\
   &c^{(+)}(z;p,w)=p^{-1}z\frac{w+w^{-1}}{p-p^{-1}}\
	   \hyp{- p^2 w^2}{-p^2 w^{-2}}{p^6}{p^4}{p^2 z}.
\end{align}

Similarly, if we set
\begin{align}
  & \mathbf{M}^{(-)}(z;p,w)^{-1} = 
   \frac{(q^4 p^2 z^{-1} ;p^2,q^4)_\infty}{(q^2 p^2 z^{-1} ;p^2,q^4)_\infty}
           \begin{pmatrix}
                         \alpha^{(-)}(z^{-1};p,w) & p^{-1/2}\beta^{(-)}(z^{-1};p,w)\\
                         p^{1/2}\gamma^{(-)}(z^{-1};p,w) & \delta^{(-)}(z^{-1};p,w)
                   \end{pmatrix},
		        \label{ev-M-}
\end{align}
we have
\begin{multline}\label{eq-a-bis}
 \alpha^{(-)}(z^{-1};p,w)=(1+w^{-2})\,\alpha^{(-)}(p^{-2}z^{-1};p,w)\\
             +w^{-2}(-1+p^4 z^{-1})\,\alpha^{(-)}(p^{-4}z^{-1};p,w),
\end{multline}
with
\begin{equation}
   \alpha^{(-)}(0;p,w)=1,
\end{equation}
and 
\begin{align}
   &\beta^{(-)}(z^{-1};p,w)=w\big[ \alpha^{(-)}(z^{-1};p,w)
                        -\alpha^{(-)}(p^{-2}z^{-1};p,w)\big],\\
   &\gamma^{(-)}(z^{-1};p,w)=p^{-1} z \beta^{(-)}\big(z^{-1};p,p w^{-1}\big),\quad
    \delta^{(-)}(z^{-1};p,w)=\alpha^{(-)}\big(z^{-1};p,p w^{-1}\big).
\end{align}
These equations can be solved as
\begin{align}
 &\alpha^{(-)}(z^{-1};p,w)=\hyper{p^2 w^{-2}}{p^2}{p^4 w^{-2} z^{-1}},\\
 &\beta^{(-)}(z^{-1};p,w)=(p w^{-1}-p^{-1}w)p z^{-1}\hyper{p^4 w^{-2}}{p^2}{p^6 w^{-2} z^{-1}}.
\end{align}

However, in order to compute the explicit expression of the $2\times 2$ matrix $\mathbf{M}(z;p,w)=\ev_z \big(M(p,w)\big)$, or of its inverse $\mathbf{S}(z;p,w)=\ev_z \big(M(p,w)^{-1}\big)$, which can be obtained from $\mathbf{M}^{(+)}$ and $\mathbf{M}^{(-)}$ as
\begin{equation}\label{eq-S}
    \mathbf{M}^{(+)}(z;p,w)\,\mathbf{S}(z;p,w)= \mathbf{M}^{(-)}(z;p,w)\,\mathbf{M}^{(0)}(z;p,w)^{-1},
\end{equation}
it is more convenient to express the matrix elements of $\mathbf{M}^{(-)}$ in a form similar to those of $\mathbf{M}^{(+)}$.
It is actually possible to obtain, for the matrix elements of $\mathbf{M}^{(-)}$, the same type of equation \eqref{eq-a+} as for $a^{(+)}(z;p,w)$. Indeed, defining
\begin{align}
 &a^{(-)}(z^{-1};p,w)=(p^4 z^{-1};p^4)_\infty^{-1}\,\alpha^{(-)}(z^{-1};p,w),
\end{align}
and iterating \eqref{eq-a-bis}, we obtain that
\begin{multline}
  (1-p^4 z^{-1})\, a^{(-)}(z^{-1};p,w)
          = \big[1+w^{-4} + (1+p^{-2}) p^6 w^{-2} z^{-1}\big]\, a^{(-)}(p^{4} z^{-1};p,w)
                          \\
                      -w^{-4}(1-p^6 z^{-1})\, a^{(-)}(p^{8} z^{-1} ;p,w),\label{eq-a-}
\end{multline}
which can be solved in the same way as \eqref{eq-a+}. 
Finally we obtain
\begin{multline}
  \mathbf{M}^{(-)}(z;p,w) = 
   \frac{(q^2 p^2 z^{-1} ;p^2,q^4)_\infty}{(q^4 p^2 z^{-1} ;p^2,q^4)_\infty}
      (p^2 z^{-1};p^4)_\infty^{-1}\\
  \times
           \begin{pmatrix}
                         d^{(-)}(z^{-1};p,w) & -p^{-1/2}b^{(-)}(z^{-1};p,w)\\
                         -p^{1/2}c^{(-)}(z^{-1};p,w) & a^{(-)}(z^{-1};p,w)
                   \end{pmatrix},
		        \label{ev-M-bis}
\end{multline}
with
\begin{align}
   &a^{(-)}(z^{-1};p,w)=
         \hyp{-w^{-2}}{-p^2 w^{-2}}{p^4 w^{-4}}{p^4}{p^4 z^{-1}},
   		  \\
   &c^{(-)}(z^{-1};p,w)=\frac{1}{w+w^{-1}}\
         \hyp{-w^{2}}{-p^2 w^{2}}{p^4 w^{4}}{p^4}{p^4 z^{-1}},
\end{align}
and
\begin{equation}
    b^{(-)}(z^{-1};p,w)=p z^{-1} c^{(-)}\big(z^{-1};p,p w^{-1}\big),\qquad
    d^{(-)}(z^{-1};p,w)=a^{(-)}\big(z^{-1};p,p w^{-1}\big).
\end{equation}
As a remark, we have in fact used the following identity:
\begin{equation}
 \hyper{a}{q}{az} = (z;q^2)_\infty \; \hyp{-q a}{-a}{a^2}{q^2}{z}.
\end{equation}

The explicit expression of the $2\times 2$ matrix $\mathbf{S}(z;p,w)=\ev_z \big(M(p,w)^{-1}\big)$ follows then from the identity \eqref{eq-S}
and from the connection formula for the basic hypergeometric series:
\begin{multline}\label{con-form}
   \hyp{a}{b}{c}{q}{z}=
         \frac{(b,c/a;q)_\infty}{(c,b/a;q)_\infty}\,
         \frac{\Theta_q(qa^{-1}z^{-1})}{\Theta_q(qz^{-1})}\
   \hyp{a}{q a c^{-1}}{q a b^{-1}}{q}{q\frac{ c}{ab}\, z^{-1}}
                     \\
   +\frac{(a,c/b;q)_\infty}{(c,a/b;q)_\infty}\,
         \frac{\Theta_q(qb^{-1}z^{-1})}{\Theta_q(qz^{-1})}\
  \hyp{b}{q b c^{-1}}{q b a^{-1}}{q}{q\frac{c}{ab}\,z^{-1}}.
\end{multline}
One obtains:
\begin{equation}\label{S-ev}
  \mathbf{S}(z;p,w)=
     \begin{pmatrix}
       \Theta_{p^4}(-p^2w^{-2}z) & p^{-1/2}w \Theta_{p^4}(-p^2w^2 z)\\
       p^{1/2}w\Theta_{p^4}(-w^{-2}z) & \Theta_{p^4}(-w^2z)
     \end{pmatrix}
     \Lambda(z;p,w),
\end{equation}
with
\begin{equation}
   \Lambda(z;p,w)=\frac{1}{\Theta_{p^2}(z)}
                  \frac{(z,q^2 p^2 z^{-1};q^4,p^2)_\infty}
                            {(q^2 z,q^4 p^2 z^{-1};q^4,p^2)_\infty}
     \begin{pmatrix}
       p^{-1/8}w^{1/4}(w^2;p^2)_\infty^{-1} &0 \\
       0 & p^{1/8}w^{-1/4}(p^2w^{-2};p^2)^{-1}_\infty
     \end{pmatrix}.
\end{equation}
This corresponds to Baxter's vertex-IRF transformation (see Appendix~\ref{append-8V-IRF}).

\section{Conclusion}
\label{sectionConcl}
In this article we have developped a method enabling us to give an explicit product formula for the Vertex-IRF transformation relating the universal IRF elliptic $A_r^{(1)}$ solution to the universal  Vertex solution of Baxter-Belavin type with $\aleph=1.$
There are open questions in this field which would be interesting to analyze.

The first problem is to which extent is the genereralized translation datum that we have introduced  sufficiently general? Is it possible to associate to each generalized Belavin-Drinfeld triple in the finite dimensional case and in the affine case a generalized translation datum?

The second problem is to generalize the theorem of   J. Balog, L. Dabrowski and L. Feher to solutions of CDYBE which are not standard and to extend this theorem in the affine case.
Once the solution to this problem is solved one can eventually extend our result to these cases.

The third problem is related to the evaluation, in specific representations, of the universal formulas we give. We have constructed a universal solution of the generalized coBoundary equation living in  $\A(\h)\otimes\left({\cal F}^{(0)}(\h)\otimes U_q({\mathfrak g})\right)^c$. Moreover this element represented in  the evaluation representation gives a well defined meromorphic function. However it would be interesting for the applications to integrable systems to evaluate it in infinite integrable representations. On this problem little is known: even the evaluation of the solutions of the Quantum Dynamical coCycle in integrable highest weight representation has not been studied.

An intriguing point is the relation between our work and one dimensional representations of
${U}_q(\n_\pm)^{\omega_\pm}.$ The one dimensional representations have been constructed previously in \cite{Sev} (see also \cite{E2}) in order to study the $q$-deformed  Toda chain by a $q$-analog of Whittaker modules. There are  link between these Whittaker modules and the construction of the Vertex-IRF solution that will be studied in a future work.

\section*{Acknowledgements}
We thank B. Enriquez,  P. Etingof, G. Halbout, O. Schiffmann for interesting discussions.

The research of the authors is supported by CNRS. 
The research of E. Buffenoir, Ph. Roche and V. Terras is also supported by the contract  
ANR-05-BLAN-0029-01. The research of V. Terras is also 
supported by the ANR program MIB-05 JC05-52749.

\newpage

\section{Appendix}
\subsection{Construction of a PBW of $U_q({\mathfrak g})$}
\label{append-PBW}

We recall here the construction of \cite{KT}, it is the easiest to read and is correct.
The difference with our definition of $U_q({\mathfrak{g}})$ is in the expression of the coproduct: they have used instead $\check{\Delta}$ acting as 
$\check{\Delta}(e_i)=e_i\otimes 1+q^{-h_{\alpha_i}}\otimes e_i$, 
$\check{\Delta}(f_i)=f_i\otimes q^{h_{\alpha_i}}+
1\otimes f_i.$
The corresponding Hopf algebra $\check{U}_q(\g)$ is isomorphic to ${U}_q(\g)$,
\begin{align*}
 &U_q(\g)\rightarrow \check{U}_q(\g)\\
 &h_{\alpha_i}\mapsto h_{\alpha_i},\quad   
  e_i\mapsto k^{-1}e_i k, \quad
  f_i\mapsto k^{-1}f_ik,
\end{align*}
and we can therefore translate all their formulas to ours via this isomorphism.

Let us first consider $U_q(A_1^{ (1)})$. A normal order is given by (\ref{normalorderaffinesl2}).
We define
\begin{align*}
 &e_{\delta}=e_{\alpha_1}e_{\alpha_0}-q^{-2}e_{\alpha_0}e_{\alpha_1},\\
 &e_{\alpha_1+k\delta}=(-[(\alpha_1,\alpha_1)]_q)^{-k}
  \ad_{(e_{\delta})^{k}}(e_{\alpha_1}),\\
 &e_{\alpha_0+k\delta}=([(\alpha_1,\alpha_1)]_q)^{-k}
  \ad_{(e_{\delta})^{k}}(e_{\alpha_0}),\\
 &e_{k\delta}'=e_{\alpha_1+(k-1)\delta}e_{\alpha_0}
  -q^{-2}e_{\alpha_0}e_{\alpha_1+(k-1)\delta}.
\end{align*}
The elements  $e_{k\delta}$ are defined through the transformation:
\begin{equation*}
(q-q^{-1})\sum_{n\geq 1}e_{n\delta}z^n=\log \Big(1
                       +(q-q^{-1})\sum_{n\geq 1}e_{n\delta}'z^n\Big).
\end{equation*}
$(e_{\alpha_1+k_1\delta}, e_{k\delta}, e_{\alpha_0+k_0\delta}),\  k_0,k_1\in \Z^{+},\, k\in \Z^{+*},$ is a PBW basis of $U_q(\n^+).$

We moreover define $f_{\gamma}=e_{\gamma}^*,\ \gamma\in \Delta^+,$ which defines a PBW basis of $U_q(\n^-).$

We know discuss the case $U_q(A_r^{(1)}),\ r\geq 2.$
A normal order on the set of positive roots can be constructed using Beck's result \cite{Bec}.
We will need an additional Lemma in order to compute explicit expressions for 
${\mathfrak C}^{[\pm]}(x).$

\begin{lemma}{}
Every normal order $<$ on   $\overset{\circ}{\Delta}{}^+$ can be extended to a normal order  $<$ on the set ${\Delta}^+$ satisfying to  
\begin{equation*}
 \alpha<\alpha'+k\delta<l \delta< \delta-\beta<\alpha_0,\quad 
 \forall\ \alpha,\alpha'\in\overset{\circ}{\Delta}{}^+,\ 
 k\geq 1,\ l\geq 1,\ 
\beta\in \overset{\circ}{\Delta}{}^+\setminus\{\theta\}. 
\end{equation*}
\end{lemma}
\proof
\cqfd

We denote $\alpha_{i,j}=\alpha_i+\cdots+\alpha_{j-1},\ 1\leq i<j\leq r+1,$ and 
we will choose the following normal order on  $\overset{\circ}{\Delta}{}^+:$
\begin{multline}
 \alpha_{r,r+1}<\alpha_{r-1,r+1}<\cdots< \alpha_{1,r+1}<
 \alpha_{r-1,r}<\alpha_{r-2,r}<\cdots< \alpha_{1,r}<\cdots\\
 \cdots<\
 \alpha_{2,3}<\alpha_{1,3} <\alpha_{1,2}.
\end{multline}

We apply the procedure of \cite{KT}, we obtain: 
\begin{alignat*}{2}
 &\text{ev}_z(e_{\alpha_{i,j}})=(-q^{-1})^{j-i-1}E_{i,j}, \quad &
 &\text{ev}_z(f_{\alpha_{i,j}})=(-q)^{j-i-1}E_{j,i},\\
 &\text{ev}_z(e_{\delta-\alpha_{i,j}})=z(-q^{-1})^{i-1}E_{j,i}, \quad &
 &\text{ev}_z(f_{\delta-\alpha_{i,j}})=z^{-1}(-q)^{i-1}E_{i,j}.
\end{alignat*}

Let us choose $\theta:\overset{\circ}{\Gamma}\rightarrow \{0,1\}$ such that 
$\{ \alpha\not=\beta,\; (\alpha,\beta)\not=0\}\Rightarrow 
\theta(\alpha)\not=\theta(\beta)$. 
We will take $\theta(\alpha_i)=i-1\pmod 2,$ and define $\epsilon_m(\alpha_i)=(-1)^{m\theta(\alpha_i)}=(-1)^{m(i-1)}.$ 
We define 
\begin{align*}
 &e_{\delta}^{(\alpha_i)}=
\epsilon_1(\alpha_i)(e_{\alpha_i}e_{\delta-\alpha_i}-
q^{-2}e_{\delta-\alpha_i}e_{\alpha_i}),\\
 &e_{\alpha_i+m\delta}=\epsilon_m(\alpha_i)(-[(\alpha_i,\alpha_i)]_q)^{-m}
 (\ad_{(e_{\delta}^{(\alpha_i)})})^m(e_{\alpha_i}),\\
 &e_{m\delta}^{(\alpha_i)}{}'=\epsilon_m(\alpha_i)
(e_{\alpha_i+(m-1)\delta}e_{\delta-\alpha_i}-q^{-2}e_{\delta-\alpha_i}e_{\alpha_i+(m-1)\delta}).
\end{align*}
The elements $e_{n\delta}^{(\alpha_i)}$ are constructed through the transformation
\begin{equation}
(q-q^{-1})\sum_{n\geq 1}e_{n\delta}^{(\alpha_i)}z^n
=\log \Big(1+(q-q^{-1})\sum_{n\geq 1}e_{n\delta}^ {(\alpha_i)}{}'z^n\Big).
\end{equation}
We still define $f_{\gamma}=e_{\gamma}^*, \gamma\in \Delta^+.$

As a result we obtain
\begin{align*}
 &\text{ev}_z(e_{\delta}^{(\alpha_i)})=zq^{1-i}(E_{i,i}-q^{-2}E_{i+1,i+1})\;\;,\\
 &\text{ev}_z(e_{\alpha_i+m\delta})=(-1)^{mi}z^mq^{-im}E_{i,i+1}\;\;,\\
 &\text{ev}_z(e_{m\delta}^{(\alpha_i)}{}')=(-1)^{m-1}z^m q^{1-im}(E_{i,i}-q^{-2}E_{i+1,i+1})\;\;,\\ 
 &\text{ev}_z(e_{m\delta}^{(\alpha_i)})=(-1)^{m-1}z^mq^{(1-i)m}\frac{[m]_q}{m} (E_{i,i}-q^{-2m}E_{i+1,i+1}).
\end{align*}
The knowledge of these elements is sufficient to compute ${\mathfrak C}^{[\pm]}(x).$

\subsection{Proof of Theorem~\ref{th-cocycle}: useful Lemmas}
\label{append-cocycle}

Let $J(x)$ be a solution of \eqref{ABRR2} and \eqref{Jhyp}, and let us define
\begin{equation*}
   Y_{123}(x)=(\id \otimes \Delta)(J(x)^{-1})\,
          (\Delta\otimes \id)(J(x))\,J_{12}(xq^{h_3}).
\end{equation*}
In this Appendix, we give the explicit proof of the properties \eqref{property2}-\eqref{property1c} of $Y_{123}(x)$ that we use in the proof of Theorem~\ref{th-cocycle}.

\begin{lemma}{\label{lem1}}
Let $\widehat{Y}_{123}(x)=S_{23}^{(1)}{}^{-1}Y_{123}(x)$. We have that
\begin{equation}
  \widehat{Y}_{123}(x)  \in \; 1^{\otimes 3} 
  \oplus (\big[\left( 1\otimes U_q^+({\mathfrak g})\otimes U_q^-({\mathfrak g}) \right) 
  \oplus \left(U_q^+({\mathfrak g}) \otimes U_q({\mathfrak g}) 
                                          \otimes U_q({\mathfrak b}^-)\right)\big]\otimes {\cal F}^{(0)}(\h))^c.
\end{equation}
\end{lemma}

\proof
It is obvious from \eqref{Jhyp} that 
$\widehat{Y}_{123}(x) \;  \in \; 
\big[ U_q({\mathfrak b}^+)\otimes U_q({\mathfrak g}) \otimes 
U_q({\mathfrak b}^-)\otimes {\cal F}^{(0)}(\h)\big]^c.$ 
To be more precise, let us express this element in the following form:
\begin{equation*}
  \widehat{Y}_{123}(x)
  =\Big[S_{23}^{(1)}{}^{-1}(\id \otimes\Delta)(\widehat{J}(x)^{-1})\,
                        S_{23}^{(1)}\Big]
   \Big[S_{12}^{(1)}{}^{-1}(\Delta \otimes \id)(\widehat{J}(x))\,
                        S_{12}^{(1)}\Big]\;\widehat{J}_{12}(xq^{h_3}).
\end{equation*}
Using the fact that $\widehat{J}(x) \in 1^{\otimes 2} 
\oplus \left[(U_q^+({\mathfrak g}) \otimes U_q^-({\mathfrak g}))\otimes {\cal F}^{(0)}(\h)\right]^c $, 
we obtain immediately that
\begin{align*}
  &(\iota_+)_1 \big(\, S_{23}^{(1)}{}^{-1}(\id \otimes\Delta)(\widehat{J}(x)^{-1})\,
                S_{23}^{(1)}\,\big)=1^{\otimes 3},\\
  &(\iota_+)_1\big(\widehat{ J}_{12}(xq^{h_3})\big)
  =(\iota_-)_2\big(\widehat{ J}_{12}(xq^{h_3})\big)
  =1^{\otimes 3}.
\end{align*}
As a result,
\begin{align*} 
 &S_{23}^{(1)}{}^{-1}(\id \otimes\Delta)(\widehat{J}(x)^{-1})\,S_{23}^{(1)}
   \ \in \ 1^{\otimes 3}\oplus \left[\left( U_q^+({\mathfrak g}) 
   \otimes U_q({\mathfrak b}^-)  \otimes  U_q({\mathfrak b}^-) \right)\otimes {\cal F}^{(0)}(\h)\right]^c, \\
 &\widehat{J}_{12}(xq^{h_3}) \ \in\ 
    1^{\otimes 3}\oplus  \left[\left( U_q^+({\mathfrak g}) 
         \otimes U_q^-({\mathfrak g})  \otimes  U_q({\mathfrak h}) \right)\otimes {\cal F}^{(0)}(\h)\right]^c.
\end{align*}
Moreover, using the expression \eqref{Jprod} of $J(x)$ as an infinit product, the quasitriangularity property of the $R$-matrix, the fact that $\widehat{R}\in \left[1^{\otimes 2} \oplus (U_q^+({\mathfrak g}) \otimes U_q^-({\mathfrak g}))\right] $ and that $(\id\otimes\Delta)(S^{(1)})=S_{12}^{(1)}S_{13}^{(1)}$, 
we have
\begin{align*}
  &(\iota_+)_1 \left( S_{12}^{(1)}{}^{-1}(\Delta \otimes \id)(\widehat{J}(x))\,
        S_{12}^{(1)}\right) \\
  &\hspace{1.5cm}
   = S_{12}^{(1)}{}^{-1}\left\{
     \prod_{k=1}^{+\infty} \big(\theta^{-}_{[x] 3}\big)^k   
     \left(
     \big(S^{(1)}_{13}S^{(1)}_{23}\big)^{-1}\;
     (\iota_+)_1 \big(K^{-1}_{23}\widehat{R}_{13}K_{23}\widehat{R}_{23}\big)\;
     S^{(1)}_{13}S^{(1)}_{23}
     \right)
     \right\} S_{12}^{(1)} \\
  &\hspace{1.5cm}
   = S_{12}^{(1)}{}^{-1}\left\{ 
     \prod_{k=1}^{+\infty}\big(\theta^{-}_{[x] 3}\big)^k   \left(
     \big(S^{(1)}_{13}S^{(1)}_{23}\big)^{-1}\;\widehat{R}_{23}\;
     S^{(1)}_{13}S^{(1)}_{23} \right)
     \right\} S_{12}^{(1)} \\
  &\hspace{1.5cm}
   = \left\{ \prod_{k=1}^{+\infty}\big(\theta^{-}_{[x] 3}\big)^k   \left(
     (S^{(1)}_{23})^{-1}\;\widehat{R}_{23}\;S^{(1)}_{23} \right) \right\}
   \ \in\ \big[1^{\otimes 3} \oplus \left( 1 \otimes U_q^+({\mathfrak g}) 
           \otimes U_q^-({\mathfrak g})\;\right)\big],
\end{align*}
which means that
\begin{equation*}
\hskip -0.4cm  S_{12}^{(1)}{}^{-1}(\Delta \otimes \id)(\widehat{J}(x))\,S_{12}^{(1)} 
  \in 1^{\otimes 3} \oplus ( \left[
  \left( 1 \otimes U_q^+({\mathfrak g}) \otimes U_q^-({\mathfrak g})\;\right) 
  \oplus \left( U_q^+({\mathfrak g}) \otimes U_q({\mathfrak b}^+)  
  \otimes  U_q({\mathfrak b}^-)\right)\right]\otimes {\cal F}^{(0)}(\h))^c. 
\end{equation*}
This concludes the proof of Lemma~\ref{lem1}.
\cqfd

\begin{lemma}{\label{lem2}}
$Y_{123}(x)$ satisfies the following linear equation:
\begin{equation}
   \big[ \Ad_{(\varphi_{12}^{0}\,\varphi_{13}^{0})^{-1}}\circ
   \theta^+_{[x] 1}\big] (Y_{123}(x))
   = Y_{123}(x).
\end{equation}
\end{lemma}

\proof
Let $\widetilde{Y}_{123}(x)=\big[ \Ad_{(\varphi_{12}^{0}\,\varphi_{13}^{0})^{-1}}\circ
   \theta^+_{[x] 1}\big](Y_{123}(x))$.
Using the quasitriangularity of the $R$-matrix, \eqref{axiomGTDb}, and the fact that $J(x)$ satisfies \eqref{ABRR1}, we have
\begin{align*}
 \widetilde{Y}_{123}(x)
  &=\big[\Ad_{(\varphi_{12}^{0}\,\varphi_{13}^{0})^{-1}}\circ
    \theta^+_{[x]1}\big]\Big( (\id\otimes\Delta)(J(x)^{-1}) \, R_{12}^{-1}\,
    (\Delta'\! \otimes \id)(J(x))\,R_{12} \, J_{12}(xq^{h_3}) \Big)
                 \\
  &=\big[\Ad_{(\varphi_{12}^{0}\,\varphi_{13}^{0})^{-1}}\circ
    \theta^+_{[x]1}\big]\Big( 
    (\id\otimes \Delta)\big( \widehat{R}\, J(x) \big)^{-1} \,
    \, K_{12}^{-1}\,\widehat{R}_{13}
                     \\
  &\hspace{1.2cm}\times
    (\Delta'\! \otimes \id)(J(x))\, R_{12} \, J_{12}(xq^{h_3}) \Big)
                  \\
  &= (\id\otimes \Delta)\Big( W^{(1)}\,\big(\Ad_{(\varphi^{0})^{-1}}\circ
     \theta^+_{[x]1}\big)\big( \widehat{R}\, J(x) \big) \Big)^{-1}
                  \\
  &\hspace{1.2cm}\times W^{(1)}_{12}\, W^{(1)}_{13}
     \big[\Ad_{(\varphi_{12}^{0}\,\varphi_{13}^{0})^{-1}}\circ
    \theta^+_{[x]1}\big]\Big(
     \, K_{12}^{-1}\,\widehat{R}_{13}\;
     (\Delta'\! \otimes \id)(J(x))\,K_{12}\, \Big) W_{12}^{(1)\; -1}
                   \\
  &\hspace{1.2cm}\times
    \left[ W^{(1)}_{12}\,\big(\Ad_{(\varphi^{0}_{12})^{-1}}\circ
    \theta^+_{[x q^{h_3}]1}\big)\big(\widehat{R}_{12}\,J_{12}(xq^{h_3}) 
    \big) \right]\, 
                   \\
  &=(\id\otimes \Delta)(J(x)^{-1})\;W^{(1)}_{12}\, W^{(1)}_{13}\,
      \big[\Ad_{(\varphi_{12}^{0}\,\varphi_{13}^{0})^{-1}}\circ
    \theta^+_{[x]1}\big]\Big(
       K_{12}^{-1}\,\widehat{R}_{13}\;
                   \\
  &\hspace{1.2cm}\times
     (\Delta'\! \otimes \id)(J(x))\,K_{12}\,W_{12}^{(0)\; -1} \Big)\;
     J_{12}(xq^{h_3}).
\end{align*}
It remains to prove that
\begin{multline}\label{eq-to-prove}
  (\Delta \otimes \id)(J(x))=W^{(1)}_{12}\, W^{(1)}_{13}\,
      \big[\Ad_{(\varphi_{12}^{0}\,\varphi_{13}^{0})^{-1}}\circ
    \theta^+_{[x]1}\big]\Big(
       K_{12}^{-1}\,\widehat{R}_{13}\;
                   \\
  \times
     (\Delta'\! \otimes \id)(J(x))\,K_{12}\, \Big) W_{12}^{(1)\; -1}.
\end{multline}
Expressing $J(x)$ as an infinite product by means of \eqref{Jprod}, we have
\begin{equation*}
    (\Delta \otimes \id)(J(x))=
    S^{(1)}_{13}\, S^{(1)}_{23}\, \prod_{k=1}^{+\infty}
    (\theta^{-}_{[x] 3})^k\circ\Ad_{(S^{(1)}_{13}\,S^{(1)}_{23})^{-1}} \left( 
            K^{-1}_{23}\; \widehat{R}_{13}\; K_{23}\; 
            \widehat{R}_{23}\right).
\end{equation*}
On the other hand,
\begin{align*}
    (\Delta'\! \otimes \id)(J(x))
 &=
    S^{(1)}_{13}\, S^{(1)}_{23}\, \prod_{k=1}^{+\infty}
       (\theta^{-}_{[x] 3})^k\circ\Ad_{(S^{(1)}_{13}\,S^{(1)}_{23})^{-1}} \left(
            K^{-1}_{13}\; \widehat{R}_{23}\; K_{13}\; \widehat{R}_{13}\right)
                  \\
 &=
    S^{(1)}_{13}\, S^{(1)}_{23}\, \prod_{k=1}^{+\infty}
       (\theta^{-}_{[x] 3})^k\circ\Ad_{(S^{(1)}_{13}\,S^{(1)}_{23})^{-1}} 
            \left( K_{12}\;
            \widehat{R}_{23}\; K_{12}^{-1}\; \widehat{R}_{13}\right),
\end{align*}
in which we have used the zero-weight property of the $R$-matrix.
The latter infinite product can be reorganized as follows:
\begin{align*}
  (\Delta'\! \otimes \id)(J(x))
  &= (S^{(1)}_{13}\,S^{(1)}_{23})\,
    K_{12}\,
      \prod_{k=1}^{+\infty}\Big\{
    (\theta^{-}_{[x] 3})^k 
    \circ\Ad_{[
(S^{(1)}_{13}\,S^{(1)}_{23})^{-1}]}
    \big(\widehat{R}_{23}\big)
                  \\
  &\hspace{1.5cm}\times
    (\theta^{-}_{[x] 3})^k 
    \circ\Ad_{[
(S^{(1)}_{13}\,S^{(1)}_{23})^{-1}]}
    \big( K_{12}^{-1}\; \widehat{R}_{13}\; K_{12} \big) \Big\}\; 
K_{12}^{-1}
                  \\
  &= \widehat{R}_{13}^{-1}\,K_{12}\,
       S^{(1)}_{13}\,S^{(1)}_{23}\;
                  \\
  &\hspace{1.5cm}\times
      \prod_{k=1}^{+\infty}\Big\{
    (\theta^{-}_{[x] 3})^{k-1} 
    \circ\Ad_{[
(S^{(1)}_{13}\,S^{(1)}_{23})^{-1}]}
    \big( K_{12}^{-1}\; \widehat{R}_{13}\; K_{12} \big)
                  \\
  &\hspace{1.5cm}\times
    (\theta^{-}_{[x] 3})^k 
    \circ\Ad_{[
(S^{(1)}_{13}\,S^{(1)}_{23})^{-1}]}
    \big(\widehat{R}_{23}\big)\Big\}\;
 K_{12}^{-1}.
\end{align*}
Therefore, the right handside of \eqref{eq-to-prove} can be expressed as
\begin{equation*}
 \text{r.h.s.}
  =S^{(1)}_{13}\,S^{(1)}_{23}\; 
     \prod_{k=1}^{+\infty} P_{123}^{(k)}(x),
\end{equation*}
with
\begin{align*}
 P_{123}^{(k)}(x)
  &= 
    W^{(1)}_{12} \theta^+_{[x]1}\circ(\theta^{-}_{[x] 3})^{k-1} 
    \circ\Ad_{[(\varphi_{12}^{0}\,\varphi_{13}^{0})^{-1}\,
(S^{(1)}_{13}\,S^{(1)}_{23})^{-1}]} 
    \big( K_{12}^{-1}\;\widehat{R}_{13}\;K_{12}\big)    
                 \\
  &\hspace{3cm}\times
      \theta^+_{[x]1}\circ(\theta^{-}_{[x] 3})^{k}
    \circ\Ad_{[(\varphi_{12}^{0}\,\varphi_{13}^{0})^{-1}\,
(S^{(1)}_{13}\,S^{(1)}_{23})^{-1}]} 
    \big( \widehat{R}_{23} \big) W^{(1)\; -1}_{12}
                \\
  &=
      (\theta^{-}_{[x] 3})^{k-1} 
    \circ\Ad_{Q_{123}^{-1}(x)}\circ
    \theta^{-}_{[x] 3}\circ\Ad_{(S^{(1)}_{13}\,S^{(1)}_{23})^{-1}}
    \big(K_{23}^{-1}\; \widehat{R}_{13}\; K_{23}\big)
                \\
  &\hspace{3cm}\times
      (\theta^{-}_{[x] 3})^{k} \circ
      \Ad_{[S^{(1)\; -1}_{13}\, W^{(1)}_{12}\,W^{(1)}_{13}]}
    \big( \widehat{R}_{23} \big)
\end{align*}
and
\begin{equation}\label{expr-Q}
  Q_{123}(x)=\varphi_{12}^{0}\;
         \theta^{-}_{[x] 3}(K_{23}^{-1})\;\theta^+_{[x]1}(K_{12})\;
         W^{(1)}_{23}\,W_{12}^{(1)\; -1}.
\end{equation}
The equality \eqref{eq-to-prove} is then ensured by the relation 
\begin{equation}\label{condition-lem2}
 \big[\;  Q_{123}(x)\;,\;\theta^{-}_{[x] 3}\big(\widehat{R}_{13}\big)\;\big]=\big[\;  Q_{123}(x)\;,\;\theta^{+}_{[x] 1}\big(\widehat{R}_{13}\big)\;\big]=0,
\end{equation} 
which, by using the zero-weight property of $\widehat{R}$, is strictly equivalent to
\begin{align*}
0&= \big[\; \varphi_{12}^{0}\, K_{21}\,\theta^+_{[x]1}(K_{12})\;
        \big(\theta^+_{[x]1}(W_{21}^{(1)})\, W_{12}^{(1)} \big)^{-1}\;,\;
\theta^+_{[x]1}(\widehat{R}_{13})\;\big]\\
&=\big[\; \varphi_{12}^{0}\, K_{12}\,S^{(1)\; -1}_{12}\,S^{(1)}_{21}\;
  \theta^+_{[x]1}(K_{12}\,S^{(1)\; -1}_{21}\,S^{(1)}_{12})\;\;
  (\theta^+_{[x]1}\theta^-_{[x]1})(S^{(1)}_{21})\,S^{(1)\; -1}_{21}\; ,\;
  \theta^+_{[x]1}(\widehat{R}_{13})\;\big]\\
&=\big[\; \varphi_{12}^{0}\, K_{12}\, S^{(1)\; -1}_{12}\, S^{(1)}_{21}\;
  \theta^+_{[x]1}(K_{12}\,S^{(1)\; -1}_{21}\, S^{(1)}_{12})\; ,\; 
 \theta^+_{[x]1}(\widehat{R}_{13})\;\big]
\end{align*}
and follows from \eqref{axiomGTDh}.
\cqfd

\begin{lemma}{\label{lem3}}
$Y_{123}(x)$ satisfies the following linear equation:
\begin{equation}
 W_{23}^{(1)}\; \theta^-_{[x] 3}\big(\widehat{R}_{23}\, Y_{123}(x)\big)
 = Y_{123}(x).
\end{equation}
\end{lemma}

\proof
Using the linear equation~\eqref{ABRR2} and the quasitriangularity of the $R$-matrix, we have that
\begin{align*}
  W_{23}^{(1)}\; \theta^-_{[x] 3}\big(\widehat{R}_{23}\, Y_{123}(x)\big)
  &= W_{23}^{(1)}\; \theta^-_{[x] 3}\big(
     \widehat{R}_{23}\; (\id \otimes \Delta)(J(x)^{-1})\big)\;\\
  &\hspace{3cm}\times 
     (\Delta \otimes \id)\big(\theta^-_{[x]2}(J(x))\big) \; 
     \theta^-_{[x] 3}\big(J_{12}(xq^{h_3})\big)\\
  &= W_{23}^{(1)}\; \theta^-_{[x] 3}
     \big(K^{-1}_{23} \;(\id \otimes \Delta')(J(x)^{-1})\; R_{23}\big)\,\\
  &\hspace{3cm}\times 
    (\Delta \otimes \id)\big(\theta^-_{[x]2}(\widehat{R}^{-1})\,
    W^{(1)\; -1}\,J(x)\big) \; 
    J_{12}(xq^{h_3})\\
  &= W_{23}^{(1)}\; \theta^-_{[x] 3}\big(
     K^{-1}_{23} \; (\id \otimes \Delta')(J(x)^{-1})\;
     \widehat{R}^{-1}_{13}\, K_{23} \big) \;
    W^{(1)\;-1}_{13}\, W^{(1)\;-1}_{23} \\
  &\hspace{3cm}\times (\Delta \otimes \id)(J(x)) \; 
    J_{12}(xq^{h_3}).
\end{align*}
It means that we have to prove the following relation:
\begin{equation}\label{eq-to-prove2}
  (\id \otimes \Delta)(J(x))=W^{(1)}_{13}\,W^{(1)}_{23}\,
  \theta^-_{[x] 3}\big(
  K^{-1}_{23}\,\widehat{R}_{13}\; (\id \otimes \Delta')(J(x))\; K_{23}\big)\; 
  W^{(1)\; -1}_{23}.
\end{equation}
The proof of this equality is very similar to the one of~\eqref{eq-to-prove}.
Indeed, from the expression \eqref{Jprod-bis} of $J(x)$ as an infinite product, 
we have
\begin{equation*}
    (\id\otimes\Delta)(J(x))=
    S^{(1)}_{12}\, S^{(1)}_{13}\, \prod_{k=1}^{+\infty} 
    (\theta^{+}_{[x] 1})^k\circ
    \Ad_{[ (\varphi_{12}^{(0)}\,\varphi_{13}^{(0)})^{-k}\,
          (S^{(1)}_{12}\,S^{(1)}_{13})^{-1} ]}
        \left( K^{-1}_{12}\; \widehat{R}_{13}\; K_{12}\; 
            \widehat{R}_{12} \right),
\end{equation*}
whereas,
\begin{align*}
    (\id\otimes\Delta')(J(x))
 &=
    S^{(1)}_{12}\, S^{(1)}_{13}\, \prod_{k=1}^{+\infty}
    (\theta^{+}_{[x] 1})^k\circ
    \Ad_{[ (\varphi_{12}^{(0)}\,\varphi_{13}^{(0)})^{-k}\,
          (S^{(1)}_{12}\,S^{(1)}_{13})^{-1} ]}
        \left( K^{-1}_{13}\; \widehat{R}_{12}\; K_{13}\; 
            \widehat{R}_{13} \right)
                  \\
 &=
    S^{(1)}_{12}\, S^{(1)}_{13}\, \prod_{k=1}^{+\infty}
    (\theta^{+}_{[x] 1})^k\circ
    \Ad_{[ (\varphi_{12}^{(0)}\,\varphi_{13}^{(0)})^{-k}\,
          (S^{(1)}_{12}\,S^{(1)}_{13})^{-1} ]}
        \left( K_{23}\; \widehat{R}_{12}\; K_{23}^{-1}\; 
            \widehat{R}_{13} \right),
\end{align*}
which can be reorganized as follows:
\begin{align*}
  (\id\otimes\Delta')(J(x))
  &= S^{(1)}_{12}\,S^{(1)}_{13}\,
    K_{23}\;
      \prod_{k=1}^{+\infty}\Big\{
    (\theta^{+}_{[x] 1})^k\circ
    \Ad_{[ (\varphi_{12}^{(0)}\,\varphi_{13}^{(0)})^{-k}\,
          (S^{(1)}_{12}\,S^{(1)}_{13})^{-1} ]}
        \big(\widehat{R}_{12}\big)
                  \\
  &\hspace{1.3cm}\times
    (\theta^{+}_{[x] 1})^k\circ
    \Ad_{[ (\varphi_{12}^{(0)}\,\varphi_{13}^{(0)})^{-k}\,
          (S^{(1)}_{12}\,S^{(1)}_{13})^{-1} ]}
        \big(K_{23}^{-1}\,\widehat{R}_{13}\,K_{23}\big)
        \Big\}\; K_{23}^{-1}
                  \\
  &= \widehat{R}_{13}^{-1}\,K_{23}\,
       S^{(1)}_{12}\,S^{(1)}_{13}\;
                  \\
  &\hspace{1.3cm}\times
      \prod_{k=1}^{+\infty}\Big\{
   (\theta^{+}_{[x] 1})^{k-1}\circ
    \Ad_{[ (\varphi_{12}^{(0)}\,\varphi_{13}^{(0)})^{-(k-1)}\,
          (S^{(1)}_{12}\,S^{(1)}_{13})^{-1} ]}
        \big(K_{23}^{-1}\,\widehat{R}_{13}\,K_{23}\big)
                  \\
  &\hspace{1.3cm}\times
   (\theta^{+}_{[x] 1})^k\circ
    \Ad_{[ (\varphi_{12}^{(0)}\,\varphi_{13}^{(0)})^{-k}\,
          (S^{(1)}_{12}\,S^{(1)}_{13})^{-1} ]}
        \big(\widehat{R}_{12}\big)
        \Big\}\; K_{23}^{-1}.
\end{align*}
Therefore, the right handside of \eqref{eq-to-prove2} can be expressed as
\begin{align*}
 \text{r.h.s.}
  &=S^{(1)}_{12}\,S^{(1)}_{13}\; 
     \prod_{k=1}^{+\infty} \widetilde{P}_{123}^{(k)}(x),
\end{align*}
where
\begin{align*}
 \widetilde{P}_{123}^{(k)}(x)
  &= W_{23}^{(1)}
(\theta^{+}_{[x] 1})^{k-1}\circ\theta^{-}_{[x] 3}\circ
    \Ad_{[ (\varphi_{12}^{(0)}\,\varphi_{13}^{(0)})^{-(k-1)}\,
          (S^{(1)}_{12}\,S^{(1)}_{13})^{-1}\,  ]}
        \big(K_{23}^{-1}\,\widehat{R}_{13}\,K_{23}\big)
                  \\
  &\hspace{3cm}\times
   (\theta^{+}_{[x] 1})^k\circ\theta^{-}_{[x] 3}\circ
    \Ad_{[ (\varphi_{12}^{(0)}\,\varphi_{13}^{(0)})^{-k}\,
          (S^{(1)}_{12}\,S^{(1)}_{13})^{-1} \, ]}
        \big(\widehat{R}_{12}\big) W_{23}^{(1)\; -1}\\
  &=(\theta^{+}_{[x] 1})^{k-1}\circ\Ad_{{Q}_{123}(x)}\circ
    \theta^{+}_{[x] 1}\circ
    \Ad_{[ (\varphi_{12}^{(0)}\,\varphi_{13}^{(0)})^{-k}\,
          (S^{(1)}_{12}\,S^{(1)}_{13})^{-1}]}
        \big(K_{12}^{-1}\,\widehat{R}_{13}\,K_{12}\big)
                  \\
  &\hspace{3cm}\times
   (\theta^{+}_{[x] 1})^k\circ
    \Ad_{[ (\varphi_{12}^{(0)}\,\varphi_{13}^{(0)})^{-k}\,
          (S^{(1)}_{12}\,S^{(1)}_{13})^{-1} \,W_{13}^{(1)}\, W_{23}^{(1)}]}
        \big(\widehat{R}_{12}\big)
\end{align*}
with $Q_{123}(x)$ given by \eqref{expr-Q}.
Therefore, \eqref{eq-to-prove2} is satisfied under the condition
\begin{equation}
   \big[\;  Q_{123}(x)\;,\;
   \theta^{+}_{[x] 1}\big(\widehat{R}_{13}\big)\;\big]=0,  
\end{equation}
which is stricly equivalent to \eqref{condition-lem2}.
\cqfd

\subsection{Expression of the Solutions of QDYBE for the
  $U_q(A_1^{(1)})$ case in the evaluation representation: the 8-Vertex
  model and its associated height model.}
\label{append-8V-IRF}

The 8-Vertex model is based on Baxter's elliptic solution of the QYBE, which is obtained as the evaluation representation 
${\bf R}^{\Ver}(z_1,z_2;p)=(\ev_{z_1}\otimes\ev_{z_2})\big[ R(p) \big]$ (with $p=x_0x_1$) of the Belavin-Baxter elliptic solution of the QDYBE for $U_q(A_1^{(1)})$,
\begin{equation}
    R(p)=J_{21}(p)^{-1} R_{12}\,J_{12}(p),
\end{equation}
where the twist $J(p)$, obtained in \cite{JKOS}, is constructed as
\begin{equation}
    J(p)= S^{(1)}\prod_{k=1}^\infty 
       \Big[ (\Ad_{D^-_2(p)}\circ\sigma^-_2)^k
                 \big(S^{(1)}{}^{-1}\widehat{R}\,S^{(1)}\big)\Big],
\end{equation}
with $S^{(1)}$, $D^-(p)$, $\sigma^-$ given as in Subsection~\ref{subsec-BelavinBaxter}.
One has
\begin{equation}
{\bf R}^{\Ver}(z_1,z_2;p)
  =q^{1/2} \rho(z;p)
   \begin{pmatrix}
  a^{\Ver}(z;p)  & 0 & 0 &  z_1^{-1} d^{\Ver}(z;p)  \\
  0 & b^{\Ver}(z;p)  &  c^{\Ver}(z;p) & 0\\
  0 & z \, c^{\Ver}(z;p)  &  b^{\Ver}(z;p) & 0\\
  z_2\, d^{\Ver}(z;p)  & 0 & 0 &  a^{\Ver}(z;p)
  \end{pmatrix},
\end{equation}
with $z=z_1/z_2$,
\begin{alignat}{2}
  &a^{\Ver}(z;p)=\frac{ \Theta_{p^{4}}(p^{2}z) \,\Theta_{p^{4}}(p^{2}q^{2}) }
                    { \Theta_{p^{4}}(p^{2})\,\Theta_{p^{4}}(p^{2}q^{-2}z) },
     & \qquad
  &b^{\Ver}(z;p)=q^{-1}\frac{ \Theta_{p^{4}}(z)\,\Theta_{p^{4}}(p^{2}q^{2}) }
                          { \Theta_{p^{4}}(p^{2})\,\Theta_{p^{4}}(q^{-2}z) },
                  \\
  &c^{\Ver}(z;p)=\frac{ \Theta_{p^{4}}(p^{2}z)\,\Theta_{p^{4}}(q^{-2}) }
                    { \Theta_{p^{4}}(p^{2}) \,\Theta_{p^{4}}(q^{-2}z) },
     & \qquad
  &d^{\Ver}(z;p)=q^{-1}p\frac{\Theta_{p^{4}}(z)\,\Theta_{p^{4}}(q^{2}) }
                           {\Theta_{p^{4}}(p^{2})\,\Theta_{p^{4}}(p^{2}q^{-2}z) },
\end{alignat}
and
\begin{equation}
  \rho(z;p)=\frac{ (z,q^{4}z,p^{2}q^{2}z^{-1},p^{2}q^{2}z^{-1}\; ;\; 
                                                   p^{2},q^{4})_{\infty} }
                 { (q^{2}z,q^{2}z,p^{2}z^{-1},p^{2}q^{4}z^{-1}\; ;\; 
                                                   p^{2},q^{4})_{\infty} }.
\end{equation}

Let us now consider the standard elliptic solution of the QDYBE for $U_q(A_1^{(1)})$,
\begin{equation}\label{R-twist-IRF}
    R(p,w)=F_{21}(p,w)^{-1} R_{12}\,F_{12}(p,w),
\end{equation}
where the twist $F(p,w)$ (with $p=x_0 x_1$, $w=x_1$) is the quantum dynamical cocycle, obtained in \cite{JKOS}, constructed from the generalized translation datum defined in Subsection~\ref{subsec-StandardIRF}:
\begin{equation}
  F(p,w)=\prod_{k=1}^{+\infty}\Big[ B(p,w)^{-k} \widehat{R}\, B(p,w)^k\Big].
\end{equation}
Its explicit expression $\mathbf{F}(z;p,w)=(\ev_{z_1}\otimes\ev_{z_2})\big[ F(p,w) \big]$, as well as the one of its associated $R$-matrix ${\bf R}^{\IRF}(z;p,w)=(\ev_{z_1}\otimes\ev_{z_2})\big[ R(p,w) \big]$, with $z=z_1/z_2$, have been obtained in \cite{JKOS}. They read,
\begin{equation}
  \mathbf{F}(z;p,w)=\varphi(z;p)
  \begin{pmatrix}
    1 & 0 & 0 & 0\\
    0 & X_{11}(z;p,w) & X_{12}(z;p,w) & 0\\
    0 & X_{12}\big(z;p,\frac{p}{w}\big) & X_{21}\big(z;p,\frac{p}{w}\big) & 0\\
    0 & 0 & 0 & 1
   \end{pmatrix},
\end{equation}
with
\begin{align}
  &X_{11}(z;p,w)=\hyp{q^2}{q^2p^2w^{-2}}{p^2w^{-2}}{p^2}{q^{-2}p^2z},\\
  &X_{12}(z;p,w)=-\frac{q-q^{-1}}{1-w^{-2}}\
          \hyp{q^2p^2}{q^2w^{2}}{p^2w^{2}}{p^2}{q^{-2}p^2z}, \\
  &\varphi(z;p)=\frac{(p^2 z,q^4 p^2 z; q^4,p^2)_\infty}
                     {(q^2 p^2 z; q^4,p^2)_\infty^2},
\end{align}
and
\begin{equation}\label{R-IRF}
 {\bf R}^{\IRF}(z;p,w)
     = q^{1/2}\rho(z;p)
  \begin{pmatrix}
    1 & 0 & 0 & 0\\
    0 & b^{\IRF}(z;p,w) & c^{\IRF}(z;p,w) & 0\\
    0 & zc^{\IRF}\big(z;p,\frac{p}{w}\big) & b^{\IRF}\big(z;p,\frac{p}{w}\big) & 0\\
    0 & 0 & 0 & 1
   \end{pmatrix},
\end{equation}
with
\begin{align}
  &b^{\IRF}(z;p,w)=q^{-1}
     \frac{(q^2 p^2 w^{-2}, q^{-2} p^2 w^{-2} ;  p^2 )_{\infty}}
          {(p^2 w^{-2} ;  p^2 )_{\infty}^2}
     \frac{\Theta_{p^2}(z)}{\Theta_{p^2}(q^{-2}z)},\\
  &c^{\IRF}(z;p,w)=
     \frac{\Theta_{p^2}(q^{-2})\, \Theta_{p^2}(w^2 z)}
          { \Theta_{p^2}(w^{2})\, \Theta_{p^2}(q^{-2}z)}.
\end{align}

This $R$-matrix defines the Boltzmann weights of an interaction-round-a-face (IRF) model, the Andrews-Baxter-Forrester model \cite{ABF}, or solid-on-solid (SOS) model.

When expressed in the two-dimensional evaluation representation, the Vertex-IRF equation between $R(w,p)$ and $R(p)$,
\begin{equation*}
     \mathbf{S}_1 (z_1;p,w)\,\mathbf{S}_2 (z_2;p,wq^{h_1})\,
        \mathbf{R}^{\IRF}(z_1/z_2;p,w) 
     ={\bf R}^{\Ver}(z_1,z_2;p)\,
        \mathbf{S}_2 (z_2;p,w)\,\mathbf{S}_1 (z_1;p,wq^{h_2}),
\end{equation*} 
with $\mathbf{S}(z;p,w)=\ev_z\big[M(p,w)^{-1}\big]$ given by \eqref{S-ev}, corresponds to Baxter's Vertex-IRF transformation, which enabled him to solve the 8-Vertex model from the solution of the SOS-model in~\cite{Bax}.


\newpage 


\end{document}